\documentclass[manuscript,screen,nonacm]{acmart}

\newif\ifarxivmode
\makeatletter
\if@ACM@nonacm
    \arxivmodetrue
\else
    \arxivmodefalse
\fi
\makeatother

\AtBeginDocument{
  \providecommand\BibTeX{{
    \normalfont B\kern-0.5em{\scshape i\kern-0.25em b}\kern-0.8em\TeX}}}

\usepackage[T1]{fontenc}
\usepackage[utf8]{inputenc}
\usepackage{graphicx}
\usepackage{xurl}
\usepackage{hyperref}
\hypersetup{breaklinks=true}
\usepackage{multirow}
\usepackage{booktabs}
\usepackage{subcaption}
\usepackage{microtype}

\usepackage{amsfonts}
\usepackage{amsmath}
\usepackage{amssymb}
\usepackage{amsthm}
\usepackage{float} 
\usepackage{caption}
\usepackage{placeins}
\usepackage{soul}
\usepackage{cleveref}
\usepackage{array}
\usepackage{colortbl}

\newcommand{\arxivboxedincludegraphics}[2]{
    \ifarxivmode
        \fbox{\includegraphics[width=\dimexpr#1-2\fboxsep-2\fboxrule\relax]{#2}}
    \else
        \fbox{\includegraphics[width=#1]{#2}}
    \fi
}

\makeatletter
\ifarxivmode
        \renewenvironment{figure*}
            {\@float{figure}}
            {\end@float}
        \renewenvironment{table*}
            {\@float{table}}
            {\end@float}
\fi
\makeatother

\usepackage{todonotes}

\definecolor{thangcolor}{rgb}{0.635,0.998,0.722}
\definecolor{heikecolor}{rgb}{0.998,0.722,0.635}
\definecolor{hendrikcolor}{rgb}{0.978,0.534,0.534}
\definecolor{seminarcolor}{rgb}{0.978,0.734,0.434}
\definecolor{tablegray}{gray}{0.9}

\newcolumntype{L}[1]{>{\raggedright\let\newline\\\arraybackslash\hspace{0pt}}p{#1}}
\newcolumntype{C}[1]{>{\centering\let\newline\\\arraybackslash\hspace{0pt}}p{#1}}
\newcolumntype{G}[1]{>{\columncolor{tablegray}[0pt][0pt]\centering\let\newline\\\arraybackslash\hspace{0pt}}m{#1}}
\newcolumntype{R}[1]{>{\raggedleft\let\newline\\\arraybackslash\hspace{0pt}}p{#1}}

\newcommand{\namewithnumber}[2]{{#1 \par \footnotesize{(#2)}}}

\ifarxivmode\else
\setcopyright{acmlicensed}
\copyrightyear{2026}
\acmYear{2026}
\acmDOI{XXXXXXX.XXXXXXX}

\acmJournal{TOCHI}
\acmVolume{0}
\acmNumber{0}
\acmArticle{0}
\acmMonth{0}
\fi

\begin{document}

\title{Perceived System Predictability: Scale Development and Application}

\author{Hendrik Schuff}
\email{hendrik.schuff@ims.uni-stuttgart.de}
\orcid{0003-0523-4538}
\affiliation{
  \institution{University of Stuttgart}
  \country{Germany}}

\author{Heike Adel}
\affiliation{
  \institution{Hochschule der Medien}
  \city{Stuttgart}
  \country{Germany}}
\email{adel-vu@hdm-stuttgart.de}

\author{Ngoc Thang Vu}
\affiliation{
  \institution{University of Stuttgart}
  \country{Germany}}
\email{thang.vu@ims.uni-stuttgart.de}

\renewcommand{\shortauthors}{Schuff, Adel, and Vu}

\begin{abstract}
How predictable users perceive an interactive system to be shapes how they interpret, trust, and rely on it, yet HCI lacks both a precise conceptualization and a validated instrument for this perception.
We address this gap by introducing \emph{perceived system predictability} (PSP) as a user-centered construct grounded in uncertainty theory, distinguishing epistemic, aleatory, and effective predictability.
We contribute (i)~a theoretical framework that situates PSP relative to adjacent constructs such as trust and understanding, (ii)~a 6-item PSP scale, derived from a 60-item pool through expert review and cognitive interviews, and validated in a shape-classifier study ($N=200$) that supports both a unidimensional and a three-factor hierarchical structure, and (iii)~a sentiment-classifier study ($N=200$) that varies explanations and stochasticity, and relates PSP to the correctness of users' predictions of system behavior, trust, subjective information processing awareness, and need for cognition.
We find that PSP and prediction correctness capture distinct aspects of users' mental models and that both can diverge: PSP itself predicts correctness, explanations shift PSP but not correctness, and increased stochasticity degrades correctness without lowering PSP.
PSP thus goes beyond existing objective and subjective measures and offers a principled foundation for designing transparent and trustworthy interactive systems.\footnote{\url{http://perceived-system-predictability.de}}
\end{abstract}

\ifarxivmode\else

\fi


\ifarxivmode\else
\received{August 2025}
\received[revised]{...}
\received[accepted]{...}
\fi

\maketitle

\section{Introduction}

Interaction between users and a system can be evaluated along a broad range of criteria.
One common categorization
divides these criteria into objective and subjective criteria.
Objective criteria are measures such as the time a user spends interacting with a system or the number of successful task completions the user achieves.
Subjective criteria typically refer to constructs that users perceive, such as usability, workload, or trust.
Reliably measuring these requires the development and validation of appropriate scales. For example, well-known scales for usability include SUS \citep{brooke1996sus} and UMUX \citep{finstad2010usability,finstad_response_2013}.‚

While a large set of scales has been proposed in prior work, evaluating users' interaction with today's increasingly complex and opaque systems can only partially be addressed with existing tools and requires the exploration of novel constructs and the development of corresponding validated instruments to measure them.

In this paper, we therefore focus on conceptualizing and measuring perceived system predictability (PSP).
In the following, we (i) argue why we need to measure PSP (\Cref{sec:humans_measurement_need}), (ii) propose a theory of perceived system predictability comprising three facets (epistemic, aleatory, and effective predictability) (\Cref{sec:psp_theory}), (iii) develop a novel 6-item scale to measure PSP (\Cref{sec:pps_development}), (iv) evaluate our scale (\Cref{sec:psp_evaluation}), (v) use our scale to explore the effects of explanations and system stochasticity, and (vi) examine how PSP relates to prediction correctness, trust, subjective information processing awareness (SIPA), and participants' need for cognition (NFC), among other constructs (\Cref{sec:humans_scale_results}).

Our findings show that (a) our scale exhibits desirable psychometric properties, (b) PSP cannot be predicted from objective measures such as users' prediction correctness but itself predicts prediction correctness, (c) enriching a system's output with an explanation of why it produced that output affects PSP but not prediction correctness, and (d) the system's level of randomness, in turn, does not affect PSP but does affect prediction correctness.

\ifarxivmode\begin{table}[t]\else\begin{table}\fi
    \centering
    \resizebox{1.0\linewidth}{!}{
    \begin{tabular}{cL{0.4\textwidth}G{0.075\textwidth}@{\hspace{0.25em}}G{0.075\textwidth}@{\hspace{0.25em}}G{0.075\textwidth}@{\hspace{0.25em}}G{0.075\textwidth}@{\hspace{0.25em}}G{0.075\textwidth}@{\hspace{0.25em}}G{0.075\textwidth}@{\hspace{0.25em}}G{0.075\textwidth}}
    \toprule
         \multicolumn{2}{c}{\textbf{Statements}} &  \multicolumn{7}{c}{\textbf{Agreement rating (1-7)}} \\
          \cmidrule(lr){3-9}
            & & \cellcolor{white}\rotatebox[origin=l]{90}{\parbox{1.5cm}{strongly disagree}} & \cellcolor{white}\rotatebox[origin=l]{90}{\parbox{1.1cm}{do not agree}}& \cellcolor{white}\rotatebox[origin=l]{90}{\parbox{1.5cm}{somewhat disagree}}& \cellcolor{white}\rotatebox[origin=l]{90}{\parbox{2.3cm}{neither agree nor disagree}}& \cellcolor{white}\rotatebox[origin=l]{90}{\parbox{1.5cm}{generally agree}} & \cellcolor{white}\rotatebox[origin=l]{90}{\parbox{1.1cm}{\hspace{1cm} agree}}& \cellcolor{white}\rotatebox[origin=l]{90}{\parbox{1.5cm}{strongly agree}} \\
         \cmidrule(lr){1-2} \cmidrule(lr){3-3} \cmidrule(lr){4-4} \cmidrule(lr){5-5} \cmidrule(lr){6-6} \cmidrule(lr){7-7} \cmidrule(lr){8-8} \cmidrule(lr){9-9}
            EF-1 & \textbf{The system behaves in a predictable \hspace{0.2cm} manner.} & & & & & & & \\
            \addlinespace
            EF-2 & \textbf{I can tell which responses the system will likely give.} & & & & & & & \\
            \addlinespace
            EP-1 & \textbf{I observed enough system responses to predict how the system behaves.} & & & & & & & \\
            \addlinespace
            EP-2 & \textbf{Based on past system responses, I know the responses the system will likely give me.} & & & & & & & \\
            \addlinespace
            AL-1 & \textbf{I can tell the reasons for the system's decisions.} & & & & & & & \\
            \addlinespace
            AL-2 & \textbf{There is a consistent pattern in the system's behavior.}  & & & & & & & \\
    \bottomrule
    \end{tabular}}
    \caption{Final PSP scale with items and rating options. EF, EP, and AL refer to the effective, epistemic, and aleatory aspects of predictability covered in the scale. The overall PSP score is computed as the mean across all six items; Appendix~\ref{sec:appendix_psp_scoring_example} provides an illustrative scoring example. The choice of the seven agreement anchors builds upon recent findings on optimal conceptual anchor distances \citep{casper2020selecting}.}
    \label{tab:psps_final}
\end{table}

\subsection{The Need to Measure Perceived System Predictability}\label{sec:humans_measurement_need}

A user's ability to predict a system's behavior can be assessed both objectively and subjectively.
While objective predictability, i.e., a user's demonstrated ability to predict how the system will behave in an unseen situation, has been assessed frequently \citep{DBLP:conf/icml/GoyalWEBPL19,hase-bansal-2020-evaluating,DBLP:conf/cvpr/WangV20,DBLP:conf/iui/WangY21}, perceived predictability received less attention \citep{DBLP:conf/cogsci/SchulzTRSG15,schrills_kargl_bickel_franke_2022}.

We argue that objective predictability alone is not enough, that existing instruments are insufficient, and that PSP therefore requires a novel measurement instrument.

\subsubsection{Objective Measures of Predictability Are Not Enough.}\label{sec:objective_understanding_not_enough}
\begin{figure}
    \centering
    \includegraphics[width=\textwidth]{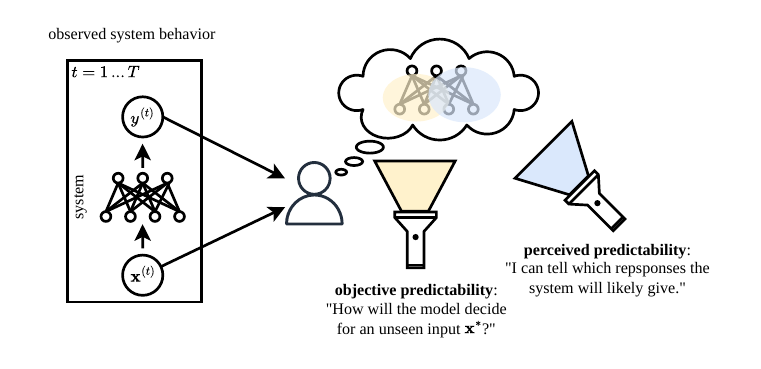}
    \caption{Objective and perceived predictability capture distinct aspects of a user's mental model of a system. The metaphor of two flashlights illustrates how these two measures illuminate different parts of that mental model, which is formed in part through the user's experience and interaction with the system, as indicated by the left box using plate notation.
    }
    \label{fig:two_flashlights}
\end{figure}

Objective predictability might appear at least as informative as subjective predictability.
However, measuring how able users \textit{feel} to predict a system's behavior can capture aspects of user interaction that are not accessible through objective predictability alone (see \Cref{fig:two_flashlights}).
To motivate this hypothesis, we draw theoretical parallels to two evaluation settings: (i) measuring system usability within human-computer interaction and (ii) measuring the feeling of learning (FOL) in educational research.
In \Cref{sec:humans_scale_predictors_psp}, we provide empirical evidence that confirms our hypothesis.

\paragraph{Subjective Versus Objective Usability}
Measuring usability is a key aspect of quantifying user experience \citep{Lewis2018MeasuringPU}.
Similarly to our context, measures related to usability can be grouped into subjective measures (e.g., obtained from questionnaires) and objective measures (e.g., task completion time) \citep{DBLP:journals/cacm/NielsenL94, HORNBAEK200679}.
\citet{DBLP:journals/cacm/NielsenL94} study how objective and subjective measures are related within a meta-analysis and found that---although there is an overall positive correlation---the two can yield contradictory conclusions.
For example, users were observed to consistently prefer an interface with which they were slower than an alternative system with which they reached faster interactions before  \citep{grudin1985adapting}.
Similarly, \citet{maclean1985evaluating} observed that users preferred a slower input method as long as it was not more than 20\%  slower than the faster alternative.
\citet{HORNBAEK200679} lists numerous arguments why both---subjective as well as objective---usability measures should be assessed.
\textit{Inter alia}, \citet{HORNBAEK200679} notes that objective and subjective measures can lead to different conclusions, for example, objective time measures and subjectively experienced time were shown to differ \citep{eisler1976experiments,TRACTINSKY2001845,czerwinski2001subjective}.
Similarly, \citet{HORNBAEK200679} mentions the dissociation of objective performance and perceived workload discussed by \citet{doi:10.1177/001872088803000110}.

\paragraph{Feeling of Learning Versus Actual Learning}
In our second example, we consider the study of active learning classroom instructions (as opposed to passive lectures) of \citet{doi:10.1073/pnas.1821936116}.
Students were evaluated on (a) what they objectively learned during a class and (b) what they \textit{felt} to have learned.
While one may expect the two measures to be strongly correlated, \citet{doi:10.1073/pnas.1821936116} find that students who participate in an active learning lecture (as opposed to a passive lecture) learn more but feel like they learn less.
The authors argue that the increased cognitive effort of active learning improves learning while also producing a cognitive disfluency that students misread as learning less.
This cognitive disfluency and the corresponding feelings of learning less can be major obstacles to the success and acceptance of active learning lectures because they threaten students' motivation and engagement \citep{doi:10.1073/pnas.1821936116}.
Measuring feeling of learning (FOL) allowed the authors to identify this disconnect and propose suitable interventions.
Without measuring FOL, they would not have been able to uncover this effect or address the resulting problems.
Similarly, measuring perceived predictability alongside objective predictability can help identify obstacles to outcomes such as user trust that would remain hidden without an adequate measurement instrument.

\begin{figure}
    \centering
    \includegraphics[width=0.9\textwidth]{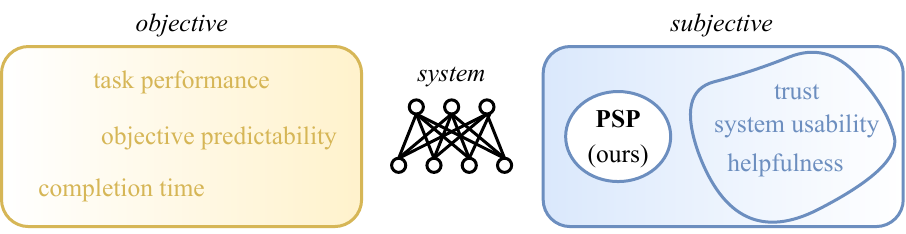}
    \caption{PSP is a subjective self-report measure. We argue that perceived predictability should be assessed alongside objective predictability and that existing subjective measures of related constructs are insufficient substitutes for PSP.}
    \label{fig:evaluation_spectrum}
\end{figure}

\subsubsection{Measuring Related Constructs Is Not Enough.}
One might argue that, instead of measuring perceived predictability directly, it is sufficient to study it indirectly via related measures of subjective constructs, such as trust, usability, or perceived helpfulness (as depicted in \Cref{fig:evaluation_spectrum}).
While a strong relation between perceived predictability and related constructs such as trust is plausible, measuring perceived predictability enables insights beyond what existing measures and constructs can capture.
Our results show that the relation between trust and objective predictability systematically differs from the relation between perceived and objective predictability (\Cref{sec:humans_scale_results}).

Prior studies on explanatory systems \citep{DBLP:conf/iui/BucincaLGG20, hase-bansal-2020-evaluating} leave open whether mismatches between subjective judgments and objective interaction outcomes stem from (a) a general dissociation between self-report and behavior or (b) commonly used subjective measures targeting constructs that are too distant from users' perceived ability to anticipate system behavior.
This second possibility deserves direct attention.
Indeed, our results show that perceived predictability is a significant predictor of objective predictability, whereas trust is negatively associated with objective predictability (\Cref{sec:humans_scale_results}).

\subsubsection{The Need for a Novel Instrument.}
Existing instruments do not yet provide an adequate validated measure of perceived predictability as a dedicated focal construct. Recent work has begun to operationalize adjacent constructs such as perceived shared understanding with AI \citep{Liang2025PerceivedSU}, but these measures target perceived shared understanding rather than users' perceived ability to anticipate how a system will behave.

\citet{DBLP:conf/cogsci/SchulzTRSG15} investigate how properties of a function (such as smoothness) affect the perceived predictability and ask participants to rate ``how well could you predict this function?'' with a 1-100 slider.
While we agree that this scale is a suitable choice within the context studied by \citet{DBLP:conf/cogsci/SchulzTRSG15}, we argue that it cannot be applied to general systems for two reasons.
First, the wording is tailored to rating ``this function,'' which is ill-suited to general systems such as chatbots.
Second, even if the wording were adapted to ``this system'', the scale lacks demonstrated psychometric validity.
Without dedicated validation, it would remain unclear whether asking ``how well could you predict this system?'' actually measures perceived predictability.
This limitation applies to any ad-hoc single-item scale (i.e., one that has not been rigorously derived from a validated multi-item instrument), as a single item in isolation does not allow estimating the item's correlation to the latent variable (perceived predictability).
When using multiple items to measure a construct, the items' correlation to the latent variable can be indirectly quantified via the item-item correlation \citep{devellis2021scale}.
Even if ultimately only a single item is retained, it is important to provide evidence that this item has a sufficiently strong relation to the construct intended to be measured.
However, subjective ratings of explanatory systems typically rely on such ad-hoc scales deployed without conducting psychometric development and validation.
While the resulting scales \textit{can} still be valid, without a dedicated analysis, there is no evidence that they actually \textit{are} valid.

An important exception to this is the SIPA scale \citep{schrills_kargl_bickel_franke_2022}. Building on the theory of situation awareness \citep{endsley1988situation}, it measures transparency, understandability, and---most importantly for our work---predictability, using two items per dimension to yield an overall SIPA score.
The SIPA scale adheres to a rigorous development process and demonstrates sound statistical properties, making it a strong point of comparison.
However, predictability constitutes only one of SIPA's three facets.
As a result, predictability is captured only through a two-item subscale, effectively treating it as unidimensional.
This leaves open whether perceived predictability could instead be measured as a multidimensional rather than a unidimensional construct, which we demonstrate to be feasible (\Cref{sec:psp_evaluation}).

\section{Related Work}\label{sec:humans_related_work}
We situate our work in relation to two strands of prior literature: (i) research on how explanations shape user judgments and behavior, and (ii) methodological work on developing validated questionnaires for user perceptions in HCI and psychology.

\subsection{Effects of Explanations on Users}\label{sec:humans_related_work_effects_on_users}
Prior work shows that explanations affect both how users judge a system and how they act on its outputs, but the size and direction of these effects vary across tasks, explanation modalities, and study designs.
In a systematic review of 73 papers, \citet{Kim2024HumancenteredEO} identify 30 components spanning in-context explanation quality, human-AI interaction, and human-AI performance in human-centered evaluations of explainable AI applications. Their review underscores that trust, understanding, usefulness, and performance are related but not interchangeable outcomes, and that evaluation practice remains methodologically heterogeneous.
Recent work further argues that static explanation formats can provide an overly simplified picture and that interactive inspection may be necessary to reveal important aspects of model behavior \citep{DBLP:journals/corr/abs-2109-07869}.

\paragraph{Calibration and Subjective Judgments}
Explanations can shape users' assessments of system performance and quality. In image classification, implausible explanations led participants to underestimate system accuracy relative to plausible explanations \citep{Nourani-Kabir-Mohseni-Ragan-2019}, whereas in stock price prediction, explanations improved users' estimates of whether classifier outputs were correct \citep{DBLP:conf/ijcai/BiranM17}. Similar over- and under-estimation effects have been observed in explainable question answering \citep{schuff-etal-2020-f1}. Explanations have also been reported to increase acceptability and perceived usefulness in some settings \citep{DBLP:conf/cscw/HerlockerKR00, DBLP:journals/umuai/CramerERSRSAW08, DBLP:conf/vl/KhuranaAC21, DBLP:conf/chi/BansalWZFNKRW21}, while evidence for trust is mixed, ranging from null findings in recommender and news settings to positive effects in chatbot settings \citep{DBLP:journals/umuai/CramerERSRSAW08, DBLP:conf/interact/RibesHPDPGS21, DBLP:conf/vl/KhuranaAC21}. Taken together, these studies suggest that explanations shape users' judgments, but not in a uniform or consistently beneficial way.

\paragraph{Agreement, Reliance, and Human-AI Performance}
A broad body of work reports improved human decisions when explanations accompany AI outputs \citep{Lundberg2018ExplainableMP, DBLP:conf/fat/LaiT19, DBLP:conf/iui/FengB19, DBLP:journals/pacmhci/GreenC19, DBLP:conf/chi/LaiLT20, DBLP:conf/fat/ZhangLB20, DBLP:conf/iui/BucincaLGG20, DBLP:conf/chi/Poursabzi-Sangdeh21}. However, later work questions how far these gains generalize and shows that explanations can also increase users' willingness to accept erroneous system outputs \citep{DBLP:conf/chi/BansalWZFNKRW21}. Related work in clinical decision support likewise documents the risk of over-reliance \citep{DBLP:conf/ichi/BussoneSO15}. This literature therefore points to a central tension: explanations may improve human-AI performance in some settings while also encouraging uncritical agreement.

\paragraph{Perceived Helpfulness Versus Actual Helpfulness}
Subjective impressions of explanations do not necessarily align with objective benefit. \citet{DBLP:conf/iui/BucincaLGG20} show that users can prefer and trust one decision-support interface even when it yields worse task performance than an alternative, and \citet{DBLP:journals/pacmhci/BucincaMG21} further identify a trade-off between subjective system quality ratings and effective human-AI performance. Comparative work likewise finds that perceived explainability correlates strongly with trust but only modestly with task accuracy, suggesting that subjective and objective effects need not move in lockstep \citep{Silva2022ExplainableAI}. Overall, the literature indicates that explanations influence both user perceptions and behavior, but their effects are heterogeneous and can diverge from, or even run counter to, objective performance.

This literature points to a question that is central to our work: to what extent do explanations help users anticipate how a system will behave? For end users, this matters because explanations are often expected to calibrate reliance, helping users avoid both uncritical acceptance and unwarranted dismissal of system outputs \citep{DBLP:conf/ichi/BussoneSO15, DBLP:conf/chi/BansalWZFNKRW21}. More broadly, explanations are commonly motivated as a means of supporting understanding and shaping users' mental models of AI systems \citep{rutjes2019-chi, DBLP:journals/csur/MadsenRC23}, and recent human-centered accounts explicitly connect understanding to prediction \citep{DBLP:journals/corr/abs-2112-04417}. Yet prior work typically addresses predictability only as a facet of broader constructs \citep{schrills_kargl_bickel_franke_2022} or via nearby outcomes such as trust and usefulness \citep{DBLP:journals/umuai/CramerERSRSAW08, DBLP:conf/vl/KhuranaAC21, DBLP:conf/chi/BansalWZFNKRW21}. We therefore argue that perceived predictability merits direct attention alongside broader outcomes such as trust, usefulness, and acceptability.

\subsection{Scale Development}\label{sec:humans_related_work_scale_development}
Validated questionnaires are a prerequisite for reliably measuring user perceptions in HCI, psychology, and adjacent fields, for example for usability, trust, cognitive load, social attribution, and user interface language quality \citep{brooke1996sus, finstad2010usability, korber2018theoretical, hart1988development, carpinella_robotic_2017, bargas2016measuring}.
Developing such instruments requires established psychometric practice, including careful item generation, expert review, target-population feedback, and quantitative validation \citep{boateng_best_2018, devellis2021scale}.
Recent scale-development work has produced validated instruments for adjacent user-centered constructs in human-AI interaction, including perceived shared understanding with AI and system trustworthiness \citep{Liang2025PerceivedSU, Alarcon2023DevelopmentAV}. Together with SIPA, these measures show growing interest in related constructs, but they do not directly operationalize perceived system predictability as the focal construct. Perceived shared understanding centers on whether users feel that they and the AI are on the same page, system trustworthiness targets evaluations of performance, purpose, and process.
However, to the best of our knowledge, no validated scale currently measures perceived system predictability or its dimensions.
Prior work therefore leaves open how perceived predictability can be assessed rigorously as a user-centered outcome.

\section{Scale Development and Validation}\label{sec:humans_scale_development}
We define perceived predictability as \textit{the degree to which a user feels able to predict how a system behaves}.
We develop a multidimensional theory of perceived predictability that distinguishes effective, epistemic, and aleatory predictability, drawing on structured target-population interviews and uncertainty theory.

\subsection{A Theory of Perceived Predictability}\label{sec:psp_theory}
\subsubsection{Target Population Interviews.}
As a first step, we empirically validate that the construct of system predictability is salient in our target population's mental models of automated, in particular, artificial intelligence (AI) systems.
We ask 20 crowdworkers on Amazon Mechanical Turk (MTurk) to broadly state what ``understanding an AI system'' means to them, without explicitly prompting for predictability. We recruit crowdworkers from the US, Australia, and the UK within this and the following studies.
We find that, besides transparency, technical details, and intended usage, perceived predictability is intuitively mentioned by the majority of participants, confirming that predictability is a natural and central aspect of users' system perception.

Even at this initial stage, participant comments point towards a nuanced understanding of predictability. For instance, one participant notes:
\begin{quote}
\textit{``I would not trust a system that behaved randomly unless it was 'controlled random.' [...] Like if a system was programmed to randomly select from a set of pictures for example''}
\end{quote}
Such remarks suggest that users intuitively distinguish between different underlying dimensions of predictability.

In our second step, we ask another 20 participants what ``having the feeling that you can predict how an AI system behaves'' means to them in order to obtain a more focused picture of perceived predictability.
Again, we observe responses suggesting multiple aspects of predictability:
\begin{quote}
\textit{``[...] it is following a set of rules [...], so given a certain input it should always produce the same output. I feel that as I gained more experience with the system it would become more and more predictable to me''}
\end{quote}
The first sentence can be interpreted as reflecting \textit{aleatory} uncertainty (i.e., uncertainty arising from a phenomenon's inherent randomness) and the second as reflecting \textit{epistemic} uncertainty (i.e., uncertainty arising from a lack of observations). We discuss both terms and their relation to our theory of PSP in \Cref{sec:uncertainty_theory} below.
\Cref{tab:target_population_comments_three_types} shows a list of target group participant comments.

\begin{table}
    \centering
    \resizebox{\linewidth}{!}{
    \begin{tabular}{C{0.05\textwidth}L{1.04\textwidth}}
    \toprule
          \multirow{6}{*}{\rotatebox[origin=c]{90}{epistemic}}  &  -- \textit{I feel like I can predict an AI system's behavior when I have interacted with it many times before.}\\
          &  -- \textit{If I use the AI frequently, I know what I can and cannot ask or do with the AI.}\\
          &  -- \textit{Being able to anticipate how it will respond after some moderate use [...].}\\
          &  -- \textit{I feel like I can predict how an AI system behaves the more I interact with it.}\\
          &  -- \textit{I feel that as I gained more experience with the system it would become more and more predictable to me.}\\
          & -- \textit{knowing that result it is likely to give me, either based on it's past responses or my own assumptions}\\
          \midrule
          \multirow{5}{*}{\rotatebox[origin=c]{90}{aleatory}} & -- \textit{[...] given a certain input it should always product [sic] the same output.}\\
          & -- \textit{means to me that the AI normally behaves in a consistent manner}\\
          & -- \textit{It may start following a certain pattern and I get an idea of how the algorithm works.}\\
          & -- \textit{I would definitely want such a system to give consistent and reliable results.}\\
          & -- \textit{I have a basic understand of the various rules or conditions that the AI system uses to make it's judgments}\\
          \midrule
          \multirow{5}{*}{\rotatebox[origin=c]{90}{effective}} & -- \textit{its more of like you can guess how it will react.}\\
          & -- \textit{It means that I can guestimate how it comes to its conclusions.}\\
          & -- \textit{I can know what to expect.}\\
          & -- \textit{I can usually predict if the AI can answer the question I have in mind for it.}\\
          & -- \textit{a good overall understanding of the AI}\\
        \bottomrule
    \end{tabular}
    }
    \caption{Target population comments on what ``having the feeling that you can predict how an AI system behaves'' means to individual participants. We identify three aspects of perceived predictability: epistemic, aleatory, and effective predictability.}
    \label{tab:target_population_comments_three_types}
\end{table}

An additional observation is that users relate a slight level of unpredictability with a preferable or more powerful system by comments, such as:
\begin{quote}
\textit{``If it's too predictable then I wonder what the point of having it is. The same as if it is too unpredictable. Sometimes like on a racing game with really good AI you can have races that you really can't tell if it is a human or a bot. The same goes for chatbots that people test on MTurk. Some of them are uncanny at how real they are. I guess to sum it up AI needs to be just unpredictable enough and in the right way for me to like it and see a need for it.''}
\end{quote}
or
\begin{quote}
\textit{``with AI currently it seems you can usually within some reason, predict how the AI will answer or what actions it will give, i think its because its not a true AI as it stands. there is no conscious thought, just the data we gave it to act like its own self, which its just a shell of that. i do think in the somewhat future, not sure on distant or close, that AI will have its own conscious thought and then be slightly unpredictable in its answers/actions''}
\end{quote}
We revisit this effect in light of our quantitative results in \Cref{sec:humans_scale_results}.

\subsubsection{Uncertainty Theory: Epistemic and Aleatory Uncertainty.}\label{sec:uncertainty_theory}
Our evaluation of target population comments indicates that perceived predictability has multiple facets.
These observed categories can be related to epistemic and aleatory uncertainty.
In uncertainty theory, \citet{fox2011distinguishing} distinguish two types of uncertainty: (a) \textit{epistemic} uncertainty that relates to the uncertainty of not knowing something that could be known, e.g. due to a lack of observations of a phenomenon and (b) \textit{aleatory} uncertainty that refers to a phenomenon's inherent stochasticity and that cannot be addressed with a higher number of observations.
For example, we can have perfect epistemic certainty about how a die functions and yet be unable to predict the outcome of rolling dice due to the aleatory uncertainty arising from the dice's randomness.

\begin{figure}
    \centering
    \includegraphics[width=0.8\textwidth]{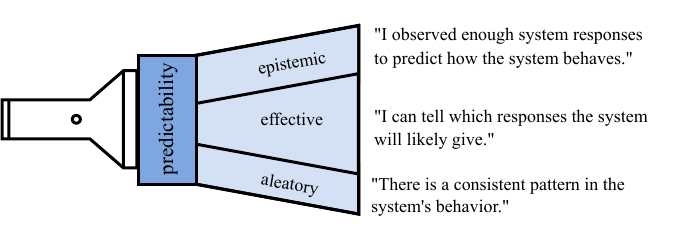}
    \caption{We decompose perceived predictability into three (partially overlapping) facets.
    In contrast to uncertainty in statistics, in which the total uncertainty is the sum of the epistemic and the aleatory uncertainty in terms of variances, our notion of ``effective'' predictability includes additional information beyond epistemic and aleatory predictability. Representative items from our final PSP scale are added to their respective facet on the right side of the figure.}
    \label{fig:three_beams}
\end{figure}

\subsubsection{Dimensions of Predictability: Epistemic, Aleatory, and Effective.}
The concepts of epistemic and aleatory uncertainty can be transferred to predictability (objective and perceived), as failing to predict a system's behavior, or feeling unable to do so, can arise from (i) a lack of system behavior observations and/or (ii) inconsistencies in the system's behavior.
Regarding epistemic uncertainty, in the extreme case, if users had no exposure to a system at all, their (perceived) prediction abilities would be reduced to their general notion of an unknown system's predictability.
On the contrary, having observed a theoretically infinite number of system decisions removes this barrier to predictability.
Regarding aleatory uncertainty, in the extreme case, the system takes completely random decisions.
In this case, users will not be able to predict the system's behavior, even with access to infinite observations.
In contrast, a completely deterministic system can (in theory) be fully predictable by observing all possible contexts before making the prediction.
For uncertainty, epistemic and aleatory uncertainty can be summed in terms of variances to yield the overall uncertainty.
For perceived predictability, however, we argue that this additive view does not hold, and \textit{effective} predictability covers more than the sum of epistemic and aleatory predictability.
We illustrate our theory in \Cref{fig:three_beams}.

\subsection{Scale Development}\label{sec:pps_development}
We develop the scale in multiple stages following established best practices for item development and validation \citep{boateng_best_2018,menold2016design,devellis2021scale}.
\Cref{fig:scale_development_process} displays an overview of the separate development stages and the respective number of participants and retained scale items.

\subsubsection{Initial Item Pool.}
We start with an initial item pool of 60 candidate items based on (i) the target population interviews described above and (ii) our proposed theory of perceived predictability.
We report the full item pool in Appendix~\ref{sec:appendix_psp_item_pool}.

\begin{figure}
    \centering
    \includegraphics[width=0.9\textwidth]{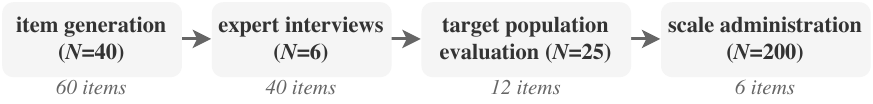}
    \caption{Overview of our scale development process. For each stage, we report the number of participants in parentheses and the resulting number of items retained in our scale in \textit{italics}.}
    \label{fig:scale_development_process}
\end{figure}

\subsubsection{Expert Ratings.}

We ask experts to review each item within our initial item pool in order to ensure content validity, i.e., that our items capture the intended domain of perceived predictability.
We had six experts rate our initial item pool in terms of relevance and clarity and collect additional textual feedback from each rater for each item.
The experts are researchers with a background in artificial intelligence, natural language processing, and human-computer interaction.
Based on the experts' ratings, we remove eight items found to be irrelevant and 13 items found to be unclear.
We additionally modify the wording of five items and add one new item based on expert feedback.
Our revised item pool thus contains 40 items.

\subsubsection{Target Population Evaluation.}
To assess the face validity of our scale (i.e., appropriateness for the target population), we conduct a crowdsourced adaptation of cognitive interviews \citep{beatty2007research} using probes from \citet{willis2004cognitive} with 25 participants.
In the first round, we present predictions of a hypothetical classification system to 20 MTurk crowdworkers (12 female, 8 male; mean age 42.0 years, SD=11.1).
We ask participants to rate their agreement with each item on a 7-point Likert scale ranging from ``strongly disagree'' (1) to ``strongly agree'' (7).
Details of the classification system and items are discussed in \Cref{sec:psp_evaluation}.
Additionally, we ask crowdworkers to (i) rephrase each item in their own words and (ii) explain how they arrived at their answer.
These responses allow us to identify ambiguous or unclear items.
Based on this feedback, we remove 15 items, modify five, and add one new item.
We merge similar items, streamline wording, and add items based on further discussion, resulting in a pool of 16 items.
\paragraph{Full Verbalization and Optimized Response Anchors}
During interviews, we observed discrepancies between participants' intended neutral ratings and their numerical selections (e.g., verbalizing neutrality but selecting 5).
To address this, we adopted a fully verbalized 7-point scale, as recommended by \citet{menold2016design}.
We selected response anchors based on \citet{casper2020selecting} to aim for minimal overlap and approximately equal intervals: ``strongly disagree'', ``do not agree'', ``somewhat disagree'', ``neither agree nor disagree'', ``generally agree'', ``agree'', and ``strongly agree''.
We then conducted a second round of cognitive interviews with five crowdworkers (3 female, 2 male; mean age 43.6 years, SD=8.4).
Based on their feedback, we removed four items and modified one, resulting in a final pool of 12 items (see Appendix~\ref{sec:appendix_psp_item_intermediate_pool}).

\subsection{Scale Evaluation}\label{sec:psp_evaluation}
To further reduce the item pool and assess the scale's psychometric properties, we collect ratings from 200 participants across five predictability scenarios of a fictional classification system.

\begin{figure}
    \centering
    \includegraphics[width=0.45\textwidth]{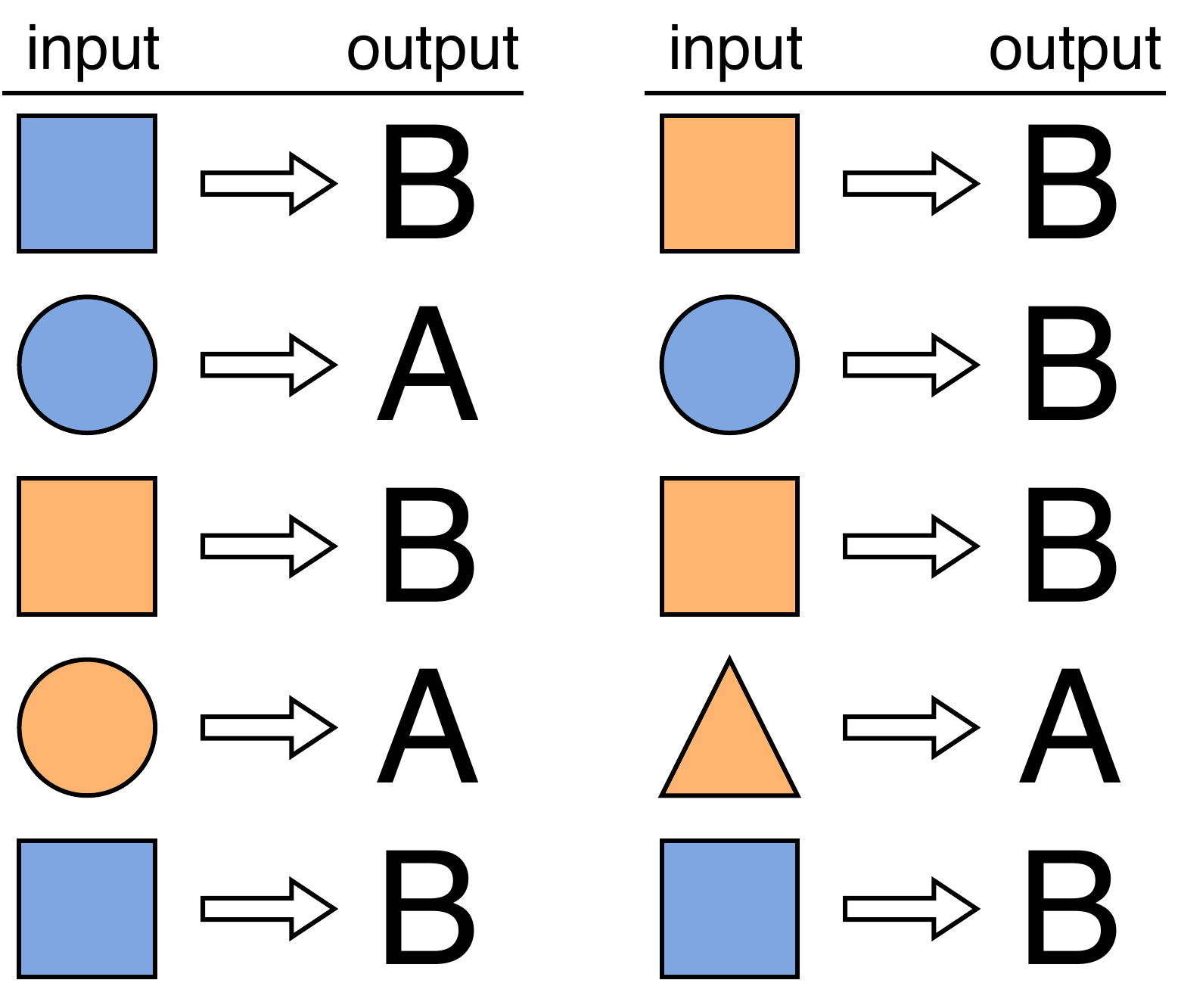}
    \caption{One of the five fictional classification system prediction scenarios we show to participants.
    The class prediction for the blue circle is inconsistent, introducing aleatory uncertainty in addition to the epistemic uncertainty caused, for example, by the lack of blue triangle inputs. We refer to this scenario as \textsc{mixed}.}
    \label{fig:scenario_mixed}
\end{figure}

\subsubsection{Scenarios.}
To elicit varying levels of perceived predictability, we designed five versions of a fictional classification system that maps colored shapes (red/orange circles, squares, triangles) to classes A or B.
We chose abstract shapes and class labels to prevent users from relying on prior knowledge or intuition about correct classifications.
We revisit scale ratings with a real AI system in \Cref{sec:humans_scale_results}.
The five scenarios vary by (a) the number of observed predictions and (b) the system's stochasticity.
\Cref{fig:scenario_mixed} illustrates the \textsc{mixed} scenario, which combines epistemic uncertainty (sparse examples) and aleatory uncertainty (inconsistent predictions for blue circles).
The remaining scenarios are detailed in Appendix~\ref{sec:appendix_psp_colored_shapes_scenarios}.
After viewing the scenario's examples, participants predicted the system's output for three test inputs (\Cref{fig:symbols_to_predict}).
This task ensured participants engaged with the system's logic to build a level of perceived predictability before completing the scale.

\begin{figure*}
    \centering
    \begin{subfigure}[t]{.31\textwidth}
        \centering
    \includegraphics[width=0.3\textwidth]{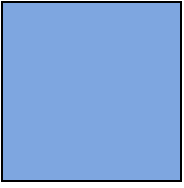}
    \caption{Blue square (predictable with high certainty).}
    \end{subfigure}
    \hfill
    \begin{subfigure}[t]{.31\textwidth}
        \centering
    \includegraphics[width=0.3\textwidth]{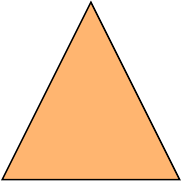}
    \caption{Orange triangle (predictable with lower certainty).}
    \end{subfigure}
    \hfill
    \begin{subfigure}[t]{.31\textwidth}
        \centering
    \includegraphics[width=0.3\textwidth]{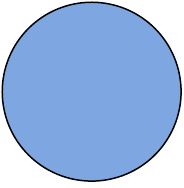}
    \caption{Blue circle (unpredictable).}
    \end{subfigure}
    \caption{The three symbols we ask users to predict the system's output for. Given the scenario shown in \Cref{fig:scenario_mixed}, the system response for the first two shapes can be predicted (with different levels of certainty) and the system response to the third input is unpredictable.}\label{fig:symbols_to_predict}
\end{figure*}

\subsubsection{Scale Rating.}
Participants then rated the twelve candidate scale items.
We randomized the item order to minimize carry-over effects and rating patterns that could confound psychometric estimates (e.g., item discrimination).
The final scale uses a fixed order, which we evaluate in \Cref{sec:humans_scale_results}.

\subsubsection{Participants.}
We recruited 200 participants from the United States, Australia, and the United Kingdom on MTurk (80 female, 119 male, 1 non-binary; mean age 40.8 years, SD=11.2).

\subsubsection{Item Reduction.}
We assess items along inter-item correlations, item-total correlations, item discrimination, and item difficulty. We reduce the scale to six items based on quantitative indicators as well as semantic overlap.
We report the reasons for each of our removal decisions in Appendix~\ref{sec:appendix_psp_item_intermediate_pool}.
\Cref{tab:item_analysis} displays item difficulty and item discrimination values for the items retained in our final scale.
To ensure the scale is applicable across diverse research contexts, we prioritized brevity while maintaining sufficient granularity.
We retained two items per dimension to allow for reliability estimation and to validate their use as independent subscales.
The resulting 6-item PSP scale is analyzed below.

\subsubsection{Reliability.}
The overall Cronbach's $\alpha$ \citep{Cronbach1951CoefficientAA} and coefficient $\omega$ \citep{Raykov2001EstimationOC} of our PSP scale both equal 0.961.\footnote{We use a CFA-based calculation of coefficient $\omega$ as described by \citet{furr2022psychometrics} (p. 515) using the R package semTools \citep{semTools}. We additionally compute McDonald's $\omega$ \citep{Mcdonald1999TestTA} indices $\omega_{h}$ and $\omega_{t}$ based on hierarchical factor analysis using the R package psych \citep{psych}, the resulting values are 0.894 and 0.967, respectively.}
Following the evaluation of the SIPA scale \citep{schrills_kargl_bickel_franke_2022}, we use Spearman-Brown coefficients to quantify the reliability of the two-item subscales.
\citet{Eisinga2013TheRO} recommend the Spearman-Brown coefficient as the reliability score of choice for two-item (sub)scales.
We obtained coefficients of $R=0.908$ for the epistemic, $R=0.889$ for the aleatory, and $R=0.881$ for the effective subscale.
These results demonstrate high internal consistency for both the overall scale and its individual dimensions.

\begin{table}[t]
    \centering
    \resizebox{1.0\linewidth}{!}{
    \begin{tabular}{cL{12cm}cc}
        \toprule
        \multicolumn{2}{l}{\textbf{Item}} & \textbf{Difficulty} & \textbf{Discrimination} \\
        \midrule
       EF-1 & The system behaves in a predictable manner.                                               & 0.72 & 0.89 \\
       EF-2 & I can tell which responses the system will likely give.                                   & 0.71 & 0.86 \\
       \addlinespace
       EP-1 & I observed enough system responses to predict how the system behaves.                     & 0.67 & 0.87 \\
       EP-2 & Based on past system responses, I know the responses the system will likely give me.      & 0.70 & 0.90 \\
       \addlinespace
       AL-1 & I can tell the reasons for the system's decisions.                                        & 0.65 & 0.84 \\
       AL-2 & There is a consistent pattern in the system's behavior.                                   & 0.70 & 0.89 \\
       \bottomrule
    \end{tabular}
    }
    \caption{Item difficulty and item discrimination values of the items in our final scale (mean inter-item-correlation = 0.804, Cronbach's $\alpha$=0.961). EF refers to the effective, EP to the epistemic, and AL to the aleatory PSP subscales.}
    \label{tab:item_analysis}
\end{table}

\subsubsection{Confirmatory Factor Analysis.}

\begin{figure*}
     \centering
     \begin{subfigure}[t]{.49\textwidth}
         \centering
     \includegraphics[width=0.8\textwidth]{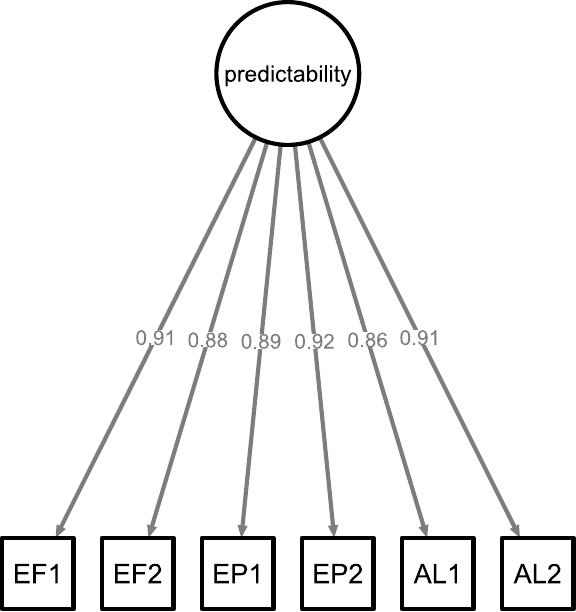}
     \caption{One-factor model.}\label{fig:models_unidimensional}
     \end{subfigure}
     \hfill
     \begin{subfigure}[t]{.49\textwidth}
         \centering
     \includegraphics[width=0.8\textwidth]{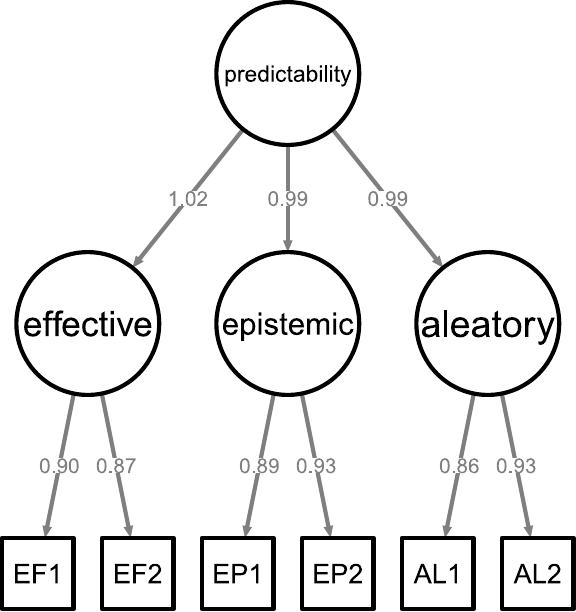}
     \caption{Three-factor model.}\label{fig:models_higher_order}
     \end{subfigure}
     \caption{The two models we compare in confirmatory factor analysis (CFA) along with standardized coefficients. The rectangular boxes correspond to the six items of our scale as denoted in \Cref{tab:item_analysis}.
     Note that standardized coefficients are not bound to be less than or equal to one as discussed by \citet{Deegan1978OnTO} and \citet{joreskog1999large}.}\label{fig:models_cfa}
 \end{figure*}

We conduct CFA to compare two models: a unidimensional model (\Cref{fig:models_unidimensional}) and a three-factor model (\Cref{fig:models_higher_order}).\footnote{Note that the three-factor model is visualized as a higher-order model. With three first-order dimensions, a higher-order model is just-identified and equivalent to a correlated factors model with three dimensions.}
We report common model fit measures in \Cref{tab:cfa_fits} and detailed model fits in Appendix~\ref{sec:appendix_psp_cfa}.
\citet{Hu1999CutoffCF} proposed the now widespread cutoff criteria for various fit indices: 0.06 for RMSEA, 0.08 for SRMR, and 0.95 for CFI and TLI.
Both models fulfill all recommended criteria.
While the three-factor model yields slightly better exact fit measures (CFI, TLI, RMSEA), the one-factor model is favored by parsimony indices (AIC, BIC).
Given the minor differences in fit and the strong correlations between the three scale dimensions (see below), statistical evidence supports a unidimensional structure.
Concretely, we observe Pearson correlations of 0.901  ($p<0.001$) between effective and epistemic predictability, 0.889 ($p<0.001$) between effective and aleatory predictability, and 0.876 ($p<0.001$) between epistemic and aleatory predictability.\footnote{Pearson correlations are calculated using list-wise deletion as implemented in sjPlot \citep{sjplot}.}
Overall, together with the identical differentiation patterns across the subscales and the total scale, our scale evaluation indicates that a unifactorial usage of our scale (i.e., measuring perceived predictability as the mean value over all six items) is warranted.
However, we retain the theoretical distinction of the three-factor model as it allows for a more granular evaluation in applications where levels of epistemic and aleatory predictability are expected to differ, for example, when users have a very high epistemic predictability (stemming from, e.g., a long usage period) but very low aleatory predictability.

 \begin{table}[t!]
     \centering
     \begin{tabular}{lcc}
        \toprule
        \textbf{Fit Measure} & \multicolumn{2}{c}{\textbf{Model}}\\
        \cmidrule(lr){2-3}
         &   One-factor  &   Three-factor  \\
        \midrule
        $\chi^2$ ($\downarrow$)   & 12.67	($p$=0.18)     & 7.14 ($p$=0.31)     \\
        RMSEA ($\downarrow$)      & 0.045 ([0.000, 0.098])    & 0.031 ([0.000, .101])     \\
        SRMR ($\downarrow$)       & 0.010                   & 0.008                    \\
        CFI ($\uparrow$)          & 0.997                   & 0.999                    \\
        TLI ($\uparrow$)          & 0.995                   & 0.998                    \\
        AIC  ($\downarrow$)       & 3354.034               & 3354.502                \\
        BIC ($\downarrow$)        & 3393.613               & 3403.977                \\
        Adj. BIC ($\downarrow$)   & 3355.596               & 3356.455                \\
        \bottomrule
     \end{tabular}
     \caption{Fit indices reported per model following \citet{Dunn2020ThePO}. Higher-is-better fit measures are marked with ($\uparrow$) while lower-is-better fit measures are marked with ($\downarrow$). While exact fit measures (RMSEA, CFI, TLI) slightly favor the three-factor model, parsimony indices (AIC, BIC) favor the one-factor model. The RMSEA value is reported along with a 90\% confidence interval.}
     \label{tab:cfa_fits}
 \end{table}

\subsubsection{Differentiation by Known Groups.}
To assess construct validity, we evaluate the scale's ability to differentiate between groups expected to differ in the construct of interest (known-groups validity).
Specifically, we test if PSP scores vary across the five classifier scenarios designed to induce different levels of perceived predictability.
We find strong evidence for the scale's known-groups validity: A one-way ANOVA reveals a statistically significant and large main effect of the scenario on total PSP scores ($F(4, 195)=16.25$, $p < 0.001$; $\eta^2 = 0.25$, 95\% CI [0.16, 1.00]).
This demonstrates that the scale is sensitive to the manipulated levels of predictability.
Post-hoc comparisons using Tukey's HSD test indicate significant differences for six out of ten scenario pairs.
The four non-significant pairs correspond to scenarios with theoretical, yet smaller, differences in predictability (e.g., \textsc{mixed} vs. \textsc{mixed-less-aleatory}).
These differences are smaller than for the other scenario contrasts.
The lack of statistical significance suggests that these subtle differences were not perceived as distinct by participants.
Detailed statistics are provided in Appendix~\ref{sec:appendix_psp_known_groups}.

\subsubsection{Concurrent and Predictive Validity.}
In an additional study, we assess how PSP scores associate with related measures.
While \Cref{sec:humans_scale_results} details the relation of PSP to trust, objective predictability, and NFC, we briefly summarize the key findings regarding concurrent and predictive validity here.
Regarding concurrent validity, SIPA scores correlate strongly with PSP scores ($r=0.856$, $p<0.001$).
However, this correlation is weaker than the internal correlations among PSP subscales (average $r=0.889$, as reported above).
This aligns with our expectation, as the two scales share an overlapping but distinct theoretical foundation.
Regarding predictive validity, we find that PSP is a significant non-linear predictor of objective predictability, whereas we observe no significant association between SIPA and objective predictability.
This confirms our theoretical expectation: While PSP scores should relate to objective predictability, SIPA is a higher-level construct that we expected to show a much weaker or no relation to objective predictability.

\section{Drivers, Consequences, and Objective Predictability}\label{sec:humans_scale_results}

Having developed and validated the PSP scale using the fictional shape classification setting, we now examine its properties in a realistic yet general application context.
Specifically, we investigate the relationship between PSP and objective predictability, trust, SIPA, and Need for Cognition (NFC) in the context of a sentiment classification system.
To explore PSP and its nomological network (i.e., its relationships with related constructs) in a range of realistic settings, we vary objective predictability by evaluating system variants with different levels of stochasticity.
Furthermore, since explanations are a key mechanism for modulating predictability, we vary the form of explanations provided (saliency maps, bar charts, or none) to examine their effect on PSP.

\subsection{Experiment Design}
We conduct an additional user study with 200 participants to investigate PSP and its relationship with related constructs in the context of a simple yet real sentiment classification system.
We deliberately use a transparent text classifier rather than a contemporary black-box model because the goal of this study is construct validation: we want to vary predictability and explanation access without simultaneously introducing uncertainty about whether the communicated explanation is faithful to the underlying model.

\subsubsection{Sentiment Classification System.}
Our sentiment classifier aggregates word polarity scores from SentiWordNet3\footnote{\url{https://github.com/aesuli/SentiWordNet}} \citep{baccianella-etal-2010-sentiwordnet} to compute an overall sentiment score.
Scores greater than zero indicate a positive prediction. Otherwise, the prediction is negative.
We selected this straightforward system for three reasons: (i) it is inherently interpretable, such that word polarity scores coincide with the model's decision rule and therefore provide faithful explanations, (ii) it produces identifiable systematic errors (e.g., failing to handle negations or contractions), and (iii) it supports on-demand execution, which is necessary for the interactive interface we explore in this study.
This choice is deliberate for a second reason as well: PSP is intended as a model-agnostic construct, and validating it on a transparent system first avoids narrowing the scale to the idiosyncrasies of one contemporary model family.
By contrast, using a large language model or another opaque system would have required a post-hoc explanation method whose own faithfulness and communicative adequacy could affect participants' judgments, making it harder to attribute observed effects to perceived predictability rather than to the explanation pipeline itself.
We therefore treat this sentiment classifier as a conservative test bed that prioritizes internal validity while still constituting a real AI system in the sense that participants interact with a functioning model and observe its actual outputs and errors.
To introduce stochasticity, we create system variants by adding independent, normally distributed noise to each word's polarity score.
Specifically, we sample noise from $\mathcal{N}(0, 1)$, scaled by 0.4 for medium noise or 0.8 for high noise, and clip the resulting sum to the range [-1, 1].

\subsubsection{Explanation Modalities.}
We compare six explanation modalities.
First, we evaluate (i) no explanations, (ii) heatmap explanations, and (iii) bar chart explanations (\Cref{fig:base_modalities}).
Heatmap and bar chart explanations display absolute importance values, meaning we do not communicate class-specific importance scores.
In addition, we include interactive versions of these three types.
Concretely, we provide an interface allowing users to enter arbitrary text and receive the sentiment prediction and—for heatmap and bar chart conditions—the corresponding explanation.
\Cref{fig:interactive_modalities} shows these interactive interfaces.

\begin{figure*}
     \centering
     \begin{subfigure}[t]{.3\textwidth}
         \centering
     \arxivboxedincludegraphics{\linewidth}{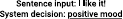}
     \caption{No explanation.}\label{fig:explanation_modality_no_explanation}
     \end{subfigure}
     \hfill
     \begin{subfigure}[t]{.3\textwidth}
         \centering
     \arxivboxedincludegraphics{\linewidth}{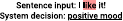}
     \caption{Heatmap explanation.}\label{fig:explanation_modality_saliency}
     \end{subfigure}
     \hfill
     \begin{subfigure}[t]{.3\textwidth}
         \centering
     \arxivboxedincludegraphics{\linewidth}{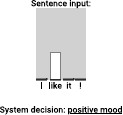}
     \caption{Bar chart explanation.}\label{fig:explanation_modality_bar_chart}
     \end{subfigure}
     \caption{The three explanations forms underlying the six explanation modalities used in our experiment.}\label{fig:base_modalities}
 \end{figure*}

  \begin{figure*}
     \centering
     \begin{subfigure}[t]{.8\textwidth}
         \centering
     \arxivboxedincludegraphics{\linewidth}{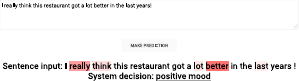}
     \caption{Interactive heatmap explanation.}\label{fig:explanation_modality_saliency_interactive}
     \end{subfigure}
     \begin{subfigure}[t]{.8\textwidth}
         \centering
     \arxivboxedincludegraphics{\linewidth}{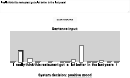}
     \caption{Interactive bar chart explanation.}\label{fig:explanation_modality_bar_chart_interactive}
     \end{subfigure}
     \caption{Two of the three additional interactive explanation modalities.}\label{fig:interactive_modalities}
 \end{figure*}

 \begin{figure}
    \centering
    \arxivboxedincludegraphics{0.5\textwidth}{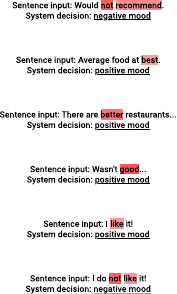}
    \caption{Subset of the system predictions shown to users in the heatmap conditions.}
    \label{fig:screenshot_study_interface_understanding_saliencies_short}
\end{figure}

\subsubsection{Procedure.}
We use a between-subjects experimental design and assign 25 participants to each of the six explanation modalities, using the noiseless prediction system.
To explore the effect of noisy systems, we additionally assign 25 participants each to the medium noise and high noise system variants, using system prediction examples without explanations or an interactive interface.
We ask each participant to complete three phases of our experiment.

\paragraph{Training Phase}
In the first step, we show each participant 20 system predictions using the respective explanation modalities.
For the interactive modalities, we present participants the same 20 examples of fictional restaurant reviews and \textit{additionally} provide them with the interactive interface to give them the same context and ensure they receive the same information regarding the system's behavior and quirks.
\Cref{fig:screenshot_study_interface_understanding_saliencies_short} depicts six of the 20 example predictions we display to participants in the heatmap conditions.
We provide the full list of examples in Appendix~\ref{sec:appendix_psp_explanation_experiments}.
Each participant receives the same 20 prediction examples but in a randomized order to mitigate carry-over effects.
We compose the 20 fictional review sentences in such a way that half of them correspond to a positive system decision, and half of them to a negative decision, and such that each of these halves contains five correct system decisions and five incorrect system decisions.
Further, the examples are chosen in a way that demonstrates that the model treats contractions, such as ``don't'' differently from ``do not'' and does not correctly resolve negations.
In addition, the examples contain two sentences that are each repeated to make participants aware of (non-)deterministic model behavior, which is relevant for the noisy systems described above.

\paragraph{Prediction Phase}
In the second step, we ask participants to predict which predictions the system will make for new, unseen sentence inputs (e.g., ``I love the food at this place!'' and ``I expected it to be better.'').
We provide the full list of sentences in Appendix~\ref{sec:appendix_psp_explanation_experiments}.
We again randomize the order of sentences across participants.

\paragraph{Questionnaire Phase}
In the third step, we ask participants for self-reports of (i) PSP using our scale, (ii) SIPA using the respective scale proposed by \citet{schrills_kargl_bickel_franke_2022}, (iii) system trust using the facets of system trustworthiness (FOST) scale \citep{DBLP:conf/automotiveUI/FrankeTGKZK15}, and (iv) participants' individual NFC using the NCS-6 scale \citep{LinsdeHolandaCoelho2018TheVE}.
We choose to include a measure of NFC based on \citet{DBLP:journals/pacmhci/BucincaMG21} who find that their intervention on explanation presentation to reduce over-reliance disproportionately affected participants who reported a high NFC.

\subsubsection{Participants.}
We recruited 200 participants from the United States, Australia, and the United Kingdom on MTurk (89 female, 110 male, 1 non-binary; mean age 43.2 years, SD=12.2).
Across the entire experiment, the population's mean PSP rating was 5.39 (SD=0.89).

\subsection{Reliability Reproduction}
Before analyzing the relation between constructs and objective performance measures, we re-assess each scale's internal reliability.
\Cref{tab:scale_reliabilities} reports Cronbach's $\alpha$ \citep{Cronbach1951CoefficientAA} and McDonald's $\omega$ \citep{Mcdonald1999TestTA}.
As in our first reliability assessment, we quantify the reliability of the PSP subscales using the Spearman-Brown coefficient and find $R=0.718$ for the epistemic subscale, $R=0.726$ for the aleatory subscale, and $R=0.698$ for the effective subscale, indicating high reliability across subscales.

\begin{table}[t]
    \centering
    \small
    \begin{tabular}{lcccc}
        \toprule
         \textbf{Reliability measure} & \multicolumn{4}{c}{\textbf{Scale}}  \\
         \cmidrule(lr){2-5}
                             & PSP      & SIPA      & FOST      & NCS-6 \\
        \midrule
         $\alpha$            & 0.874     & 0.852      & 0.780      & 0.766  \\
         $\omega_h$          & 0.713     & 0.593      & 0.100      & 0.392  \\
         $\omega_t$          & 0.896     & 0.886      & 0.830      & 0.834  \\
         \bottomrule
    \end{tabular}
    \caption{All four scales demonstrate adequate internal reliability ($N=200$), with Cronbach's $\alpha$ \citep{Cronbach1951CoefficientAA} exceeding 0.75. Notably, our PSP scale achieves the highest reliability across all measures. $\omega_h$ and $\omega_t$ denote hierarchical and total McDonald's $\omega$ \citep{Mcdonald1999TestTA}.}
    \label{tab:scale_reliabilities}
\end{table}

\subsection{Predictors of Perceived Predictability}\label{sec:humans_scale_predictors_psp}
Next, we examine which factors predict PSP scores.
In particular, we test whether PSP scores can be predicted from objective measures such as completion time or prediction correctness.
To this end, we model PSP scores using a generalized additive model (GAM).

\subsubsection{Model.}
GAMs allow us to model additive, smooth, non-linear effects of numeric covariates.
We include smooth terms for prediction correctness, completion time during the prediction phase, FOST trust scores, SIPA scores, NCS-6 NFC scores, and participant age.
We also include parametric terms for explanation format (none, saliency, and bar charts), interactivity\footnote{We find that 21.3\% of participants in the interactive conditions did not use the interactive prediction interface. We therefore define the ``interactivity'' factor to distinguish between participants who interacted with a prediction interface and those who did not, including participants who could have used the interface but chose not to.}, noise level, and participant identification to account for the experiment design and control for potential confounds.
We include gender identification and participant age solely as control covariates to account for potential confounds; consistent with our expectations, neither showed a significant effect in any model.
Finally, we include an interaction term between explanation format and interactivity, as we expect different explanation formats to induce different levels of interaction motivation.

\begin{figure*}
     \centering
     \begin{subfigure}[t]{.32\textwidth}
         \centering
     \includegraphics[width=\textwidth]{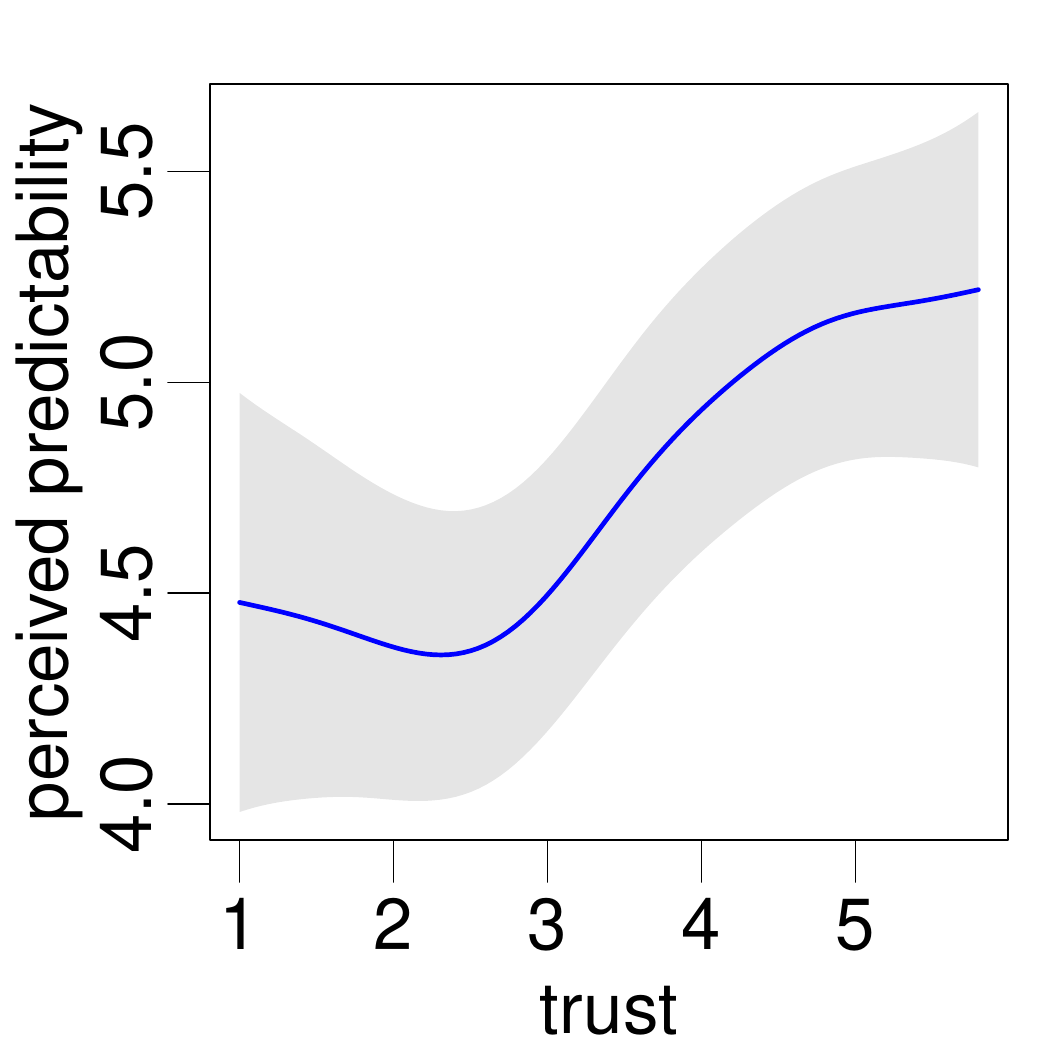}
     \caption{Trust (FOST)}\label{fig:gam_psp_trust}
     \end{subfigure}
     \hspace{2cm}
     \begin{subfigure}[t]{.32\textwidth}
         \centering
     \includegraphics[width=\textwidth]{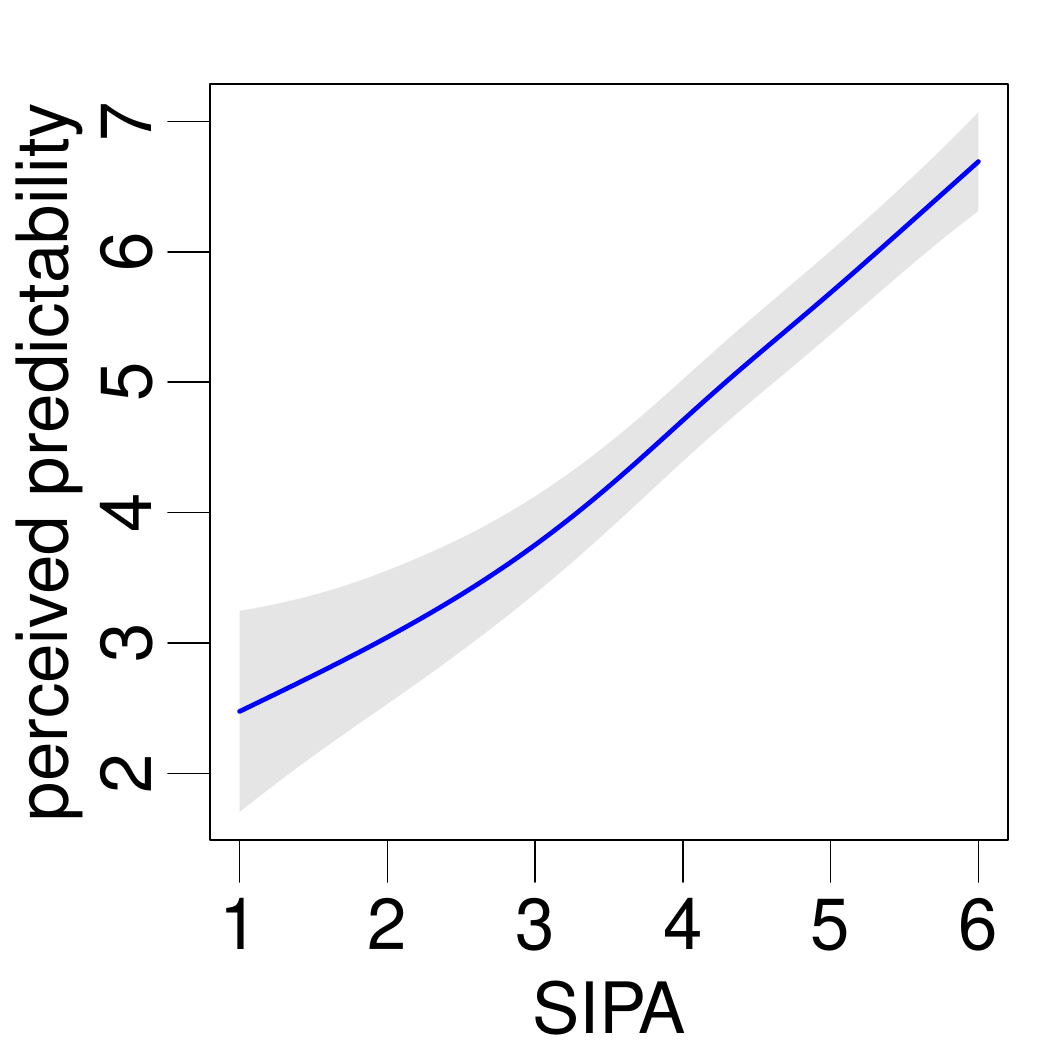}
     \caption{Situational Information Processing Awareness (SIPA)}\label{fig:gam_psp_sipa}
     \end{subfigure}
    \caption{Partial effect plots for factors with significant effects on perceived predictability in our GAM analysis. The plots show all significant smooth effects while accounting for the remaining parametric and smooth terms. Note that the y-axes are scaled separately for each plot. Notably, no objective score, such as prediction correctness or completion time, is a significant predictor of PSP, highlighting the complementary nature of subjective and objective measures of predictability.}\label{fig:gam_psp}
 \end{figure*}

\subsubsection{Results.}
We report Wald tests for the parametric and smooth terms in \Cref{tab:gam_results_psp_paramtric} and \Cref{tab:gam_results_psp_smooth}, respectively.

\paragraph{Objective Scores Do Not Predict PSP}
Notably, we do not find significant effects of the studied objective scores (i.e., prediction correctness and completion time) on PSP scores, supporting our assumption that PSP captures a construct distinct from objective predictability.
Separate Pearson correlations between PSP and prediction correctness ($r=0.050$, $p=0.478$) and between PSP and completion time ($r=0.109$, $p=0.125$) yield the same conclusion.
This result indicates that PSP scores capture information that cannot be substituted by automatic scores and, in general, suggests that (AI) systems cannot be evaluated adequately without subjective human evaluation, as has also been argued in the context of AI and natural language processing (NLP) \citep{callison-burch-etal-2006-evaluating,liu-etal-2016-evaluate,novikova-etal-2017-need,sulem-etal-2018-bleu,reiter-2018-structured}, and explainability in particular \citep{DBLP:conf/iui/RiberaL19,DBLP:journals/corr/abs-2007-12248,DBLP:conf/acl/GonzalezRS21,thomas_reliance_2022, DBLP:journals/corr/abs-2112-04417,DBLP:conf/icwsm/SchlegelGB22,liao2022connecting, https://doi.org/10.48550/arxiv.2210.07126}.

Conversely, we will show that, in the \textit{opposite direction} (i.e., predicting prediction correctness from PSP scores), subjective ratings are significant predictors of prediction correctness in \Cref{sec:humans_scale_results_objective_subjectiv}.

\paragraph{Strong Association Between PSP and SIPA}
We find that higher trust and SIPA scores are associated with higher PSP scores, consistent with the Pearson correlations reported in \Cref{tab:scale_correlations}.
\Cref{fig:gam_psp} displays the respective partial effects of trust and SIPA ratings.
Again, the strong association between PSP and SIPA is consistent with their overlapping theoretical background: SIPA models predictability as one facet of situational information processing awareness, whereas PSP ``zooms in'' on predictability and models it using the three proposed facets.

\paragraph{Effects of Explanation Modality and No Effect of Noise Level}
For the parametric terms shown in \Cref{tab:gam_results_psp_paramtric}, we find a significant main effect of explanation format as well as a significant interaction effect between explanation format and interactivity.
Detailed parametric estimates are reported in Appendix~\ref{sec:appendix_psp_explanation_experiments}.
A post hoc Wald comparison of the contrasts for explanation formats revealed a significant difference between saliency and bar chart explanation formats ($\chi^2(1) = 6.619$, $p = 0.010$), with saliency explanations associated with significantly higher PSP ratings.
A joint post hoc test of explanation formats and interactivity further revealed differences between non-interactive bar charts and non-interactive no-explanation sentences ($\chi^2(1) = 3.963$, $p = 0.047$), non-interactive bar charts and interactive bar charts ($\chi^2(1) = 4.759$, $p = 0.029$), and non-interactive bar charts and non-interactive saliencies ($\chi^2(1) = 6.619$, $p = 0.010$).
Interestingly, we do not find a significant effect of noise level on PSP scores.
We revisit this observation below in the context of the effect of noise level on prediction correctness.

\begin{table}
    \centering
    \begin{tabular}{lrrr}
      \toprule
       & \textbf{df} & \textbf{F} & \textbf{p} \\ 
      \midrule
      explanation format & 2.00 & 3.44 & \textbf{0.03} \\ 
      interactivity & 1.00 & 1.80 & 0.18 \\
      explanation format:interactivity & 2.00 & 3.18 & \textbf{0.04} \\
      noise level & 2.00 & 0.41 & 0.67 \\ 
      identification & 2.00 & 0.37 & 0.69 \\  
      \bottomrule
    \end{tabular}
    \caption{Wald tests for the parametric terms in our model of PSP scores. Explanation format (none, saliency, or bar chart) and its interaction with interactivity have significant effects on PSP scores.}
    \label{tab:gam_results_psp_paramtric}
\end{table}

\begin{table}
    \centering
    \begin{tabular}{rrrrr}
      \toprule
         & \textbf{edf} & \textbf{Ref.df} & \textbf{F} & \textbf{p} \\ 
          \midrule
          s(prediction correctness) & 0.69 & 9.00 & 0.15 & 0.14 \\ 
          s(trust) & 3.20 & 9.00 & 5.82 & \textbf{$<$0.001} \\ 
          s(completion time) & 0.00 & 9.00 & 0.00 & 0.39 \\ 
          s(SIPA) & 2.79 & 9.00 & 44.13 & \textbf{$<$0.001} \\ 
          s(NFC) & 0.00 & 9.00 & 0.00 & 0.53 \\
          s(age) &  0.00 & 9.00 & 0.00 & 0.66 \\
       \bottomrule
    \end{tabular}
    \caption{Wald tests for the smooth terms in our model of PSP scores.
    Trust and SIPA ratings have significant effects on PSP scores.
    Notably, no objective score (correctness or completion time) is a significant predictor of PSP scores.}
    \label{tab:gam_results_psp_smooth}
\end{table}

\subsection{Perceived Predictability Versus Prediction Correctness}\label{sec:humans_scale_results_objective_subjectiv}
In the previous analysis, we examined which factors predict PSP scores.
We now investigate whether and how PSP scores and additional factors relate to objective prediction correctness.

\subsubsection{Model.}
We analyze the relation between perceived predictability and prediction correctness using another GAM.
We consider smooth terms for PSP scores, FOST trust scores, SIPA scores, NCS-6 NFC scores, completion time, and participant age.
In addition, we consider parametric terms for explanation format, interactivity, noise level, and participant identification.
As before, we include gender identification and participant age solely as control covariates; neither showed a significant effect in this model either.
As in the GAM discussed above, we also include an interaction term between explanation format and interactivity.

\subsubsection{Results.}
We report Wald tests for the parametric and smooth terms in \Cref{tab:gam_results_correctness_paramtric} and \Cref{tab:gam_results_correctness_smooth}, respectively.
Detailed results are reported in Appendix~\ref{sec:appendix_psp_explanation_experiments}.

\paragraph{Subjective Ratings Are Predictive of Objective Correctness}
While objective prediction correctness did not have a significant effect on subjective PSP scores in \Cref{sec:humans_scale_predictors_psp}, PSP scores have a significant effect on prediction correctness.
As shown in \Cref{fig:gam_correctness_psp}, our model associates higher PSP scores with a moderate increase in objective prediction correctness.
In addition, we find a strong effect in the opposite direction for trust.
As shown in \Cref{fig:gam_correctness_trust}, higher levels of participant trust in the system correspond to lower prediction correctness.
\citet{hase-bansal-2020-evaluating} found that subjective explanation quality ratings of ``Does this explanation show me why the system thought what it did?'' are not predictive of user correctness.
Similarly, \citet{DBLP:conf/iui/BucincaLGG20} found that trust ratings are not predictive of prediction performance.
In contrast to \citet{hase-bansal-2020-evaluating,DBLP:conf/iui/BucincaLGG20}, we find that PSP and trust ratings are predictive of prediction correctness.
Although differences in the particular scale-score combinations and usage contexts do not allow us to draw general conclusions in either direction, we note that, unlike their work, we model non-linear effects of subjective ratings on prediction correctness using GAMs and likewise find the resulting function estimates to be non-linear (\Cref{fig:gam_correctness_psp,fig:gam_correctness_trust}).

\paragraph{Better Predictions Need Time}
\Cref{fig:gam_correctness_completion_time} shows that, within our model, longer prediction completion times are associated with a moderate increase in prediction correctness.
We attribute this effect to substantial differences in participants' interest in the prediction task and, correspondingly, in their willingness to think about the system's behavior.
Although this finding is not surprising, it supports time-based detection of insufficient-effort responding, consistent with the findings of \citet{Bowling2021TheQA}.

\paragraph{Noise Level Affects Prediction Correctness}
Among the parametric terms, only noise level has a significant effect on prediction correctness.
In particular, explanation format does not have a significant effect.
A post hoc Wald comparison of the contrasts for noise level revealed significant differences between a noise level of 0.0 and 0.4 ($\chi^2(1) = 28.196$, $p < 0.001$) and 0.0 and 0.8 ($\chi^2(1) = 34.539$, $p < 0.001$).
The difference between 0.4 and 0.8 was not significant ($\chi^2(1) = 0.347$, $p = 0.556$).
These findings are consistent with the box plots of prediction correctness (i.e., objective predictability) shown in \Cref{fig:boxplot_noise_levels}.

\begin{table}
\centering
    \begin{tabular}{lrrr}
      \toprule
       & \textbf{df} & \textbf{F} & \textbf{p} \\
      \midrule
      explanation format & 2.00 & 1.44 & 0.24 \\ 
      interactivity & 1.00 & 0.55 & 0.46 \\ 
      explanation format:interactivity & 2.00 & 0.85 & 0.43 \\ 
      noise level & 2.00 & 22.13 & \textbf{$<$0.01} \\ 
      identification & 2.00 & 0.66 & 0.52 \\ 
    \bottomrule
    \end{tabular}
    \caption{Wald tests for the parametric terms in our model of prediction correctness scores. Noise level (i.e., the level of system stochasticity) has a significant effect on prediction correctness.}
    \label{tab:gam_results_correctness_paramtric}
\end{table}

\begin{table}
\centering
    \begin{tabular}{rrrrr}
      \toprule
        & \textbf{edf} & \textbf{Ref.df} & \textbf{F} & \textbf{p} \\ 
        \midrule
        s(PSP) & 1.66 & 9.00 & 0.79 & \textbf{0.01} \\ 
        s(trust) & 2.34 & 9.00 & 2.45 & \textbf{$<$0.01} \\ 
        s(completion time) & 0.80 & 9.00 & 0.45 & \textbf{0.02} \\ 
        s(SIPA) & 0.00 & 9.00 & 0.00 & 0.85 \\ 
        s(NFC) & 0.86 & 9.00 & 0.20 & 0.11 \\ 
        s(age) & 0.48 & 9.00 & 0.08 & 0.22 \\ 
        \bottomrule
    \end{tabular}
    \caption{Wald tests for the smooth terms in our model of prediction correctness. PSP scores, trust scores, and completion time have significant effects on prediction correctness.}
    \label{tab:gam_results_correctness_smooth}
\end{table}

\begin{figure*}
     \centering
     \begin{subfigure}[t]{.32\textwidth}
         \centering
     \includegraphics[width=\textwidth]{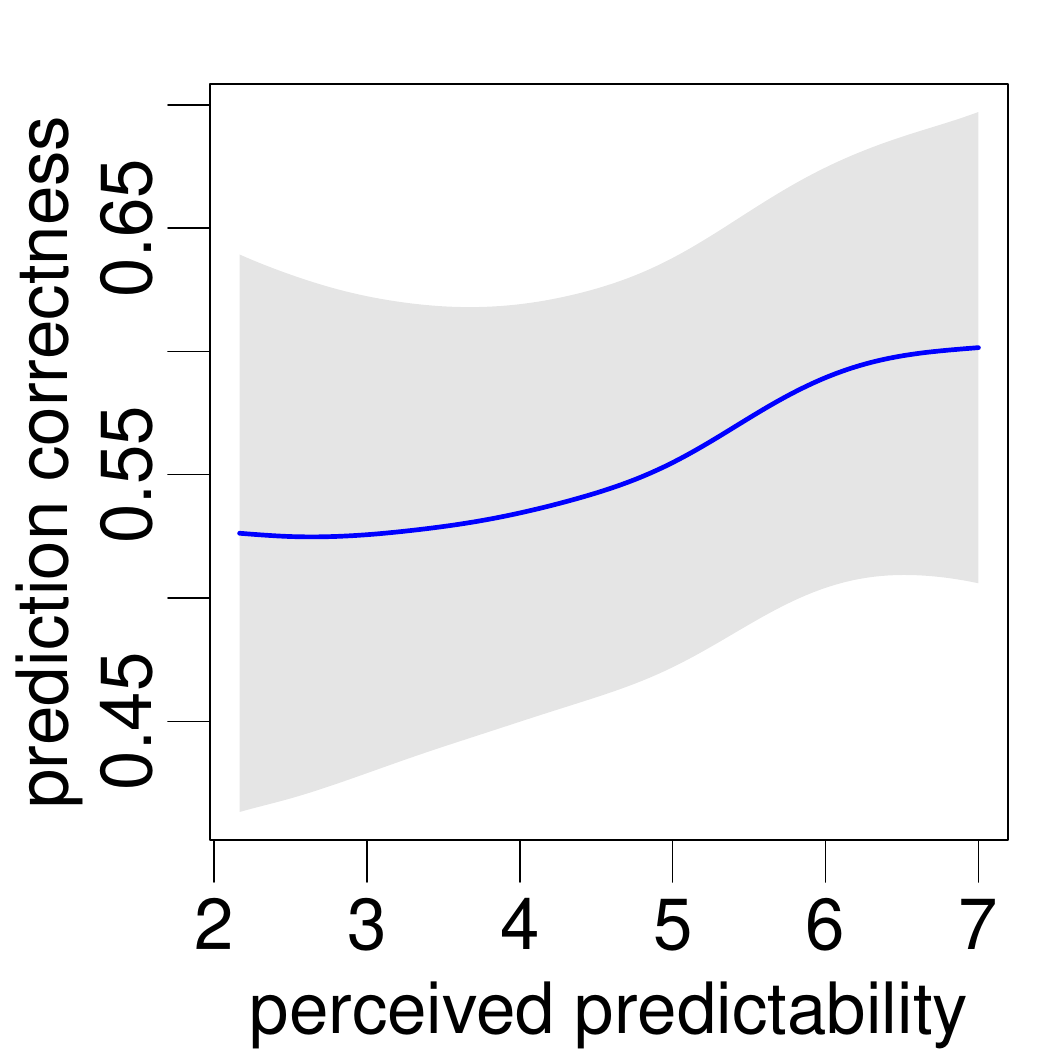}
     \caption{PSP}\label{fig:gam_correctness_psp}
     \end{subfigure}
     \hfill
     \begin{subfigure}[t]{.32\textwidth}
         \centering
     \includegraphics[width=\textwidth]{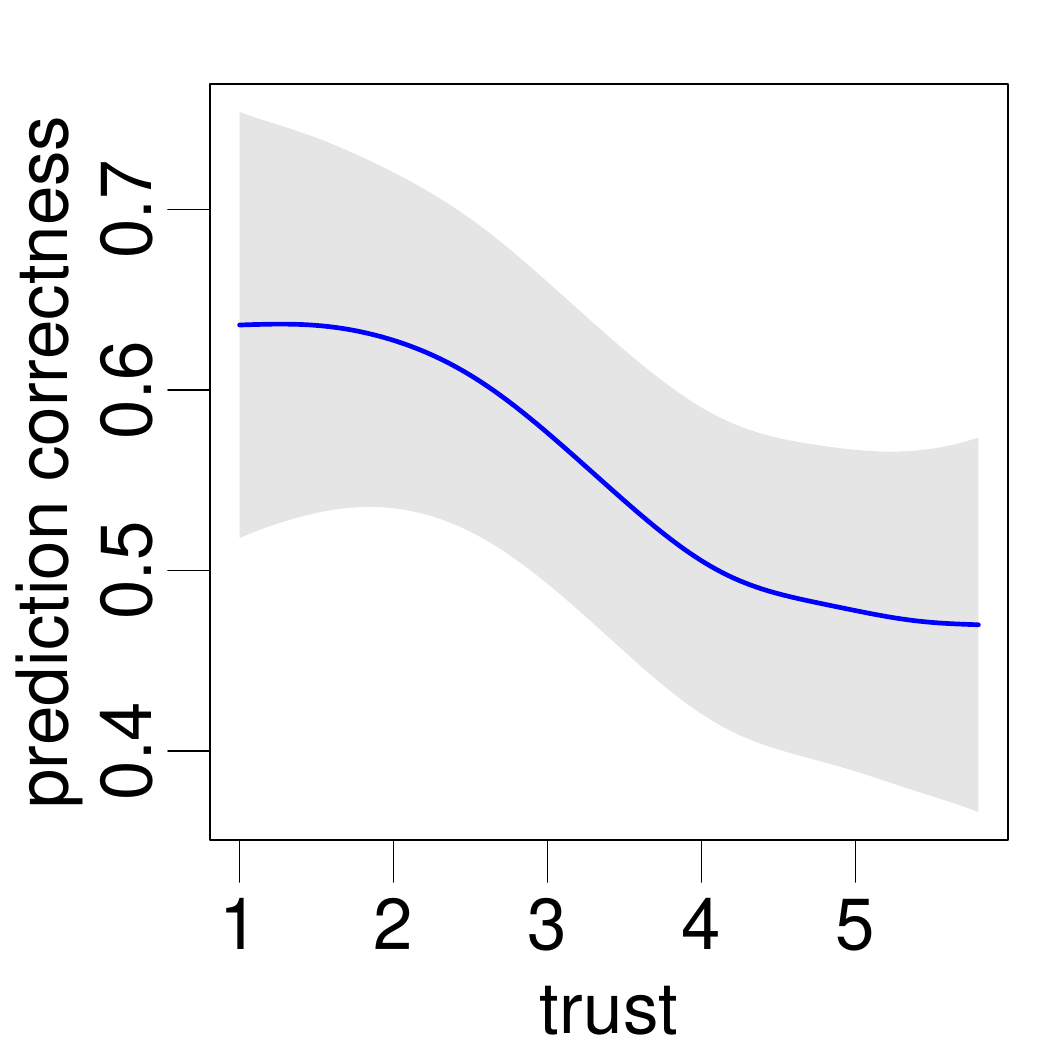}
     \caption{Trust}\label{fig:gam_correctness_trust}
     \end{subfigure}
     \hfill
     \begin{subfigure}[t]{.32\textwidth}
         \centering
     \includegraphics[width=\textwidth]{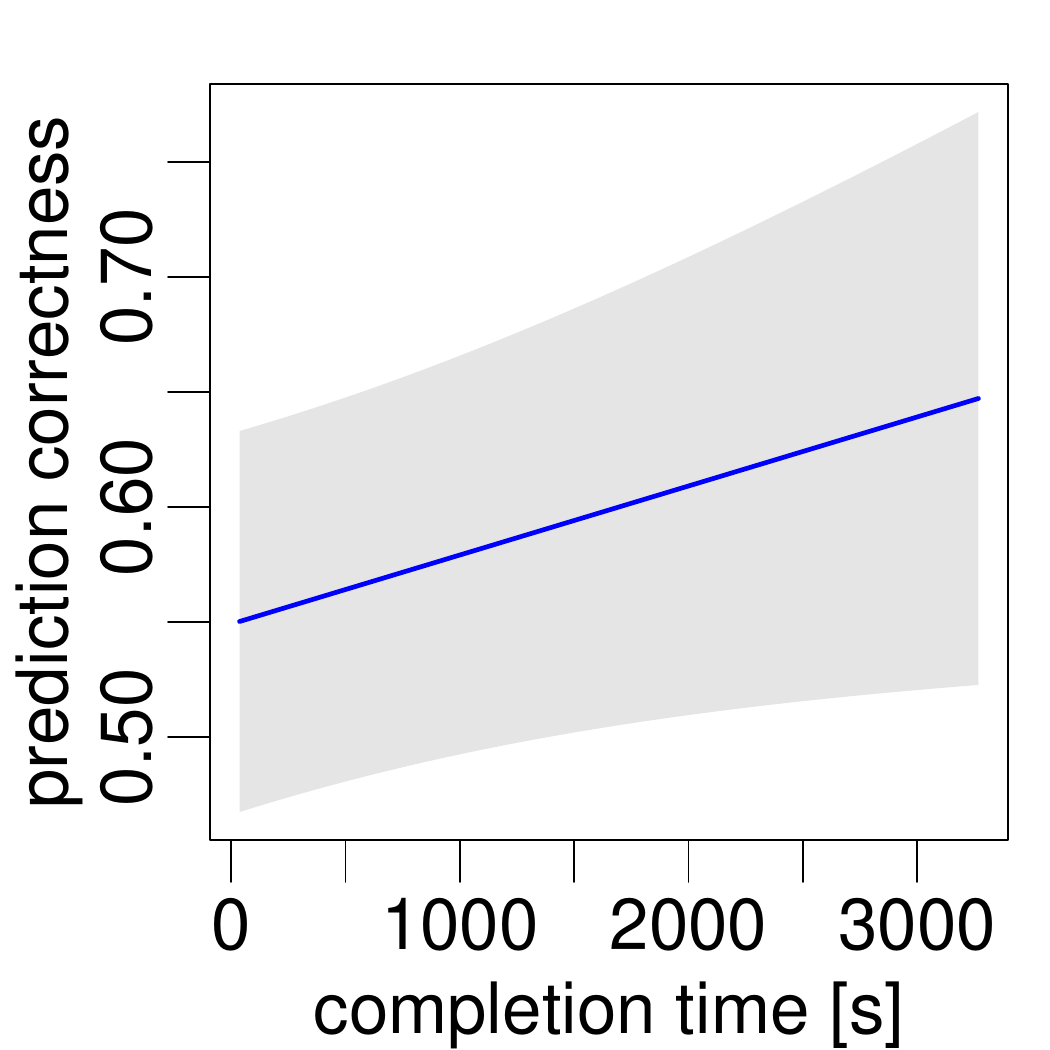}
     \caption{Prediction time}\label{fig:gam_correctness_completion_time}
     \end{subfigure}
     \caption{Partial effect plots for factors with significant effects on objective prediction correctness in our analysis. The plots show all significant smooth effects while accounting for the remaining parametric and smooth terms. Note that the y-axes are scaled separately for each plot.}\label{fig:gam_correctness}
 \end{figure*}

 \begin{figure}
     \centering
     \includegraphics[width=0.45\textwidth]{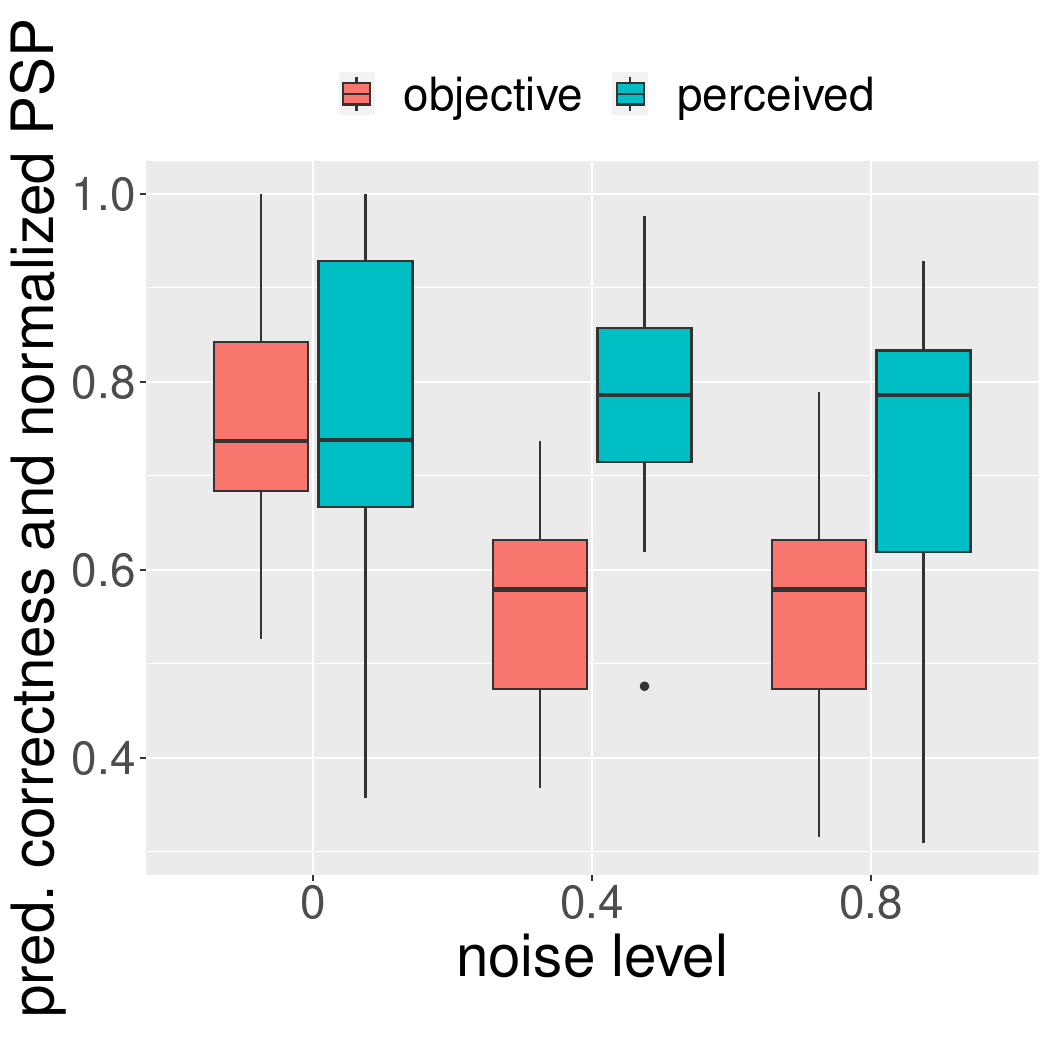}
     \caption{Boxplot showing the distributions of prediction correctness and normalized PSP scores for different levels of system stochasticity and no additional explanations.}
     \label{fig:boxplot_noise_levels}
 \end{figure}

\paragraph{Subjective $\neq$ Objective Predictability}
Comparing our findings on subjective PSP scores and objective prediction correctness, we observe that PSP scores are affected by explanation modality, whereas prediction correctness is affected by noise level.
While using saliency visualizations instead of bar charts results in a significant increase in PSP ratings, we find no effect of explanation format on prediction correctness.
This observation is consistent with Haag's recent meta-analysis of the effect of explainable AI in decision support systems on task performance \citep{Haag2025TheEO}, which finds no significant difference between explanation types.

Similarly, adding noise to the system's word polarity estimates corresponds to a significant drop in prediction correctness without affecting PSP ratings.
This observation supports our hypothesis that objective predictability and subjective predictability measure distinct characteristics of the user's mental model of a system (as illustrated in \Cref{fig:two_flashlights}).
Our results further suggest that seemingly minor visualization decisions can affect users' perception of explanations, consistent with prior work on biased perception of saliency scores \citep{10.1145/3531146.3533127,jacovi-etal-2023-neighboring}.

\paragraph{Hallucinated (Lack of) Predictability}
Although heatmaps and bar charts communicate the same objective information, we hypothesize that heatmaps have properties that induce a stronger sense of information gain than bar charts.
Note that the results of this experiment do not allow us to judge whether this increase in perceived predictability is beneficial or misleading.
Similarly, our results do not show whether the observed difference reflects an increase induced by heatmaps or a decrease induced by bar charts.
Related work on the perceptual misinterpretation of bar charts identified the ``within-the-bar bias,'' users' tendency to perceive values contained within the bar (i.e., below the top edge) as more probable when, for example, inspecting means depicted using bars and being asked about the probability of equidistant values above and below the bar's edge \citep{Newman2012BarGD}.
We hypothesize that part of our observations can be explained by this effect and the supposedly corresponding underestimation of importance scores, which, in turn, leads to a less concise mental model of the system.
\citet{Kang2021EstimatingBG} investigate modifications of the bar chart visualization and find that cumulative bar charts (i.e., also filling the upper part using a different color) can reduce this bias.
Consequently, we recommend using cumulative bar charts as shown in \Cref{fig:explanation_modality_bar_chart_cumulative}.

\paragraph{Illusion of Explanatory Depth}
The invariance of PSP scores across noise levels, alongside the observed drop in prediction correctness and the demonstrated discrimination of known groups in \Cref{sec:psp_evaluation}, raises the question of why participants did not notice their reduced ability to predict the system's behavior while remaining sensitive to the choice of explanation visualization.
We hypothesize that this phenomenon may be explained by the illusion of explanatory depth explored by \citet{Rozenblit2002TheML}.
Concretely, we hypothesize that participants' own judgments of the sentences' sentiment predispose them to form an illusion of explanatory depth. In contrast, the fictional shape classification task does not allow participants to fall back on their own judgments or purported familiarity with the domain and instead forces them to reflect on what they actually know.
Similarly, \citet{DBLP:conf/acl/GonzalezRS21} explore how explainees are affected by belief bias and argue that fictional domains might mitigate its distorting influence.

\begin{figure}
    \centering
    \includegraphics[width=0.2\textwidth]{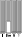}
    \caption{Enhanced bar chart explanation visualization using cumulative bars as proposed by \citet{Kang2021EstimatingBG} to mitigate ``within-the-bar bias'' \citep{Newman2012BarGD} and the supposedly associated underestimation of importance scores.}
    \label{fig:explanation_modality_bar_chart_cumulative}
\end{figure}
\subsection{Relation to Trust, SIPA, and Need for Cognition}
In \Cref{sec:humans_scale_predictors_psp}, we explored the extent to which PSP scores can be predicted from other subjective ratings (such as trust ratings), objective performance measures (such as completion time), and properties of the system and its explanations (such as explanation type).
The focus of that GAM analysis was to \textit{combine} these factors into a prediction of PSP scores.
Among the three related scale scores collected for trust, SIPA, and NFC, we found trust and SIPA scores to be significantly related to PSP scores.
In contrast, once all other terms were accounted for, NFC did not have a significant effect.

In the following analysis, we investigate a related but slightly different question: how strongly are ratings on one scale related, in \textit{isolation}, to ratings on another scale, without accounting for their relations to additional scales?
To answer this question, we examine pairwise Pearson correlations between the paired responses collected with the four scales.

\begin{table}
    \centering
    \begin{tabular}{lC{1.5cm}C{1.5cm}C{1.5cm}C{1.5cm}}
      \toprule
            & \textbf{SIPA} & \textbf{FOST} & \textbf{NCS-6}\\ 
      \midrule
        \textbf{PSP}       & \namewithnumber{\textbf{0.856}}{$<0.001$} & \namewithnumber{\textbf{0.558}}{$<0.001$} & \namewithnumber{\textbf{0.248}}{$<0.001$} \\ 
        \textbf{SIPA}     & & \namewithnumber{\textbf{0.424}}{$<0.001$} & \namewithnumber{\textbf{0.272}}{$<0.001$} \\ 
        \textbf{FOST}      &  &  & \namewithnumber{0.118}{0.096} \\ 
       \bottomrule
    \end{tabular}
    \caption{Pearson correlation coefficients between PSP, SIPA, trust (FOST), and NFC (NCS-6). Numbers in parentheses correspond to Holm-adjusted $p$ values. Significant correlation coefficients are highlighted in \textbf{bold} font.}
    \label{tab:scale_correlations}
\end{table}

\subsubsection{Results.}
\Cref{tab:scale_correlations} displays Pearson correlations and the respective Holm-adjusted $p$ values.

\paragraph{Confirming the Association Between PSP and SIPA, as Well as Trust}
We observe a strong linear correlation between PSP and SIPA ratings ($r=0.856$, $p<0.001$), consistent with our theory-driven expectations (see \Cref{sec:humans_scale_predictors_psp}) as well as our empirical results from the GAM analysis (see \Cref{fig:gam_psp_sipa}).
Similarly, our correlation analysis confirms the association between PSP and trust ($r=0.558$, $p<0.001$), which we also observed in the GAM analysis.

\paragraph{Predictors of Trust}
\citet{schrills_kargl_bickel_franke_2022} also evaluate the correlations between SIPA scores and FOST trust scores and find that, across three different samples, correlations vary between 0.55 and 0.84.
We confirm their finding that SIPA and trust have a significant linear association.
We statistically compare the strength of the SIPA-trust correlation with that of the PSP-trust correlation using the cocor R package \citep{Diedenhofen2015cocorAC}.\footnote{The cocor package provides numerous tests including Dunn and Clark's z \citep{Dunn1969CorrelationCM} and Zou's confidence interval \citep{Zou2007TowardUC}. All implemented tests indicate a significant difference between the two correlation coefficients.}
We find that, in our study, the correlation between PSP and trust is significantly stronger than the correlation between SIPA and trust.
We argue that this difference can be explained by the theoretical foundations of the PSP and SIPA scales.
While SIPA measures the facets transparency, understandability, and predictability, our PSP scale focuses on (three facets of) predictability.
We therefore hypothesize that trust depends more strongly on predictability and less on potentially preceding effects such as perceived understandability.
To test this hypothesis, we analyze the correlations between trust and the respective SIPA subscales.
Consistent with this hypothesis, the SIPA predictability subscale should show significantly higher correlations with trust than the transparency and understandability subscales.
We find that the SIPA predictability subscale has a stronger correlation to trust ($r=0.447$, $p<0.001$) than the transparency subscale ($r=0.318$, $p<0.001$) and the understandability subscale ($r=0.351$, $p<0.001$).
The corresponding cocor tests of differences in correlation strengths all indicate that the predictability subscale has higher correlations with trust than the remaining two subscales.
This supports our hypothesis that the stronger correlation between our PSP scale and trust, relative to the SIPA scale and trust, can be attributed to the focused construct of the PSP scale and the stronger association between predictability and trust.
In another experiment, \citet{DBLP:journals/corr/abs-2009-06349} found that an increased rate of system misclassifications was associated with lower self-reported trust.
We fit another GAM model to assess whether we can replicate their observations.\footnote{We include the same smooth and parametric terms as for our analysis of PSP ratings and swap PSP and trust.}
In contrast to their findings, we do not find a significant effect of noise level on trust.
However, we do find a significant effect of explanation modality on trust.
A post hoc Wald comparison of the contrasts of explanation modality reveals significant pairwise differences between bar charts and saliency explanations ($\chi^2(1) = 8.007$, $p = 0.005$) as well as between no explanation and saliency explanations ($\chi^2(1) = 8.161$, $p = 0.004$).
For both pairs, saliency explanations are associated with significantly lower trust scores.
We report detailed test statistics for the discussed and additional effects in Appendix~\ref{sec:appendix_psp_explanation_experiments}.
This pattern contrasts with our finding for PSP, for which saliency explanations were associated with a significant increase in PSP relative to bar chart explanations (see \Cref{sec:humans_scale_predictors_psp}).
This conflict indicates that PSP ratings not only offer an additional perspective alongside objective scores, but also capture information beyond related subjective constructs whose association with PSP should be investigated in more depth in future work.

\paragraph{Effect of High Need for Cognition}
Our correlation analysis identifies a significant linear association between PSP and NFC.
As the results of \citet{DBLP:journals/pacmhci/BucincaMG21} indicate that users' NFC affects the effect that explanations have on their decision behavior, it is plausible to assume that NFC has an effect on perceived predictability as well.
Interestingly, however, we do not find an effect of NFC scores on prediction correctness in our scenario.\footnote{We additionally evaluate a binarized NFC covariate following a partition of subjects above and under the median NFC as applied by \citet{DBLP:journals/pacmhci/BucincaMG21} and still do not find a significant effect of NFC.}
An analysis of the correlation between NFC and prediction correctness confirms this result ($r=0.071$, $p=0.318$).
At the same time, we do not observe a significant correlation between NFC and trust.
We therefore hypothesize that participants' need for cognition affects their system perception in a way that is captured neither by prediction correctness nor trust, yet still positively affects PSP.

\section{Overall Discussion}
In this paper, we developed and evaluated a novel 6-item scale and applied it to explore the effects of explanations and system stochasticity on perceived system predictability, as well as the relation of PSP to, \textit{inter alia}, prediction correctness, trust, SIPA, and participants' NFC.

Overall, we collected opinions from 40 participants to guide our theory development grounded in uncertainty theory, incorporated feedback from six experts to improve our initial item pool, conducted written cognitive interviews with 25 participants to further refine and filter our items, ran a study with a functional shape classification system involving 200 participants to distill and evaluate the final version of our scale, and conducted an additional study with different variants of a sentiment classification system to confirm our evaluation and explore the relation of PSP to related scales and prediction correctness.

Our scale evaluations demonstrate that the PSP scale exhibits desirable psychometric properties, including consistently high internal reliability, and indicate that it can be used both as (a) a unidimensional measure of perceived predictability as well as (b) a hierarchical measure of perceived predictability with three subscales for epistemic, aleatory, and effective predictability.

Our results suggest that (i) PSP cannot be predicted from automatic measures such as prediction correctness or completion time, (ii) conversely, prediction correctness is significantly affected by subjective scores, with higher PSP scores being associated with higher prediction correctness and higher trust scores being associated with notably lower prediction correctness, (iii) explanation format affects PSP but not prediction correctness, and (iv) higher system stochasticity affects prediction correctness but not PSP.

Overall, we find that subjective PSP and objective prediction correctness capture distinct aspects of users' mental models of a system and can therefore diverge, highlighting the need to examine both subjective and objective predictability.
We relate our observations to the ``within-the-bar bias'' \citep{Newman2012BarGD} and the illusion of explanatory depth \citep{Rozenblit2002TheML}, and recommend using cumulative bars in bar chart visualizations as shown in \Cref{fig:explanation_modality_bar_chart_cumulative}.

\section{Limitations}
Our empirical validation relies on deliberately constrained task settings: (i) a fictional shape classification task for scale development and (ii) a sentiment classification task for the second study.
These settings let us tightly control prediction rules, stochasticity, and error patterns, but they do not cover the full diversity of contemporary AI applications.
At the same time, the contrast between the two studies is informative: the fictional shape task deliberately suppresses participants' prior beliefs about what the system should output, whereas the sentiment task reveals how readily such pre-existing expectations can shape perceived predictability judgments.
Accordingly, we do not claim that the magnitude of the observed explanation or noise effects will transfer unchanged to higher-stakes domains, longer-term use, or more complex deployed systems.

A second limitation is that the second study uses a transparent sentiment classifier rather than an opaque model such as a large language model.
We consider this a deliberate trade-off in favor of construct validation and explanation faithfulness: because PSP is intended as a model-agnostic user perception, we first tested it in a setting where the communicated explanation coincides with the actual decision rule by design.
Using an opaque model would have required a post-hoc explanation method whose own faithfulness and communicative adequacy could affect participants' judgments, making it difficult to separate perceived predictability from effects of the explanation pipeline itself.
This choice therefore strengthens internal validity, but it also limits the external validity of our findings for contemporary black-box systems.
Future work should extend the scale to LLMs and other opaque models, ideally in designs that either treat explanation method as an explicit independent factor or otherwise characterize explanation faithfulness sufficiently to keep these influences analytically distinct.

Finally, our nomological validation covers trust, SIPA, NFC, and prediction correctness, but not the full range of adjacent human-AI constructs.
Additional work should therefore investigate how PSP relates to constructs such as mental demand or reliance calibration and should complement questionnaire data with qualitative methods, such as think-aloud studies, to better explain why PSP can diverge from trust or objective correctness.

\section{Conclusion}
We introduced \emph{perceived system predictability} as a user-centered construct grounded in uncertainty theory, developed and validated a 6-item PSP scale that supports both unidimensional use and a three-factor hierarchical interpretation across epistemic, aleatory, and effective predictability, and demonstrated the scale's empirical value in two complementary studies.
Our findings show that (a) PSP cannot be inferred from automatic measures such as completion time or prediction correctness, yet it itself predicts the correctness of users' predictions of system behavior, (b) explanations shift PSP without affecting correctness, and (c) increased system stochasticity degrades correctness without lowering PSP.
Together, these results establish PSP as a measurable construct that captures aspects of users' mental models which existing objective and subjective measures leave unaddressed.
By providing both a theoretical framework and a validated instrument, our work enables HCI researchers and practitioners to study perceived predictability systematically and to design interactive systems whose behavior users can meaningfully anticipate -- a prerequisite for calibrated reliance and accountable human-AI interaction.

\bibliographystyle{ACM-Reference-Format}
\bibliography{anthology,custom}


\begin{thebibliography}{95}

%%% the Web address.

\ifx \showCODEN    \undefined \def \showCODEN     #1{\unskip}     \fi
\ifx \showDOI      \undefined \def \showDOI       #1{#1}\fi
\ifx \showISBNx    \undefined \def \showISBNx     #1{\unskip}     \fi
\ifx \showISBNxiii \undefined \def \showISBNxiii  #1{\unskip}     \fi
\ifx \showISSN     \undefined \def \showISSN      #1{\unskip}     \fi
\ifx \showLCCN     \undefined \def \showLCCN      #1{\unskip}     \fi
\ifx \shownote     \undefined \def \shownote      #1{#1}          \fi
\ifx \showarticletitle \undefined \def \showarticletitle #1{#1}   \fi
\ifx \showURL      \undefined \def \showURL       {\relax}        \fi
% invisible to TeX
\providecommand\bibfield[2]{#2}
\providecommand\bibinfo[2]{#2}
\providecommand\natexlab[1]{#1}
\providecommand\showeprint[2][]{arXiv:#2}

\bibitem[Alarcon et~al\mbox{.}(2024)]%
        {Alarcon2023DevelopmentAV}
\bibfield{author}{\bibinfo{person}{Gene~M. Alarcon}, \bibinfo{person}{August~A.
  Capiola}, \bibinfo{person}{Michael~A. Lee}, \bibinfo{person}{Sasha~M.
  Willis}, \bibinfo{person}{Izz~Aldin Hamdan}, \bibinfo{person}{Sarah~A.
  Jessup}, {and} \bibinfo{person}{Krista~N. Harris}.}
  \bibinfo{year}{2024}\natexlab{}.
\newblock \showarticletitle{Development and Validation of the System
  Trustworthiness Scale}.
\newblock \bibinfo{journal}{\emph{Hum. Factors}} \bibinfo{volume}{66},
  \bibinfo{number}{7} (\bibinfo{year}{2024}), \bibinfo{pages}{1893--1913}.
\newblock
\urldef\tempurl%
\url{https://doi.org/10.1177/00187208231189000}
\showDOI{\tempurl}


\bibitem[Baccianella et~al\mbox{.}(2010)]%
        {baccianella-etal-2010-sentiwordnet}
\bibfield{author}{\bibinfo{person}{Stefano Baccianella},
  \bibinfo{person}{Andrea Esuli}, {and} \bibinfo{person}{Fabrizio Sebastiani}.}
  \bibinfo{year}{2010}\natexlab{}.
\newblock \showarticletitle{{S}enti{W}ord{N}et 3.0: An Enhanced Lexical
  Resource for Sentiment Analysis and Opinion Mining}. In
  \bibinfo{booktitle}{\emph{Proceedings of the Seventh International Conference
  on Language Resources and Evaluation ({LREC}'10)}}.
  \bibinfo{publisher}{European Language Resources Association (ELRA)},
  \bibinfo{address}{Valletta, Malta}.
\newblock
\urldef\tempurl%
\url{http://www.lrec-conf.org/proceedings/lrec2010/pdf/769_Paper.pdf}
\showURL{%
\tempurl}


\bibitem[Bansal et~al\mbox{.}(2021)]%
        {DBLP:conf/chi/BansalWZFNKRW21}
\bibfield{author}{\bibinfo{person}{Gagan Bansal}, \bibinfo{person}{Tongshuang
  Wu}, \bibinfo{person}{Joyce Zhou}, \bibinfo{person}{Raymond Fok},
  \bibinfo{person}{Besmira Nushi}, \bibinfo{person}{Ece Kamar},
  \bibinfo{person}{Marco~T{\'{u}}lio Ribeiro}, {and} \bibinfo{person}{Daniel~S.
  Weld}.} \bibinfo{year}{2021}\natexlab{}.
\newblock \showarticletitle{Does the Whole Exceed its Parts? The Effect of {AI}
  Explanations on Complementary Team Performance}. In
  \bibinfo{booktitle}{\emph{{CHI} '21: {CHI} Conference on Human Factors in
  Computing Systems, Virtual Event / Yokohama, Japan, May 8-13, 2021}},
  \bibfield{editor}{\bibinfo{person}{Yoshifumi Kitamura},
  \bibinfo{person}{Aaron Quigley}, \bibinfo{person}{Katherine Isbister},
  \bibinfo{person}{Takeo Igarashi}, \bibinfo{person}{Pernille Bj{\o}rn}, {and}
  \bibinfo{person}{Steven~Mark Drucker}} (Eds.). \bibinfo{publisher}{{ACM}},
  \bibinfo{pages}{81:1--81:16}.
\newblock
\urldef\tempurl%
\url{https://doi.org/10.1145/3411764.3445717}
\showDOI{\tempurl}


\bibitem[Bargas-Avila and Br{\"u}hlmann(2016)]%
        {bargas2016measuring}
\bibfield{author}{\bibinfo{person}{Javier~A Bargas-Avila} {and}
  \bibinfo{person}{Florian Br{\"u}hlmann}.} \bibinfo{year}{2016}\natexlab{}.
\newblock \showarticletitle{Measuring user rated language quality: development
  and validation of the user interface Language Quality Survey (LQS)}.
\newblock \bibinfo{journal}{\emph{International Journal of Human-Computer
  Studies}}  \bibinfo{volume}{86} (\bibinfo{year}{2016}),
  \bibinfo{pages}{1--10}.
\newblock


\bibitem[Beatty and Willis(2007)]%
        {beatty2007research}
\bibfield{author}{\bibinfo{person}{Paul~C Beatty} {and}
  \bibinfo{person}{Gordon~B Willis}.} \bibinfo{year}{2007}\natexlab{}.
\newblock \showarticletitle{Research synthesis: The practice of cognitive
  interviewing}.
\newblock \bibinfo{journal}{\emph{Public opinion quarterly}}
  \bibinfo{volume}{71}, \bibinfo{number}{2} (\bibinfo{year}{2007}),
  \bibinfo{pages}{287--311}.
\newblock


\bibitem[Biran and McKeown(2017)]%
        {DBLP:conf/ijcai/BiranM17}
\bibfield{author}{\bibinfo{person}{Or Biran} {and} \bibinfo{person}{Kathleen~R.
  McKeown}.} \bibinfo{year}{2017}\natexlab{}.
\newblock \showarticletitle{Human-Centric Justification of Machine Learning
  Predictions}. In \bibinfo{booktitle}{\emph{Proceedings of the Twenty-Sixth
  International Joint Conference on Artificial Intelligence, {IJCAI} 2017,
  Melbourne, Australia, August 19-25, 2017}},
  \bibfield{editor}{\bibinfo{person}{Carles Sierra}} (Ed.).
  \bibinfo{publisher}{ijcai.org}, \bibinfo{pages}{1461--1467}.
\newblock
\urldef\tempurl%
\url{https://doi.org/10.24963/ijcai.2017/202}
\showDOI{\tempurl}


\bibitem[Boateng et~al\mbox{.}(2018)]%
        {boateng_best_2018}
\bibfield{author}{\bibinfo{person}{Godfred~O. Boateng},
  \bibinfo{person}{Torsten~B. Neilands}, \bibinfo{person}{Edward~A. Frongillo},
  \bibinfo{person}{Hugo~R. Melgar-Quiñonez}, {and} \bibinfo{person}{Sera~L.
  Young}.} \bibinfo{year}{2018}\natexlab{}.
\newblock \showarticletitle{Best {Practices} for {Developing} and {Validating}
  {Scales} for {Health}, {Social}, and {Behavioral} {Research}: {A} {Primer}}.
\newblock \bibinfo{journal}{\emph{Frontiers in Public Health}}
  \bibinfo{volume}{6} (\bibinfo{year}{2018}), \bibinfo{pages}{149}.
\newblock
\showISSN{2296-2565}
\urldef\tempurl%
\url{https://doi.org/10.3389/fpubh.2018.00149}
\showDOI{\tempurl}


\bibitem[Bowling et~al\mbox{.}(2021)]%
        {Bowling2021TheQA}
\bibfield{author}{\bibinfo{person}{Nathan~A. Bowling},
  \bibinfo{person}{Jason~L. Huang}, \bibinfo{person}{Cheyna~K. Brower}, {and}
  \bibinfo{person}{Caleb~B. Bragg}.} \bibinfo{year}{2021}\natexlab{}.
\newblock \showarticletitle{The Quick and the Careless: The Construct Validity
  of Page Time as a Measure of Insufficient Effort Responding to Surveys}.
\newblock \bibinfo{journal}{\emph{Organizational Research Methods}}
  \bibinfo{volume}{26} (\bibinfo{year}{2021}), \bibinfo{pages}{323 -- 352}.
\newblock


\bibitem[Brooke(1996)]%
        {brooke1996sus}
\bibfield{author}{\bibinfo{person}{John Brooke}.}
  \bibinfo{year}{1996}\natexlab{}.
\newblock \showarticletitle{SUS: a “quick and dirty'usability}.
\newblock \bibinfo{journal}{\emph{Usability evaluation in industry}}
  (\bibinfo{year}{1996}), \bibinfo{pages}{189}.
\newblock


\bibitem[Bu{\c{c}}inca et~al\mbox{.}(2020)]%
        {DBLP:conf/iui/BucincaLGG20}
\bibfield{author}{\bibinfo{person}{Zana Bu{\c{c}}inca}, \bibinfo{person}{Phoebe
  Lin}, \bibinfo{person}{Krzysztof~Z. Gajos}, {and} \bibinfo{person}{Elena~L.
  Glassman}.} \bibinfo{year}{2020}\natexlab{}.
\newblock \showarticletitle{Proxy tasks and subjective measures can be
  misleading in evaluating explainable {AI} systems}. In
  \bibinfo{booktitle}{\emph{{IUI} '20: 25th International Conference on
  Intelligent User Interfaces, Cagliari, Italy, March 17-20, 2020}},
  \bibfield{editor}{\bibinfo{person}{Fabio Patern{\`{o}}},
  \bibinfo{person}{Nuria Oliver}, \bibinfo{person}{Cristina Conati},
  \bibinfo{person}{Lucio~Davide Spano}, {and} \bibinfo{person}{Nava Tintarev}}
  (Eds.). \bibinfo{publisher}{{ACM}}, \bibinfo{pages}{454--464}.
\newblock
\urldef\tempurl%
\url{https://doi.org/10.1145/3377325.3377498}
\showDOI{\tempurl}


\bibitem[Bu{\c{c}}inca et~al\mbox{.}(2021)]%
        {DBLP:journals/pacmhci/BucincaMG21}
\bibfield{author}{\bibinfo{person}{Zana Bu{\c{c}}inca},
  \bibinfo{person}{Maja~Barbara Malaya}, {and} \bibinfo{person}{Krzysztof~Z.
  Gajos}.} \bibinfo{year}{2021}\natexlab{}.
\newblock \showarticletitle{To Trust or to Think: Cognitive Forcing Functions
  Can Reduce Overreliance on {AI} in AI-assisted Decision-making}.
\newblock \bibinfo{journal}{\emph{Proc. {ACM} Hum. Comput. Interact.}}
  \bibinfo{volume}{5}, \bibinfo{number}{{CSCW1}} (\bibinfo{year}{2021}),
  \bibinfo{pages}{188:1--188:21}.
\newblock
\urldef\tempurl%
\url{https://doi.org/10.1145/3449287}
\showDOI{\tempurl}


\bibitem[Bussone et~al\mbox{.}(2015)]%
        {DBLP:conf/ichi/BussoneSO15}
\bibfield{author}{\bibinfo{person}{Adrian Bussone}, \bibinfo{person}{Simone
  Stumpf}, {and} \bibinfo{person}{Dympna O'Sullivan}.}
  \bibinfo{year}{2015}\natexlab{}.
\newblock \showarticletitle{The Role of Explanations on Trust and Reliance in
  Clinical Decision Support Systems}. In \bibinfo{booktitle}{\emph{2015
  International Conference on Healthcare Informatics, {ICHI} 2015, Dallas, TX,
  USA, October 21-23, 2015}}, \bibfield{editor}{\bibinfo{person}{Prabhakaran
  Balakrishnan}, \bibinfo{person}{Jaideep Srivatsava},
  \bibinfo{person}{Wai{-}Tat Fu}, \bibinfo{person}{Sanda~M. Harabagiu}, {and}
  \bibinfo{person}{Fei Wang}} (Eds.). \bibinfo{publisher}{{IEEE} Computer
  Society}, \bibinfo{pages}{160--169}.
\newblock
\urldef\tempurl%
\url{https://doi.org/10.1109/ICHI.2015.26}
\showDOI{\tempurl}


\bibitem[Callison-Burch et~al\mbox{.}(2006)]%
        {callison-burch-etal-2006-evaluating}
\bibfield{author}{\bibinfo{person}{Chris Callison-Burch},
  \bibinfo{person}{Miles Osborne}, {and} \bibinfo{person}{Philipp Koehn}.}
  \bibinfo{year}{2006}\natexlab{}.
\newblock \showarticletitle{Re-evaluating the Role of {B}leu in Machine
  Translation Research}. In \bibinfo{booktitle}{\emph{11th Conference of the
  {E}uropean Chapter of the Association for Computational Linguistics}}.
  \bibinfo{publisher}{Association for Computational Linguistics},
  \bibinfo{address}{Trento, Italy}, \bibinfo{pages}{249--256}.
\newblock
\urldef\tempurl%
\url{https://aclanthology.org/E06-1032}
\showURL{%
\tempurl}


\bibitem[Carpinella et~al\mbox{.}(2017)]%
        {carpinella_robotic_2017}
\bibfield{author}{\bibinfo{person}{Colleen~M. Carpinella},
  \bibinfo{person}{Alisa~B. Wyman}, \bibinfo{person}{Michael~A. Perez}, {and}
  \bibinfo{person}{Steven~J. Stroessner}.} \bibinfo{year}{2017}\natexlab{}.
\newblock \showarticletitle{The {Robotic} {Social} {Attributes} {Scale}
  ({RoSAS}): {Development} and {Validation}}. In
  \bibinfo{booktitle}{\emph{Proceedings of the 2017 {ACM}/{IEEE}
  {International} {Conference} on {Human}-{Robot} {Interaction}}}.
  \bibinfo{publisher}{ACM}, \bibinfo{address}{Vienna Austria},
  \bibinfo{pages}{254--262}.
\newblock
\showISBNx{978-1-4503-4336-7}
\urldef\tempurl%
\url{https://doi.org/10.1145/2909824.3020208}
\showDOI{\tempurl}


\bibitem[Casper et~al\mbox{.}(2020)]%
        {casper2020selecting}
\bibfield{author}{\bibinfo{person}{Wm Casper}, \bibinfo{person}{Bryan~D
  Edwards}, \bibinfo{person}{J~Craig Wallace}, \bibinfo{person}{Ronald~S
  Landis}, \bibinfo{person}{Dustin~A Fife}, {et~al\mbox{.}}}
  \bibinfo{year}{2020}\natexlab{}.
\newblock \showarticletitle{Selecting response anchors with equal intervals for
  summated rating scales.}
\newblock \bibinfo{journal}{\emph{Journal of Applied Psychology}}
  \bibinfo{volume}{105}, \bibinfo{number}{4} (\bibinfo{year}{2020}),
  \bibinfo{pages}{390}.
\newblock


\bibitem[Chu et~al\mbox{.}(2020)]%
        {DBLP:journals/corr/abs-2007-12248}
\bibfield{author}{\bibinfo{person}{Eric Chu}, \bibinfo{person}{Deb Roy}, {and}
  \bibinfo{person}{Jacob Andreas}.} \bibinfo{year}{2020}\natexlab{}.
\newblock \showarticletitle{Are Visual Explanations Useful? {A} Case Study in
  Model-in-the-Loop Prediction}.
\newblock \bibinfo{journal}{\emph{CoRR}}  \bibinfo{volume}{abs/2007.12248}
  (\bibinfo{year}{2020}).
\newblock
\showeprint[arXiv]{2007.12248}
\urldef\tempurl%
\url{https://arxiv.org/abs/2007.12248}
\showURL{%
\tempurl}


\bibitem[Colin et~al\mbox{.}(2022)]%
        {DBLP:journals/corr/abs-2112-04417}
\bibfield{author}{\bibinfo{person}{Julien Colin}, \bibinfo{person}{Thomas Fel},
  \bibinfo{person}{R{\'{e}}mi Cad{\`{e}}ne}, {and} \bibinfo{person}{Thomas
  Serre}.} \bibinfo{year}{2022}\natexlab{}.
\newblock \showarticletitle{What {I} Cannot Predict, {I} Do Not Understand: {A}
  Human-Centered Evaluation Framework for Explainability Methods}. In
  \bibinfo{booktitle}{\emph{Advances in Neural Information Processing Systems
  35: Annual Conference on Neural Information Processing Systems 2022, NeurIPS
  2022, New Orleans, LA, USA, November 28 - December 9, 2022}},
  \bibfield{editor}{\bibinfo{person}{Sanmi Koyejo},
  \bibinfo{person}{S.~Mohamed}, \bibinfo{person}{A.~Agarwal},
  \bibinfo{person}{Danielle Belgrave}, \bibinfo{person}{K.~Cho}, {and}
  \bibinfo{person}{A.~Oh}} (Eds.).
\newblock
\urldef\tempurl%
\url{http://papers.nips.cc/paper\_files/paper/2022/hash/13113e938f2957891c0c5e8df811dd01-Abstract-Conference.html}
\showURL{%
\tempurl}


\bibitem[Cramer et~al\mbox{.}(2008)]%
        {DBLP:journals/umuai/CramerERSRSAW08}
\bibfield{author}{\bibinfo{person}{Henriette S.~M. Cramer},
  \bibinfo{person}{Vanessa Evers}, \bibinfo{person}{Satyan Ramlal},
  \bibinfo{person}{Maarten van Someren}, \bibinfo{person}{Lloyd Rutledge},
  \bibinfo{person}{Natalia Stash}, \bibinfo{person}{Lora Aroyo}, {and}
  \bibinfo{person}{Bob~J. Wielinga}.} \bibinfo{year}{2008}\natexlab{}.
\newblock \showarticletitle{The effects of transparency on trust in and
  acceptance of a content-based art recommender}.
\newblock \bibinfo{journal}{\emph{User Model. User Adapt. Interact.}}
  \bibinfo{volume}{18}, \bibinfo{number}{5} (\bibinfo{year}{2008}),
  \bibinfo{pages}{455--496}.
\newblock
\urldef\tempurl%
\url{https://doi.org/10.1007/s11257-008-9051-3}
\showDOI{\tempurl}


\bibitem[Cronbach(1951)]%
        {Cronbach1951CoefficientAA}
\bibfield{author}{\bibinfo{person}{Lee~Joseph Cronbach}.}
  \bibinfo{year}{1951}\natexlab{}.
\newblock \showarticletitle{Coefficient alpha and the internal structure of
  tests}.
\newblock \bibinfo{journal}{\emph{Psychometrika}}  \bibinfo{volume}{16}
  (\bibinfo{year}{1951}), \bibinfo{pages}{297--334}.
\newblock


\bibitem[Czerwinski et~al\mbox{.}(2001)]%
        {czerwinski2001subjective}
\bibfield{author}{\bibinfo{person}{Mary Czerwinski}, \bibinfo{person}{Eric
  Horvitz}, {and} \bibinfo{person}{Edward Cutrell}.}
  \bibinfo{year}{2001}\natexlab{}.
\newblock \showarticletitle{Subjective duration assessment: An implicit probe
  for software usability}. In \bibinfo{booktitle}{\emph{Proceedings of IHM-HCI
  2001 conference}}, Vol.~\bibinfo{volume}{2}. \bibinfo{pages}{167--170}.
\newblock


\bibitem[de~Holanda~Coelho et~al\mbox{.}(2018)]%
        {LinsdeHolandaCoelho2018TheVE}
\bibfield{author}{\bibinfo{person}{Gabriel~Lins de Holanda~Coelho},
  \bibinfo{person}{Paul H.~P. Hanel}, {and} \bibinfo{person}{Lukas~J Wolf}.}
  \bibinfo{year}{2018}\natexlab{}.
\newblock \showarticletitle{The Very Efficient Assessment of Need for
  Cognition: Developing a Six-Item Version*}.
\newblock \bibinfo{journal}{\emph{Assessment}}  \bibinfo{volume}{27}
  (\bibinfo{year}{2018}), \bibinfo{pages}{1870 -- 1885}.
\newblock


\bibitem[Deegan(1978)]%
        {Deegan1978OnTO}
\bibfield{author}{\bibinfo{person}{John~P. Deegan}.}
  \bibinfo{year}{1978}\natexlab{}.
\newblock \showarticletitle{On the Occurrence of Standardized Regression
  Coefficients Greater Than One}.
\newblock \bibinfo{journal}{\emph{Educational and Psychological Measurement}}
  \bibinfo{volume}{38} (\bibinfo{year}{1978}), \bibinfo{pages}{873 -- 888}.
\newblock


\bibitem[Deslauriers et~al\mbox{.}(2019)]%
        {doi:10.1073/pnas.1821936116}
\bibfield{author}{\bibinfo{person}{Louis Deslauriers},
  \bibinfo{person}{Logan~S. McCarty}, \bibinfo{person}{Kelly Miller},
  \bibinfo{person}{Kristina Callaghan}, {and} \bibinfo{person}{Greg Kestin}.}
  \bibinfo{year}{2019}\natexlab{}.
\newblock \showarticletitle{Measuring actual learning versus feeling of
  learning in response to being actively engaged in the classroom}.
\newblock \bibinfo{journal}{\emph{Proceedings of the National Academy of
  Sciences}} \bibinfo{volume}{116}, \bibinfo{number}{39}
  (\bibinfo{year}{2019}), \bibinfo{pages}{19251--19257}.
\newblock
\urldef\tempurl%
\url{https://doi.org/10.1073/pnas.1821936116}
\showDOI{\tempurl}
\showeprint{https://www.pnas.org/doi/pdf/10.1073/pnas.1821936116}


\bibitem[DeVellis and Thorpe(2021)]%
        {devellis2021scale}
\bibfield{author}{\bibinfo{person}{Robert~F DeVellis} {and}
  \bibinfo{person}{Carolyn~T Thorpe}.} \bibinfo{year}{2021}\natexlab{}.
\newblock \bibinfo{booktitle}{\emph{Scale development: Theory and
  applications}}.
\newblock \bibinfo{publisher}{Sage publications}.
\newblock


\bibitem[Diedenhofen and Musch(2015)]%
        {Diedenhofen2015cocorAC}
\bibfield{author}{\bibinfo{person}{Birk Diedenhofen} {and}
  \bibinfo{person}{Jochen Musch}.} \bibinfo{year}{2015}\natexlab{}.
\newblock \showarticletitle{cocor: A Comprehensive Solution for the Statistical
  Comparison of Correlations}.
\newblock \bibinfo{journal}{\emph{PLoS ONE}}  \bibinfo{volume}{10}
  (\bibinfo{year}{2015}).
\newblock


\bibitem[Dunn and McCray(2020)]%
        {Dunn2020ThePO}
\bibfield{author}{\bibinfo{person}{Karen Dunn} {and} \bibinfo{person}{Gareth
  McCray}.} \bibinfo{year}{2020}\natexlab{}.
\newblock \showarticletitle{The Place of the Bifactor Model in Confirmatory
  Factor Analysis Investigations Into Construct Dimensionality in Language
  Testing}.
\newblock \bibinfo{journal}{\emph{Frontiers in Psychology}}
  \bibinfo{volume}{11} (\bibinfo{year}{2020}).
\newblock


\bibitem[Dunn and Clark(1969)]%
        {Dunn1969CorrelationCM}
\bibfield{author}{\bibinfo{person}{Olive~Jean Dunn} {and}
  \bibinfo{person}{Virginia~A. Clark}.} \bibinfo{year}{1969}\natexlab{}.
\newblock \showarticletitle{Correlation Coefficients Measured on the Same
  Individuals}.
\newblock \bibinfo{journal}{\emph{J. Amer. Statist. Assoc.}}
  \bibinfo{volume}{64} (\bibinfo{year}{1969}), \bibinfo{pages}{366--377}.
\newblock


\bibitem[Eisinga et~al\mbox{.}(2013)]%
        {Eisinga2013TheRO}
\bibfield{author}{\bibinfo{person}{Rob Eisinga}, \bibinfo{person}{Manfred te
  Grotenhuis}, {and} \bibinfo{person}{Ben Pelzer}.}
  \bibinfo{year}{2013}\natexlab{}.
\newblock \showarticletitle{The reliability of a two-item scale: Pearson,
  Cronbach, or Spearman-Brown?}
\newblock \bibinfo{journal}{\emph{International Journal of Public Health}}
  \bibinfo{volume}{58} (\bibinfo{year}{2013}), \bibinfo{pages}{637--642}.
\newblock


\bibitem[Eisler(1976)]%
        {eisler1976experiments}
\bibfield{author}{\bibinfo{person}{Hannes Eisler}.}
  \bibinfo{year}{1976}\natexlab{}.
\newblock \showarticletitle{Experiments on subjective duration 1868-1975: A
  collection of power function exponents.}
\newblock \bibinfo{journal}{\emph{Psychological Bulletin}}
  \bibinfo{volume}{83}, \bibinfo{number}{6} (\bibinfo{year}{1976}),
  \bibinfo{pages}{1154}.
\newblock


\bibitem[Endsley(1988)]%
        {endsley1988situation}
\bibfield{author}{\bibinfo{person}{Mica~R Endsley}.}
  \bibinfo{year}{1988}\natexlab{}.
\newblock \showarticletitle{Situation awareness global assessment technique
  (SAGAT)}. In \bibinfo{booktitle}{\emph{Proceedings of the IEEE 1988 national
  aerospace and electronics conference}}. IEEE, \bibinfo{pages}{789--795}.
\newblock


\bibitem[Feng and Boyd{-}Graber(2019)]%
        {DBLP:conf/iui/FengB19}
\bibfield{author}{\bibinfo{person}{Shi Feng} {and} \bibinfo{person}{Jordan~L.
  Boyd{-}Graber}.} \bibinfo{year}{2019}\natexlab{}.
\newblock \showarticletitle{What can {AI} do for me?: evaluating machine
  learning interpretations in cooperative play}. In
  \bibinfo{booktitle}{\emph{Proceedings of the 24th International Conference on
  Intelligent User Interfaces, {IUI} 2019, Marina del Ray, CA, USA, March
  17-20, 2019}}, \bibfield{editor}{\bibinfo{person}{Wai{-}Tat Fu},
  \bibinfo{person}{Shimei Pan}, \bibinfo{person}{Oliver Brdiczka},
  \bibinfo{person}{Polo Chau}, {and} \bibinfo{person}{Gaelle Calvary}} (Eds.).
  \bibinfo{publisher}{{ACM}}, \bibinfo{pages}{229--239}.
\newblock
\urldef\tempurl%
\url{https://doi.org/10.1145/3301275.3302265}
\showDOI{\tempurl}


\bibitem[Finstad(2010)]%
        {finstad2010usability}
\bibfield{author}{\bibinfo{person}{Kraig Finstad}.}
  \bibinfo{year}{2010}\natexlab{}.
\newblock \showarticletitle{The usability metric for user experience}.
\newblock \bibinfo{journal}{\emph{Interacting with Computers}}
  \bibinfo{volume}{22}, \bibinfo{number}{5} (\bibinfo{year}{2010}),
  \bibinfo{pages}{323--327}.
\newblock


\bibitem[Finstad(2013)]%
        {finstad_response_2013}
\bibfield{author}{\bibinfo{person}{Kraig Finstad}.}
  \bibinfo{year}{2013}\natexlab{}.
\newblock \showarticletitle{Response to commentaries on '{The} {Usability}
  {Metric} for {User} {Experience}'}.
\newblock \bibinfo{journal}{\emph{Interact. Comput.}} \bibinfo{volume}{25},
  \bibinfo{number}{4} (\bibinfo{year}{2013}), \bibinfo{pages}{327--330}.
\newblock
\urldef\tempurl%
\url{https://doi.org/10.1093/iwc/iwt005}
\showDOI{\tempurl}


\bibitem[Ford et~al\mbox{.}(2020)]%
        {DBLP:journals/corr/abs-2009-06349}
\bibfield{author}{\bibinfo{person}{Courtney Ford}, \bibinfo{person}{Eoin~M.
  Kenny}, {and} \bibinfo{person}{Mark~T. Keane}.}
  \bibinfo{year}{2020}\natexlab{}.
\newblock \showarticletitle{Play {MNIST} For Me! User Studies on the Effects of
  Post-Hoc, Example-Based Explanations {\&} Error Rates on Debugging a Deep
  Learning, Black-Box Classifier}.
\newblock \bibinfo{journal}{\emph{CoRR}}  \bibinfo{volume}{abs/2009.06349}
  (\bibinfo{year}{2020}).
\newblock
\showeprint[arXiv]{2009.06349}
\urldef\tempurl%
\url{https://arxiv.org/abs/2009.06349}
\showURL{%
\tempurl}


\bibitem[Fox and {\"U}lk{\"u}men(2011)]%
        {fox2011distinguishing}
\bibfield{author}{\bibinfo{person}{Craig~R Fox} {and}
  \bibinfo{person}{G{\"u}lden {\"U}lk{\"u}men}.}
  \bibinfo{year}{2011}\natexlab{}.
\newblock \showarticletitle{Distinguishing two dimensions of uncertainty}.
\newblock \bibinfo{journal}{\emph{Fox, Craig R. and G{\"u}lden {\"U}lk{\"u}men
  (2011),“Distinguishing Two Dimensions of Uncertainty,” in Essays in
  Judgment and Decision Making, Brun, W., Kirkeb{\o}en, G. and Montgomery, H.,
  eds. Oslo: Universitetsforlaget}} (\bibinfo{year}{2011}).
\newblock


\bibitem[Franke et~al\mbox{.}(2015)]%
        {DBLP:conf/automotiveUI/FrankeTGKZK15}
\bibfield{author}{\bibinfo{person}{Thomas Franke}, \bibinfo{person}{Maria
  Trantow}, \bibinfo{person}{Madlen G{\"{u}}nther}, \bibinfo{person}{Josef~F.
  Krems}, \bibinfo{person}{Viktoria Zott}, {and} \bibinfo{person}{Andreas
  Keinath}.} \bibinfo{year}{2015}\natexlab{}.
\newblock \showarticletitle{Advancing electric vehicle range displays for
  enhanced user experience: the relevance of trust and adaptability}. In
  \bibinfo{booktitle}{\emph{Proceedings of the 7th International Conference on
  Automotive User Interfaces and Interactive Vehicular Applications,
  AutomotiveUI 2015, Nottingham, United Kingdom, September 1-3, 2015}},
  \bibfield{editor}{\bibinfo{person}{Gary~E. Burnett},
  \bibinfo{person}{Joseph~L. Gabbard}, \bibinfo{person}{Paul~A. Green}, {and}
  \bibinfo{person}{Sebastian Osswald}} (Eds.). \bibinfo{publisher}{{ACM}},
  \bibinfo{pages}{249--256}.
\newblock
\urldef\tempurl%
\url{https://doi.org/10.1145/2799250.2799283}
\showDOI{\tempurl}


\bibitem[Furr(2022)]%
        {furr2022psychometrics}
\bibfield{author}{\bibinfo{person}{R~Michael Furr}.}
  \bibinfo{year}{2022}\natexlab{}.
\newblock \bibinfo{booktitle}{\emph{Psychometrics: an introduction}}.
\newblock \bibinfo{publisher}{SAGE publications}.
\newblock


\bibitem[Gonzalez et~al\mbox{.}(2021)]%
        {DBLP:conf/acl/GonzalezRS21}
\bibfield{author}{\bibinfo{person}{Ana~Valeria Gonzalez}, \bibinfo{person}{Anna
  Rogers}, {and} \bibinfo{person}{Anders S{\o}gaard}.}
  \bibinfo{year}{2021}\natexlab{}.
\newblock \showarticletitle{On the Interaction of Belief Bias and
  Explanations}. In \bibinfo{booktitle}{\emph{Findings of the Association for
  Computational Linguistics: {ACL/IJCNLP} 2021, Online Event, August 1-6,
  2021}} \emph{(\bibinfo{series}{Findings of {ACL}},
  Vol.~\bibinfo{volume}{{ACL/IJCNLP} 2021})},
  \bibfield{editor}{\bibinfo{person}{Chengqing Zong}, \bibinfo{person}{Fei
  Xia}, \bibinfo{person}{Wenjie Li}, {and} \bibinfo{person}{Roberto Navigli}}
  (Eds.). \bibinfo{publisher}{Association for Computational Linguistics},
  \bibinfo{pages}{2930--2942}.
\newblock
\urldef\tempurl%
\url{https://doi.org/10.18653/v1/2021.findings-acl.259}
\showDOI{\tempurl}


\bibitem[Goyal et~al\mbox{.}(2019)]%
        {DBLP:conf/icml/GoyalWEBPL19}
\bibfield{author}{\bibinfo{person}{Yash Goyal}, \bibinfo{person}{Ziyan Wu},
  \bibinfo{person}{Jan Ernst}, \bibinfo{person}{Dhruv Batra},
  \bibinfo{person}{Devi Parikh}, {and} \bibinfo{person}{Stefan Lee}.}
  \bibinfo{year}{2019}\natexlab{}.
\newblock \showarticletitle{Counterfactual Visual Explanations}. In
  \bibinfo{booktitle}{\emph{Proceedings of the 36th International Conference on
  Machine Learning, {ICML} 2019, 9-15 June 2019, Long Beach, California,
  {USA}}} \emph{(\bibinfo{series}{Proceedings of Machine Learning Research},
  Vol.~\bibinfo{volume}{97})}, \bibfield{editor}{\bibinfo{person}{Kamalika
  Chaudhuri} {and} \bibinfo{person}{Ruslan Salakhutdinov}} (Eds.).
  \bibinfo{publisher}{{PMLR}}, \bibinfo{pages}{2376--2384}.
\newblock
\urldef\tempurl%
\url{http://proceedings.mlr.press/v97/goyal19a.html}
\showURL{%
\tempurl}


\bibitem[Green and Chen(2019)]%
        {DBLP:journals/pacmhci/GreenC19}
\bibfield{author}{\bibinfo{person}{Ben Green} {and} \bibinfo{person}{Yiling
  Chen}.} \bibinfo{year}{2019}\natexlab{}.
\newblock \showarticletitle{The Principles and Limits of Algorithm-in-the-Loop
  Decision Making}.
\newblock \bibinfo{journal}{\emph{Proc. {ACM} Hum. Comput. Interact.}}
  \bibinfo{volume}{3}, \bibinfo{number}{{CSCW}} (\bibinfo{year}{2019}),
  \bibinfo{pages}{50:1--50:24}.
\newblock
\urldef\tempurl%
\url{https://doi.org/10.1145/3359152}
\showDOI{\tempurl}


\bibitem[Grudin and MacLean(1985)]%
        {grudin1985adapting}
\bibfield{author}{\bibinfo{person}{Jonathan Grudin} {and}
  \bibinfo{person}{Allan MacLean}.} \bibinfo{year}{1985}\natexlab{}.
\newblock \showarticletitle{Adapting A Psychophysical Method To Measure
  Performance And Preference Tradeoffs In Human-Computer Interaction}. In
  \bibinfo{booktitle}{\emph{Human-Computer Interaction - INTERACT '84}
  (\bibinfo{edition}{human-computer interaction - interact '84} ed.)}.
  \bibinfo{publisher}{Elsevier Science Publishers B.V. (North-Holland)},
  \bibinfo{pages}{737--741}.
\newblock
\urldef\tempurl%
\url{https://www.microsoft.com/en-us/research/publication/adapting-psychophysical-method-measure-performance-preference-tradeoffs-human-computer-interaction/}
\showURL{%
\tempurl}


\bibitem[Haag(2025)]%
        {Haag2025TheEO}
\bibfield{author}{\bibinfo{person}{Felix Haag}.}
  \bibinfo{year}{2025}\natexlab{}.
\newblock \showarticletitle{The Effect of Explainable AI-based Decision Support
  on Human Task Performance: A Meta-Analysis}.
\newblock


\bibitem[Hart and Staveland(1988)]%
        {hart1988development}
\bibfield{author}{\bibinfo{person}{Sandra~G Hart} {and}
  \bibinfo{person}{Lowell~E Staveland}.} \bibinfo{year}{1988}\natexlab{}.
\newblock \showarticletitle{Development of NASA-TLX (Task Load Index): Results
  of empirical and theoretical research}.
\newblock In \bibinfo{booktitle}{\emph{Advances in psychology}}.
  Vol.~\bibinfo{volume}{52}. \bibinfo{publisher}{Elsevier},
  \bibinfo{pages}{139--183}.
\newblock


\bibitem[Hase and Bansal(2020)]%
        {hase-bansal-2020-evaluating}
\bibfield{author}{\bibinfo{person}{Peter Hase} {and} \bibinfo{person}{Mohit
  Bansal}.} \bibinfo{year}{2020}\natexlab{}.
\newblock \showarticletitle{Evaluating Explainable {AI}: Which Algorithmic
  Explanations Help Users Predict Model Behavior?}. In
  \bibinfo{booktitle}{\emph{Proceedings of the 58th Annual Meeting of the
  Association for Computational Linguistics}}. \bibinfo{publisher}{Association
  for Computational Linguistics}, \bibinfo{address}{Online},
  \bibinfo{pages}{5540--5552}.
\newblock
\urldef\tempurl%
\url{https://doi.org/10.18653/v1/2020.acl-main.491}
\showDOI{\tempurl}


\bibitem[Herlocker et~al\mbox{.}(2000)]%
        {DBLP:conf/cscw/HerlockerKR00}
\bibfield{author}{\bibinfo{person}{Jonathan~L. Herlocker},
  \bibinfo{person}{Joseph~A. Konstan}, {and} \bibinfo{person}{John Riedl}.}
  \bibinfo{year}{2000}\natexlab{}.
\newblock \showarticletitle{Explaining collaborative filtering
  recommendations}. In \bibinfo{booktitle}{\emph{{CSCW} 2000, Proceeding on the
  {ACM} 2000 Conference on Computer Supported Cooperative Work, Philadelphia,
  PA, USA, December 2-6, 2000}}, \bibfield{editor}{\bibinfo{person}{Wendy~A.
  Kellogg} {and} \bibinfo{person}{Steve Whittaker}} (Eds.).
  \bibinfo{publisher}{{ACM}}, \bibinfo{pages}{241--250}.
\newblock
\urldef\tempurl%
\url{https://doi.org/10.1145/358916.358995}
\showDOI{\tempurl}


\bibitem[Hornbæk(2006)]%
        {HORNBAEK200679}
\bibfield{author}{\bibinfo{person}{Kasper Hornbæk}.}
  \bibinfo{year}{2006}\natexlab{}.
\newblock \showarticletitle{Current practice in measuring usability: Challenges
  to usability studies and research}.
\newblock \bibinfo{journal}{\emph{International Journal of Human-Computer
  Studies}} \bibinfo{volume}{64}, \bibinfo{number}{2} (\bibinfo{year}{2006}),
  \bibinfo{pages}{79--102}.
\newblock
\showISSN{1071-5819}
\urldef\tempurl%
\url{https://doi.org/10.1016/j.ijhcs.2005.06.002}
\showDOI{\tempurl}


\bibitem[Hu and Bentler(1999)]%
        {Hu1999CutoffCF}
\bibfield{author}{\bibinfo{person}{{Li-tze} Hu} {and} \bibinfo{person}{Peter~M.
  Bentler}.} \bibinfo{year}{1999}\natexlab{}.
\newblock \showarticletitle{Cutoff criteria for fit indexes in covariance
  structure analysis : Conventional criteria versus new alternatives}.
\newblock \bibinfo{journal}{\emph{Structural Equation Modeling}}
  \bibinfo{volume}{6} (\bibinfo{year}{1999}), \bibinfo{pages}{1--55}.
\newblock


\bibitem[Jacovi et~al\mbox{.}(2023)]%
        {jacovi-etal-2023-neighboring}
\bibfield{author}{\bibinfo{person}{Alon Jacovi}, \bibinfo{person}{Hendrik
  Schuff}, \bibinfo{person}{Heike Adel}, \bibinfo{person}{Ngoc~Thang Vu}, {and}
  \bibinfo{person}{Yoav Goldberg}.} \bibinfo{year}{2023}\natexlab{}.
\newblock \showarticletitle{Neighboring Words Affect Human Interpretation of
  Saliency Explanations}. In \bibinfo{booktitle}{\emph{Findings of the
  Association for Computational Linguistics: ACL 2023}},
  \bibfield{editor}{\bibinfo{person}{Anna Rogers}, \bibinfo{person}{Jordan
  Boyd-Graber}, {and} \bibinfo{person}{Naoaki Okazaki}} (Eds.).
  \bibinfo{publisher}{Association for Computational Linguistics},
  \bibinfo{address}{Toronto, Canada}, \bibinfo{pages}{11816--11833}.
\newblock
\urldef\tempurl%
\url{https://doi.org/10.18653/v1/2023.findings-acl.750}
\showDOI{\tempurl}


\bibitem[J{\"o}reskog(1999)]%
        {joreskog1999large}
\bibfield{author}{\bibinfo{person}{Karl~G J{\"o}reskog}.}
  \bibinfo{year}{1999}\natexlab{}.
\newblock \showarticletitle{How large can a standardized coefficient be}.
\newblock  (\bibinfo{year}{1999}).
\newblock


\bibitem[Jorgensen et~al\mbox{.}(2022)]%
        {semTools}
\bibfield{author}{\bibinfo{person}{Terrence~D. Jorgensen},
  \bibinfo{person}{Sunthud Pornprasertmanit}, \bibinfo{person}{Alexander~M.
  Schoemann}, {and} \bibinfo{person}{Yves Rosseel}.}
  \bibinfo{year}{2022}\natexlab{}.
\newblock \bibinfo{booktitle}{\emph{\texttt{semTools}: {U}seful tools for
  structural equation modeling}}.
\newblock
\urldef\tempurl%
\url{https://CRAN.R-project.org/package=semTools}
\showURL{%
\tempurl}
\newblock
\shownote{R package version 0.5-6}.


\bibitem[Kang et~al\mbox{.}(2021)]%
        {Kang2021EstimatingBG}
\bibfield{author}{\bibinfo{person}{Hyun~Seung Kang}, \bibinfo{person}{Jeayeong
  Ji}, \bibinfo{person}{Yeji Yun}, {and} \bibinfo{person}{Kwang~Hee Han}.}
  \bibinfo{year}{2021}\natexlab{}.
\newblock \showarticletitle{Estimating Bar Graph Averages: Overcoming
  Within-the-Bar Bias}.
\newblock \bibinfo{journal}{\emph{i-Perception}}  \bibinfo{volume}{12}
  (\bibinfo{year}{2021}).
\newblock


\bibitem[Khurana et~al\mbox{.}(2021)]%
        {DBLP:conf/vl/KhuranaAC21}
\bibfield{author}{\bibinfo{person}{Anjali Khurana}, \bibinfo{person}{Parsa
  Alamzadeh}, {and} \bibinfo{person}{Parmit~K. Chilana}.}
  \bibinfo{year}{2021}\natexlab{}.
\newblock \showarticletitle{ChatrEx: Designing Explainable Chatbot Interfaces
  for Enhancing Usefulness, Transparency, and Trust}. In
  \bibinfo{booktitle}{\emph{{IEEE} Symposium on Visual Languages and
  Human-Centric Computing, {VL/HCC} 2021, St Louis, MO, USA, October 10-13,
  2021}}, \bibfield{editor}{\bibinfo{person}{Kyle~J. Harms},
  \bibinfo{person}{J{\'{a}}come Cunha}, \bibinfo{person}{Steve Oney}, {and}
  \bibinfo{person}{Caitlin Kelleher}} (Eds.). \bibinfo{publisher}{{IEEE}},
  \bibinfo{pages}{1--11}.
\newblock
\urldef\tempurl%
\url{https://doi.org/10.1109/VL/HCC51201.2021.9576440}
\showDOI{\tempurl}


\bibitem[Kim et~al\mbox{.}(2024)]%
        {Kim2024HumancenteredEO}
\bibfield{author}{\bibinfo{person}{Jenia Kim}, \bibinfo{person}{Henry
  Maathuis}, {and} \bibinfo{person}{Danielle Sent}.}
  \bibinfo{year}{2024}\natexlab{}.
\newblock \showarticletitle{Human-centered evaluation of explainable {AI}
  applications: a systematic review}.
\newblock \bibinfo{journal}{\emph{Frontiers Artif. Intell.}}
  \bibinfo{volume}{7} (\bibinfo{year}{2024}).
\newblock
\urldef\tempurl%
\url{https://doi.org/10.3389/FRAI.2024.1456486}
\showDOI{\tempurl}


\bibitem[Kirchler et~al\mbox{.}(2021)]%
        {DBLP:journals/corr/abs-2109-07869}
\bibfield{author}{\bibinfo{person}{Matthias Kirchler}, \bibinfo{person}{Martin
  Graf}, \bibinfo{person}{Marius Kloft}, {and} \bibinfo{person}{Christoph
  Lippert}.} \bibinfo{year}{2021}\natexlab{}.
\newblock \showarticletitle{Explainability Requires Interactivity}.
\newblock \bibinfo{journal}{\emph{CoRR}}  \bibinfo{volume}{abs/2109.07869}
  (\bibinfo{year}{2021}).
\newblock
\showeprint[arXiv]{2109.07869}
\urldef\tempurl%
\url{https://arxiv.org/abs/2109.07869}
\showURL{%
\tempurl}


\bibitem[K{\"o}rber(2018)]%
        {korber2018theoretical}
\bibfield{author}{\bibinfo{person}{Moritz K{\"o}rber}.}
  \bibinfo{year}{2018}\natexlab{}.
\newblock \showarticletitle{Theoretical considerations and development of a
  questionnaire to measure trust in automation}. In
  \bibinfo{booktitle}{\emph{Congress of the International Ergonomics
  Association}}. Springer, \bibinfo{pages}{13--30}.
\newblock


\bibitem[Lai et~al\mbox{.}(2020)]%
        {DBLP:conf/chi/LaiLT20}
\bibfield{author}{\bibinfo{person}{Vivian Lai}, \bibinfo{person}{Han Liu},
  {and} \bibinfo{person}{Chenhao Tan}.} \bibinfo{year}{2020}\natexlab{}.
\newblock \showarticletitle{"Why is 'Chicago' deceptive?" Towards Building
  Model-Driven Tutorials for Humans}. In \bibinfo{booktitle}{\emph{{CHI} '20:
  {CHI} Conference on Human Factors in Computing Systems, Honolulu, HI, USA,
  April 25-30, 2020}}, \bibfield{editor}{\bibinfo{person}{Regina Bernhaupt},
  \bibinfo{person}{Florian~'Floyd' Mueller}, \bibinfo{person}{David Verweij},
  \bibinfo{person}{Josh Andres}, \bibinfo{person}{Joanna McGrenere},
  \bibinfo{person}{Andy Cockburn}, \bibinfo{person}{Ignacio Avellino},
  \bibinfo{person}{Alix Goguey}, \bibinfo{person}{Pernille Bj{\o}n},
  \bibinfo{person}{Shengdong Zhao}, \bibinfo{person}{Briane~Paul Samson}, {and}
  \bibinfo{person}{Rafal Kocielnik}} (Eds.). \bibinfo{publisher}{{ACM}},
  \bibinfo{pages}{1--13}.
\newblock
\urldef\tempurl%
\url{https://doi.org/10.1145/3313831.3376873}
\showDOI{\tempurl}


\bibitem[Lai and Tan(2019)]%
        {DBLP:conf/fat/LaiT19}
\bibfield{author}{\bibinfo{person}{Vivian Lai} {and} \bibinfo{person}{Chenhao
  Tan}.} \bibinfo{year}{2019}\natexlab{}.
\newblock \showarticletitle{On Human Predictions with Explanations and
  Predictions of Machine Learning Models: {A} Case Study on Deception
  Detection}. In \bibinfo{booktitle}{\emph{Proceedings of the Conference on
  Fairness, Accountability, and Transparency, FAT* 2019, Atlanta, GA, USA,
  January 29-31, 2019}}, \bibfield{editor}{\bibinfo{person}{danah boyd} {and}
  \bibinfo{person}{Jamie~H. Morgenstern}} (Eds.). \bibinfo{publisher}{{ACM}},
  \bibinfo{pages}{29--38}.
\newblock
\urldef\tempurl%
\url{https://doi.org/10.1145/3287560.3287590}
\showDOI{\tempurl}


\bibitem[Lewis(2018)]%
        {Lewis2018MeasuringPU}
\bibfield{author}{\bibinfo{person}{James~R. Lewis}.}
  \bibinfo{year}{2018}\natexlab{}.
\newblock \showarticletitle{Measuring Perceived Usability: The CSUQ, SUS, and
  UMUX}.
\newblock \bibinfo{journal}{\emph{International Journal of Human–Computer
  Interaction}}  \bibinfo{volume}{34} (\bibinfo{year}{2018}),
  \bibinfo{pages}{1148 -- 1156}.
\newblock


\bibitem[Liang and Banks(2025)]%
        {Liang2025PerceivedSU}
\bibfield{author}{\bibinfo{person}{Qingyu Liang} {and} \bibinfo{person}{Jaime
  Banks}.} \bibinfo{year}{2025}\natexlab{}.
\newblock \showarticletitle{Perceived shared understanding between humans and
  artificial intelligence: Development and validation of a self-report scale}.
\newblock \bibinfo{journal}{\emph{Technology, Mind, and Behavior}}
  (\bibinfo{year}{2025}).
\newblock


\bibitem[Liao et~al\mbox{.}(2022)]%
        {liao2022connecting}
\bibfield{author}{\bibinfo{person}{Q~Vera Liao}, \bibinfo{person}{Yunfeng
  Zhang}, \bibinfo{person}{Ronny Luss}, \bibinfo{person}{Finale Doshi-Velez},
  {and} \bibinfo{person}{Amit Dhurandhar}.} \bibinfo{year}{2022}\natexlab{}.
\newblock \showarticletitle{Connecting Algorithmic Research and Usage Contexts:
  A Perspective of Contextualized Evaluation for Explainable AI}. In
  \bibinfo{booktitle}{\emph{Proceedings of the AAAI Conference on Human
  Computation and Crowdsourcing}}, Vol.~\bibinfo{volume}{10}.
  \bibinfo{pages}{147--159}.
\newblock


\bibitem[Liu et~al\mbox{.}(2016)]%
        {liu-etal-2016-evaluate}
\bibfield{author}{\bibinfo{person}{Chia-Wei Liu}, \bibinfo{person}{Ryan Lowe},
  \bibinfo{person}{Iulian Serban}, \bibinfo{person}{Mike Noseworthy},
  \bibinfo{person}{Laurent Charlin}, {and} \bibinfo{person}{Joelle Pineau}.}
  \bibinfo{year}{2016}\natexlab{}.
\newblock \showarticletitle{How {NOT} To Evaluate Your Dialogue System: An
  Empirical Study of Unsupervised Evaluation Metrics for Dialogue Response
  Generation}. In \bibinfo{booktitle}{\emph{Proceedings of the 2016 Conference
  on Empirical Methods in Natural Language Processing}}.
  \bibinfo{publisher}{Association for Computational Linguistics},
  \bibinfo{address}{Austin, Texas}, \bibinfo{pages}{2122--2132}.
\newblock
\urldef\tempurl%
\url{https://doi.org/10.18653/v1/D16-1230}
\showDOI{\tempurl}


\bibitem[Lundberg et~al\mbox{.}(2018)]%
        {Lundberg2018ExplainableMP}
\bibfield{author}{\bibinfo{person}{Scott~M. Lundberg}, \bibinfo{person}{Bala~G.
  Nair}, \bibinfo{person}{Monica~S. Vavilala}, \bibinfo{person}{Mayumi Horibe},
  \bibinfo{person}{Michael~J. Eisses}, \bibinfo{person}{Trevor Adams},
  \bibinfo{person}{David Liston}, \bibinfo{person}{Daniel King-Wai Low},
  \bibinfo{person}{Shu-Fang Newman}, \bibinfo{person}{Jerry~H. Kim}, {and}
  \bibinfo{person}{Su-In Lee}.} \bibinfo{year}{2018}\natexlab{}.
\newblock \showarticletitle{Explainable machine-learning predictions for the
  prevention of hypoxaemia during surgery}.
\newblock \bibinfo{journal}{\emph{Nature biomedical engineering}}
  \bibinfo{volume}{2} (\bibinfo{year}{2018}), \bibinfo{pages}{749 -- 760}.
\newblock


\bibitem[Lüdecke(2023)]%
        {sjplot}
\bibfield{author}{\bibinfo{person}{Daniel Lüdecke}.}
  \bibinfo{year}{2023}\natexlab{}.
\newblock \bibinfo{booktitle}{\emph{sjPlot: Data Visualization for Statistics
  in Social Science}}.
\newblock
\urldef\tempurl%
\url{https://CRAN.R-project.org/package=sjPlot}
\showURL{%
\tempurl}
\newblock
\shownote{R package version 2.8.14}.


\bibitem[MacLean et~al\mbox{.}(1985)]%
        {maclean1985evaluating}
\bibfield{author}{\bibinfo{person}{A MacLean}, \bibinfo{person}{PJ Barnard},
  {and} \bibinfo{person}{MD Wilson}.} \bibinfo{year}{1985}\natexlab{}.
\newblock \showarticletitle{Evaluating the human interface of a data entry
  system: user choice and performance measures yield different tradeoff
  functions}.
\newblock \bibinfo{journal}{\emph{People and computers: Designing the
  interface}} \bibinfo{volume}{5}, \bibinfo{number}{7} (\bibinfo{year}{1985}),
  \bibinfo{pages}{45--61}.
\newblock


\bibitem[Madsen et~al\mbox{.}(2023)]%
        {DBLP:journals/csur/MadsenRC23}
\bibfield{author}{\bibinfo{person}{Andreas Madsen}, \bibinfo{person}{Siva
  Reddy}, {and} \bibinfo{person}{Sarath Chandar}.}
  \bibinfo{year}{2023}\natexlab{}.
\newblock \showarticletitle{Post-hoc Interpretability for Neural {NLP:} {A}
  Survey}.
\newblock \bibinfo{journal}{\emph{{ACM} Comput. Surv.}} \bibinfo{volume}{55},
  \bibinfo{number}{8} (\bibinfo{year}{2023}), \bibinfo{pages}{155:1--155:42}.
\newblock
\urldef\tempurl%
\url{https://doi.org/10.1145/3546577}
\showDOI{\tempurl}


\bibitem[Mcdonald(1999)]%
        {Mcdonald1999TestTA}
\bibfield{author}{\bibinfo{person}{Roderick~P. Mcdonald}.}
  \bibinfo{year}{1999}\natexlab{}.
\newblock \showarticletitle{Test Theory: A Unified Treatment}.
\newblock


\bibitem[Menold and Bogner(2016)]%
        {menold2016design}
\bibfield{author}{\bibinfo{person}{N Menold} {and} \bibinfo{person}{K Bogner}.}
  \bibinfo{year}{2016}\natexlab{}.
\newblock \showarticletitle{Design of rating scales in questionnaires}.
\newblock \bibinfo{journal}{\emph{GESIS survey guidelines}}
  \bibinfo{volume}{4} (\bibinfo{year}{2016}).
\newblock


\bibitem[Newman and Scholl(2012)]%
        {Newman2012BarGD}
\bibfield{author}{\bibinfo{person}{George~E. Newman} {and}
  \bibinfo{person}{Brian~J. Scholl}.} \bibinfo{year}{2012}\natexlab{}.
\newblock \showarticletitle{Bar graphs depicting averages are perceptually
  misinterpreted: The within-the-bar bias}.
\newblock \bibinfo{journal}{\emph{Psychonomic Bulletin \& Review}}
  \bibinfo{volume}{19} (\bibinfo{year}{2012}), \bibinfo{pages}{601--607}.
\newblock


\bibitem[Nielsen and Levy(1994)]%
        {DBLP:journals/cacm/NielsenL94}
\bibfield{author}{\bibinfo{person}{Jakob Nielsen} {and}
  \bibinfo{person}{Jonathan Levy}.} \bibinfo{year}{1994}\natexlab{}.
\newblock \showarticletitle{Measuring Usability: Preference vs. Performance}.
\newblock \bibinfo{journal}{\emph{Commun. {ACM}}} \bibinfo{volume}{37},
  \bibinfo{number}{4} (\bibinfo{year}{1994}), \bibinfo{pages}{66--75}.
\newblock
\urldef\tempurl%
\url{https://doi.org/10.1145/175276.175282}
\showDOI{\tempurl}


\bibitem[Nourani et~al\mbox{.}(2019)]%
        {Nourani-Kabir-Mohseni-Ragan-2019}
\bibfield{author}{\bibinfo{person}{Mahsan Nourani}, \bibinfo{person}{Samia
  Kabir}, \bibinfo{person}{Sina Mohseni}, {and} \bibinfo{person}{Eric~D.
  Ragan}.} \bibinfo{year}{2019}\natexlab{}.
\newblock \showarticletitle{The Effects of Meaningful and Meaningless
  Explanations on Trust and Perceived System Accuracy in Intelligent Systems}.
\newblock \bibinfo{journal}{\emph{Proceedings of the AAAI Conference on Human
  Computation and Crowdsourcing}} \bibinfo{volume}{7}, \bibinfo{number}{1}
  (\bibinfo{date}{Oct.} \bibinfo{year}{2019}), \bibinfo{pages}{97--105}.
\newblock
\urldef\tempurl%
\url{https://ojs.aaai.org/index.php/HCOMP/article/view/5284}
\showURL{%
\tempurl}


\bibitem[Novikova et~al\mbox{.}(2017)]%
        {novikova-etal-2017-need}
\bibfield{author}{\bibinfo{person}{Jekaterina Novikova},
  \bibinfo{person}{Ond{\v{r}}ej Du{\v{s}}ek}, \bibinfo{person}{Amanda
  Cercas~Curry}, {and} \bibinfo{person}{Verena Rieser}.}
  \bibinfo{year}{2017}\natexlab{}.
\newblock \showarticletitle{Why We Need New Evaluation Metrics for {NLG}}. In
  \bibinfo{booktitle}{\emph{Proceedings of the 2017 Conference on Empirical
  Methods in Natural Language Processing}}. \bibinfo{publisher}{Association for
  Computational Linguistics}, \bibinfo{address}{Copenhagen, Denmark},
  \bibinfo{pages}{2241--2252}.
\newblock
\urldef\tempurl%
\url{https://doi.org/10.18653/v1/D17-1238}
\showDOI{\tempurl}


\bibitem[Poursabzi{-}Sangdeh et~al\mbox{.}(2021)]%
        {DBLP:conf/chi/Poursabzi-Sangdeh21}
\bibfield{author}{\bibinfo{person}{Forough Poursabzi{-}Sangdeh},
  \bibinfo{person}{Daniel~G. Goldstein}, \bibinfo{person}{Jake~M. Hofman},
  \bibinfo{person}{Jennifer~Wortman Vaughan}, {and} \bibinfo{person}{Hanna~M.
  Wallach}.} \bibinfo{year}{2021}\natexlab{}.
\newblock \showarticletitle{Manipulating and Measuring Model Interpretability}.
  In \bibinfo{booktitle}{\emph{{CHI} '21: {CHI} Conference on Human Factors in
  Computing Systems, Virtual Event / Yokohama, Japan, May 8-13, 2021}},
  \bibfield{editor}{\bibinfo{person}{Yoshifumi Kitamura},
  \bibinfo{person}{Aaron Quigley}, \bibinfo{person}{Katherine Isbister},
  \bibinfo{person}{Takeo Igarashi}, \bibinfo{person}{Pernille Bj{\o}rn}, {and}
  \bibinfo{person}{Steven~Mark Drucker}} (Eds.). \bibinfo{publisher}{{ACM}},
  \bibinfo{pages}{237:1--237:52}.
\newblock
\urldef\tempurl%
\url{https://doi.org/10.1145/3411764.3445315}
\showDOI{\tempurl}


\bibitem[Raykov(2001)]%
        {Raykov2001EstimationOC}
\bibfield{author}{\bibinfo{person}{Tenko Raykov}.}
  \bibinfo{year}{2001}\natexlab{}.
\newblock \showarticletitle{Estimation of congeneric scale reliability using
  covariance structure analysis with nonlinear constraints.}
\newblock \bibinfo{journal}{\emph{The British journal of mathematical and
  statistical psychology}}  \bibinfo{volume}{54 Pt 2} (\bibinfo{year}{2001}),
  \bibinfo{pages}{315--23}.
\newblock


\bibitem[Reiter(2018)]%
        {reiter-2018-structured}
\bibfield{author}{\bibinfo{person}{Ehud Reiter}.}
  \bibinfo{year}{2018}\natexlab{}.
\newblock \showarticletitle{A Structured Review of the Validity of {BLEU}}.
\newblock \bibinfo{journal}{\emph{Computational Linguistics}}
  \bibinfo{volume}{44}, \bibinfo{number}{3} (\bibinfo{date}{Sept.}
  \bibinfo{year}{2018}), \bibinfo{pages}{393--401}.
\newblock
\urldef\tempurl%
\url{https://doi.org/10.1162/coli_a_00322}
\showDOI{\tempurl}


\bibitem[Revelle(2022)]%
        {psych}
\bibfield{author}{\bibinfo{person}{William Revelle}.}
  \bibinfo{year}{2022}\natexlab{}.
\newblock \bibinfo{booktitle}{\emph{psych: Procedures for Psychological,
  Psychometric, and Personality Research}}.
\newblock Northwestern University, Evanston, Illinois.
\newblock
\urldef\tempurl%
\url{https://CRAN.R-project.org/package=psych}
\showURL{%
\tempurl}
\newblock
\shownote{R package version 2.2.9}.


\bibitem[Ribera and Lapedriza(2019)]%
        {DBLP:conf/iui/RiberaL19}
\bibfield{author}{\bibinfo{person}{Mireia Ribera} {and}
  \bibinfo{person}{{\`{A}}gata Lapedriza}.} \bibinfo{year}{2019}\natexlab{}.
\newblock \showarticletitle{Can we do better explanations? {A} proposal of
  user-centered explainable {AI}}. In \bibinfo{booktitle}{\emph{Joint
  Proceedings of the {ACM} {IUI} 2019 Workshops co-located with the 24th {ACM}
  Conference on Intelligent User Interfaces {(ACM} {IUI} 2019), Los Angeles,
  USA, March 20, 2019}} \emph{(\bibinfo{series}{{CEUR} Workshop Proceedings},
  Vol.~\bibinfo{volume}{2327})}, \bibfield{editor}{\bibinfo{person}{Christoph
  Trattner}, \bibinfo{person}{Denis Parra}, {and} \bibinfo{person}{Nathalie
  Riche}} (Eds.). \bibinfo{publisher}{CEUR-WS.org}.
\newblock
\urldef\tempurl%
\url{http://ceur-ws.org/Vol-2327/IUI19WS-ExSS2019-12.pdf}
\showURL{%
\tempurl}


\bibitem[Ribes et~al\mbox{.}(2021)]%
        {DBLP:conf/interact/RibesHPDPGS21}
\bibfield{author}{\bibinfo{person}{Delphine Ribes}, \bibinfo{person}{Nicolas
  Henchoz}, \bibinfo{person}{H{\'{e}}l{\`{e}}ne Portier}, \bibinfo{person}{Lara
  D{\'{e}}fayes}, \bibinfo{person}{Thanh{-}Trung Phan}, \bibinfo{person}{Daniel
  Gatica{-}Perez}, {and} \bibinfo{person}{Andreas Sonderegger}.}
  \bibinfo{year}{2021}\natexlab{}.
\newblock \showarticletitle{Trust Indicators and Explainable {AI:} {A} Study on
  User Perceptions}. In \bibinfo{booktitle}{\emph{Human-Computer Interaction -
  {INTERACT} 2021 - 18th {IFIP} {TC} 13 International Conference, Bari, Italy,
  August 30 - September 3, 2021, Proceedings, Part {II}}}
  \emph{(\bibinfo{series}{Lecture Notes in Computer Science},
  Vol.~\bibinfo{volume}{12933})}, \bibfield{editor}{\bibinfo{person}{Carmelo
  Ardito}, \bibinfo{person}{Rosa Lanzilotti}, \bibinfo{person}{Alessio
  Malizia}, \bibinfo{person}{Helen Petrie}, \bibinfo{person}{Antonio Piccinno},
  \bibinfo{person}{Giuseppe Desolda}, {and} \bibinfo{person}{Kori Inkpen}}
  (Eds.). \bibinfo{publisher}{Springer}, \bibinfo{pages}{662--671}.
\newblock
\urldef\tempurl%
\url{https://doi.org/10.1007/978-3-030-85616-8\_39}
\showDOI{\tempurl}


\bibitem[Rozenblit and Keil(2002)]%
        {Rozenblit2002TheML}
\bibfield{author}{\bibinfo{person}{Leon Rozenblit} {and}
  \bibinfo{person}{Frank~C. Keil}.} \bibinfo{year}{2002}\natexlab{}.
\newblock \showarticletitle{The misunderstood limits of folk science: an
  illusion of explanatory depth}.
\newblock \bibinfo{journal}{\emph{Cognitive science}}  \bibinfo{volume}{26 5}
  (\bibinfo{year}{2002}), \bibinfo{pages}{521--562}.
\newblock


\bibitem[Rutjes et~al\mbox{.}(2019)]%
        {rutjes2019-chi}
\bibfield{author}{\bibinfo{person}{Heleen Rutjes}, \bibinfo{person}{Martijn
  Willemsen}, {and} \bibinfo{person}{Wijnand IJsselsteijn}.}
  \bibinfo{year}{2019}\natexlab{}.
\newblock \showarticletitle{Considerations on explainable AI and
  users{\textquoteright} mental models}. In \bibinfo{booktitle}{\emph{Where is
  the Human? Bridging the Gap Between AI and HCI}}.
  \bibinfo{publisher}{Association for Computing Machinery, Inc},
  \bibinfo{address}{United States}.
\newblock
\newblock
\shownote{CHI 2019 Workshop : Where is the Human? Bridging the Gap Between AI
  and HCI ; Conference date: 04-05-2019 Through 04-05-2019}.


\bibitem[Schlegel et~al\mbox{.}(2022)]%
        {DBLP:conf/icwsm/SchlegelGB22}
\bibfield{author}{\bibinfo{person}{Viktor Schlegel},
  \bibinfo{person}{Erick~Mendez Guzman}, {and} \bibinfo{person}{Riza
  Batista{-}Navarro}.} \bibinfo{year}{2022}\natexlab{}.
\newblock \showarticletitle{Towards Human-Centred Explainability Benchmarks For
  Text Classification}. In \bibinfo{booktitle}{\emph{Workshop Proceedings of
  the 16th International {AAAI} Conference on Web and Social Media, {ICWSM}
  2022 Workshops, Atlanta, Georgia, {USA} [hybrid], June 6, 2022}},
  \bibfield{editor}{\bibinfo{person}{Pedro O. S.~Vaz de~Melo},
  \bibinfo{person}{Wei Jeng}, {and} \bibinfo{person}{Cody Buntain}} (Eds.).
\newblock
\urldef\tempurl%
\url{https://doi.org/10.36190/2022.93}
\showDOI{\tempurl}


\bibitem[Schrills et~al\mbox{.}(2022)]%
        {schrills_kargl_bickel_franke_2022}
\bibfield{author}{\bibinfo{person}{Tim P~P Schrills}, \bibinfo{person}{Susanne
  Kargl}, \bibinfo{person}{Mona Bickel}, {and} \bibinfo{person}{Thomas
  Franke}.} \bibinfo{year}{2022}\natexlab{}.
\newblock \bibinfo{title}{Perceive, Understand \& Predict - Empirical
  Indication for Facets in Subjective Information Processing Awareness}.
\newblock
\newblock
\urldef\tempurl%
\url{https://doi.org/10.31234/osf.io/3n95u}
\showDOI{\tempurl}


\bibitem[Schuff et~al\mbox{.}(2022a)]%
        {https://doi.org/10.48550/arxiv.2210.07126}
\bibfield{author}{\bibinfo{person}{Hendrik Schuff}, \bibinfo{person}{Heike
  Adel}, \bibinfo{person}{Peng Qi}, {and} \bibinfo{person}{Ngoc~Thang Vu}.}
  \bibinfo{year}{2022}\natexlab{a}.
\newblock \showarticletitle{Challenges in Explanation Quality Evaluation}.
\newblock \bibinfo{journal}{\emph{CoRR}}  \bibinfo{volume}{abs/2210.07126}
  (\bibinfo{year}{2022}).
\newblock
\urldef\tempurl%
\url{https://doi.org/10.48550/ARXIV.2210.07126}
\showDOI{\tempurl}
\showeprint[arXiv]{2210.07126}


\bibitem[Schuff et~al\mbox{.}(2020)]%
        {schuff-etal-2020-f1}
\bibfield{author}{\bibinfo{person}{Hendrik Schuff}, \bibinfo{person}{Heike
  Adel}, {and} \bibinfo{person}{Ngoc~Thang Vu}.}
  \bibinfo{year}{2020}\natexlab{}.
\newblock \showarticletitle{{F}1 is {N}ot {E}nough! {M}odels and {E}valuation
  {T}owards {U}ser-{C}entered {E}xplainable {Q}uestion {A}nswering}. In
  \bibinfo{booktitle}{\emph{Proceedings of the 2020 Conference on Empirical
  Methods in Natural Language Processing (EMNLP)}}.
  \bibinfo{publisher}{Association for Computational Linguistics},
  \bibinfo{address}{Online}, \bibinfo{pages}{7076--7095}.
\newblock
\urldef\tempurl%
\url{https://doi.org/10.18653/v1/2020.emnlp-main.575}
\showDOI{\tempurl}


\bibitem[Schuff et~al\mbox{.}(2022b)]%
        {10.1145/3531146.3533127}
\bibfield{author}{\bibinfo{person}{Hendrik Schuff}, \bibinfo{person}{Alon
  Jacovi}, \bibinfo{person}{Heike Adel}, \bibinfo{person}{Yoav Goldberg}, {and}
  \bibinfo{person}{Ngoc~Thang Vu}.} \bibinfo{year}{2022}\natexlab{b}.
\newblock \showarticletitle{Human Interpretation of Saliency-Based Explanation
  Over Text}. In \bibinfo{booktitle}{\emph{2022 ACM Conference on Fairness,
  Accountability, and Transparency}} (Seoul, Republic of Korea)
  \emph{(\bibinfo{series}{FAccT '22})}. \bibinfo{publisher}{Association for
  Computing Machinery}, \bibinfo{address}{New York, NY, USA},
  \bibinfo{pages}{611–636}.
\newblock
\showISBNx{9781450393522}
\urldef\tempurl%
\url{https://doi.org/10.1145/3531146.3533127}
\showDOI{\tempurl}


\bibitem[Schulz et~al\mbox{.}(2015)]%
        {DBLP:conf/cogsci/SchulzTRSG15}
\bibfield{author}{\bibinfo{person}{Eric Schulz}, \bibinfo{person}{Joshua~B.
  Tenenbaum}, \bibinfo{person}{David~N. Reshef}, \bibinfo{person}{Maarten
  Speekenbrink}, {and} \bibinfo{person}{Samuel Gershman}.}
  \bibinfo{year}{2015}\natexlab{}.
\newblock \showarticletitle{Assessing the Perceived Predictability of
  Functions}. In \bibinfo{booktitle}{\emph{Proceedings of the 37th Annual
  Meeting of the Cognitive Science Society, CogSci 2015, Pasadena, California,
  USA, July 22-25, 2015}}, \bibfield{editor}{\bibinfo{person}{David~C. Noelle},
  \bibinfo{person}{Rick Dale}, \bibinfo{person}{Anne~S. Warlaumont},
  \bibinfo{person}{Jeff Yoshimi}, \bibinfo{person}{Teenie Matlock},
  \bibinfo{person}{Carolyn~D. Jennings}, {and} \bibinfo{person}{Paul~P.
  Maglio}} (Eds.). \bibinfo{publisher}{cognitivesciencesociety.org}.
\newblock
\urldef\tempurl%
\url{https://mindmodeling.org/cogsci2015/papers/0365/index.html}
\showURL{%
\tempurl}


\bibitem[Silva et~al\mbox{.}(2023)]%
        {Silva2022ExplainableAI}
\bibfield{author}{\bibinfo{person}{Andrew Silva}, \bibinfo{person}{Mariah
  Schrum}, \bibinfo{person}{Erin Hedlund{-}Botti}, \bibinfo{person}{Nakul
  Gopalan}, {and} \bibinfo{person}{Matthew~C. Gombolay}.}
  \bibinfo{year}{2023}\natexlab{}.
\newblock \showarticletitle{Explainable Artificial Intelligence: Evaluating the
  Objective and Subjective Impacts of xAI on Human-Agent Interaction}.
\newblock \bibinfo{journal}{\emph{Int. J. Hum. Comput. Interact.}}
  \bibinfo{volume}{39}, \bibinfo{number}{7} (\bibinfo{year}{2023}),
  \bibinfo{pages}{1390--1404}.
\newblock
\urldef\tempurl%
\url{https://doi.org/10.1080/10447318.2022.2101698}
\showDOI{\tempurl}


\bibitem[Sulem et~al\mbox{.}(2018)]%
        {sulem-etal-2018-bleu}
\bibfield{author}{\bibinfo{person}{Elior Sulem}, \bibinfo{person}{Omri Abend},
  {and} \bibinfo{person}{Ari Rappoport}.} \bibinfo{year}{2018}\natexlab{}.
\newblock \showarticletitle{{BLEU} is Not Suitable for the Evaluation of Text
  Simplification}. In \bibinfo{booktitle}{\emph{Proceedings of the 2018
  Conference on Empirical Methods in Natural Language Processing}}.
  \bibinfo{publisher}{Association for Computational Linguistics},
  \bibinfo{address}{Brussels, Belgium}, \bibinfo{pages}{738--744}.
\newblock
\urldef\tempurl%
\url{https://doi.org/10.18653/v1/D18-1081}
\showDOI{\tempurl}


\bibitem[Thomas and Uminsky(2022)]%
        {thomas_reliance_2022}
\bibfield{author}{\bibinfo{person}{Rachel~L Thomas} {and}
  \bibinfo{person}{David Uminsky}.} \bibinfo{year}{2022}\natexlab{}.
\newblock \showarticletitle{Reliance on metrics is a fundamental challenge for
  {AI}}.
\newblock \bibinfo{journal}{\emph{Patterns}} \bibinfo{volume}{3},
  \bibinfo{number}{5} (\bibinfo{year}{2022}), \bibinfo{pages}{100476}.
\newblock
\newblock
\shownote{Publisher: Elsevier}.


\bibitem[Tractinsky and Meyer(2001)]%
        {TRACTINSKY2001845}
\bibfield{author}{\bibinfo{person}{Noam Tractinsky} {and}
  \bibinfo{person}{Joachim Meyer}.} \bibinfo{year}{2001}\natexlab{}.
\newblock \showarticletitle{Task structure and the apparent duration of
  hierarchical search}.
\newblock \bibinfo{journal}{\emph{International Journal of Human-Computer
  Studies}} \bibinfo{volume}{55}, \bibinfo{number}{5} (\bibinfo{year}{2001}),
  \bibinfo{pages}{845--860}.
\newblock
\showISSN{1071-5819}
\urldef\tempurl%
\url{https://doi.org/10.1006/ijhc.2001.0506}
\showDOI{\tempurl}


\bibitem[Wang and Vasconcelos(2020)]%
        {DBLP:conf/cvpr/WangV20}
\bibfield{author}{\bibinfo{person}{Pei Wang} {and} \bibinfo{person}{Nuno
  Vasconcelos}.} \bibinfo{year}{2020}\natexlab{}.
\newblock \showarticletitle{{SCOUT:} Self-Aware Discriminant Counterfactual
  Explanations}. In \bibinfo{booktitle}{\emph{2020 {IEEE/CVF} Conference on
  Computer Vision and Pattern Recognition, {CVPR} 2020, Seattle, WA, USA, June
  13-19, 2020}}. \bibinfo{publisher}{Computer Vision Foundation / {IEEE}},
  \bibinfo{pages}{8978--8987}.
\newblock
\urldef\tempurl%
\url{https://doi.org/10.1109/CVPR42600.2020.00900}
\showDOI{\tempurl}


\bibitem[Wang and Yin(2021)]%
        {DBLP:conf/iui/WangY21}
\bibfield{author}{\bibinfo{person}{Xinru Wang} {and} \bibinfo{person}{Ming
  Yin}.} \bibinfo{year}{2021}\natexlab{}.
\newblock \showarticletitle{Are Explanations Helpful? {A} Comparative Study of
  the Effects of Explanations in AI-Assisted Decision-Making}. In
  \bibinfo{booktitle}{\emph{{IUI} '21: 26th International Conference on
  Intelligent User Interfaces, College Station, TX, USA, April 13-17, 2021}},
  \bibfield{editor}{\bibinfo{person}{Tracy Hammond}, \bibinfo{person}{Katrien
  Verbert}, \bibinfo{person}{Dennis Parra}, \bibinfo{person}{Bart~P.
  Knijnenburg}, \bibinfo{person}{John O'Donovan}, {and} \bibinfo{person}{Paul
  Teale}} (Eds.). \bibinfo{publisher}{{ACM}}, \bibinfo{pages}{318--328}.
\newblock
\urldef\tempurl%
\url{https://doi.org/10.1145/3397481.3450650}
\showDOI{\tempurl}


\bibitem[Willis(2004)]%
        {willis2004cognitive}
\bibfield{author}{\bibinfo{person}{Gordon~B Willis}.}
  \bibinfo{year}{2004}\natexlab{}.
\newblock \bibinfo{booktitle}{\emph{Cognitive interviewing: A tool for
  improving questionnaire design}}.
\newblock \bibinfo{publisher}{sage publications}.
\newblock


\bibitem[Yeh and Wickens(1988)]%
        {doi:10.1177/001872088803000110}
\bibfield{author}{\bibinfo{person}{Yei-Yu Yeh} {and}
  \bibinfo{person}{Christopher~D. Wickens}.} \bibinfo{year}{1988}\natexlab{}.
\newblock \showarticletitle{Dissociation of Performance and Subjective Measures
  of Workload}.
\newblock \bibinfo{journal}{\emph{Human Factors}} \bibinfo{volume}{30},
  \bibinfo{number}{1} (\bibinfo{year}{1988}), \bibinfo{pages}{111--120}.
\newblock
\urldef\tempurl%
\url{https://doi.org/10.1177/001872088803000110}
\showDOI{\tempurl}
\showeprint{https://doi.org/10.1177/001872088803000110}


\bibitem[Zhang et~al\mbox{.}(2020)]%
        {DBLP:conf/fat/ZhangLB20}
\bibfield{author}{\bibinfo{person}{Yunfeng Zhang}, \bibinfo{person}{Q.~Vera
  Liao}, {and} \bibinfo{person}{Rachel K.~E. Bellamy}.}
  \bibinfo{year}{2020}\natexlab{}.
\newblock \showarticletitle{Effect of confidence and explanation on accuracy
  and trust calibration in AI-assisted decision making}. In
  \bibinfo{booktitle}{\emph{FAT* '20: Conference on Fairness, Accountability,
  and Transparency, Barcelona, Spain, January 27-30, 2020}},
  \bibfield{editor}{\bibinfo{person}{Mireille Hildebrandt},
  \bibinfo{person}{Carlos Castillo}, \bibinfo{person}{L.~Elisa Celis},
  \bibinfo{person}{Salvatore Ruggieri}, \bibinfo{person}{Linnet Taylor}, {and}
  \bibinfo{person}{Gabriela Zanfir{-}Fortuna}} (Eds.).
  \bibinfo{publisher}{{ACM}}, \bibinfo{pages}{295--305}.
\newblock
\urldef\tempurl%
\url{https://doi.org/10.1145/3351095.3372852}
\showDOI{\tempurl}


\bibitem[Zou(2007)]%
        {Zou2007TowardUC}
\bibfield{author}{\bibinfo{person}{Guang~Yong Zou}.}
  \bibinfo{year}{2007}\natexlab{}.
\newblock \showarticletitle{Toward using confidence intervals to compare
  correlations.}
\newblock \bibinfo{journal}{\emph{Psychological methods}}  \bibinfo{volume}{12
  4} (\bibinfo{year}{2007}), \bibinfo{pages}{399--413}.
\newblock


\end{thebibliography}

\newpage
\appendix

\section{Item Generation}\label{sec:appendix_psp_item_generation}
\subsection{Initial Item Pool}\label{sec:appendix_psp_item_pool}
Our initial item pool contains 60 items.
Concretely, these items are:
\begin{itemize}
        \item "My knowledge about the system behavior is complete."
        \item "I know a lot about the system's behavior."
        \item "I do not need to learn more about the system's behavior."
        \item "I understand how the system functions."
        \item "The system behaves as expected (including 'controlled random')."
        \item "I can explain how the system functions."
        \item "I have a lot of experience with this system."
        \item "I observed the system's behavior in many different situations."
        \item "I know enough about the system to predict how it behaves."
        \item "Seeing more of the system's behavior will not surprise me."
        \item "Based on past responses, I know the responses the system will likely give me."
        \item "I have interacted with the system many times."
        \item "I am able to anticipate how the system will respond after having used it."
        \item "I have an understanding of the system based on its responses to the given input."
        \item "I have a comfortable feeling of knowing."
        \item "I engaged with the system a lot."
        \item "I have personal experiences with the system."
        \item "I have an educated guess on how the system will behave."
        \item "I have experience with the system."
        \item "I can identify rules and patterns in the system's decisions.
        \item "The system does not take random decisions."
        \item "The system takes consistent decisions."
        \item "I feel a sense of order and direction."
        \item "There is a consistent pattern in the system's decisions."
        \item "Given a fixed input, the system always takes the same decision"
        \item "I know the reasons for the system's decisions."
        \item "I know what brought the system to its decisions."
        \item "The system's logic is similar to mine."
        \item "The system's knowledge is similar to mine."
        \item "The system is following a certain pattern."
        \item "I get an idea of how the algorithm works."
        \item "I know how the system is likely to interpret input."
        \item "I feel like the results the system gives are reliable."
        \item "The system behaves in a predictable manner."
        \item "The system gives consistent results."
        \item "The system gives reliable results."
        \item "The system normally behaves in a consistent manner."
        \item "I know how the system will respond in a given situation."
        \item "I am able to predict how the system will react."
        \item "The system's decisions are predictable."
        \item "I have an understanding how the system makes its judgements."
        \item "The system's decision process is straightforward."
        \item "I know how the system 'thinks'."
        \item "I have a good overall understanding of the system."
        \item "I know the responses the system will likely give."
        \item "I can guess how the system will behave."
        \item "I know if the system is biased."
        \item "I know how the system was created."
        \item "I know the system's quirks."
        \item "I have an understanding of why the systems responded in the way it did."
        \item "I can guess how the system will react."
        \item "I feel like I can predict how the system will behave."
        \item "I have knowledge about what the system is supposed to do."
        \item "I know what I can expect from the system."
        \item "I am certain about the system's behavior."
        \item "I can guess how the system comes to its conclusions."
        \item "I can estimate how the system comes to its conclusions."
        \item "I can usually predict if the system can answer the question I have in mind for it."
        \item "I know what I can and cannot do with the system."
        \item "I know how to use the system efficiently."
\end{itemize}

\subsection{Intermediate Item Pool}\label{sec:appendix_psp_item_intermediate_pool}
After the expert ratings and two rounds of cognitive interviews with target population participants, we filtered and refined the item pool used in our first large-scale evaluation to the following items.
We mark items included in our final scale version with ``$\ast$'' and report reasons for our removal decisions in parentheses:
\begin{itemize}
    \item $\ast$"I can tell which responses the system will likely give."
    \item "I can predict how the system will behave most of the time." ($>0.85$ inter-item correlation to previous item)
    \item $\ast$"The system behaves in a predictable manner."
    \item "The system's responses are predictable for me." (strong content similarity to previous item, removed for brevity)
    \item "I can identify rules and patterns in the system's responses." (lower item discrimination value than all other items)
    \item "I understand the system well enough to predict how it behaves." (least categorizable item regarding our three-facet theory)
    \item $\ast$"There is a consistent pattern in the system's behavior."
    \item $\ast$"I can tell the reasons for the system's decisions."
    \item "I have an understanding of why the systems responded in the way it did." (strong similarity to previous item but more complex wording, removed for brevity)
    \item $\ast$"I observed enough system responses to predict how the system behaves."
    \item $\ast$"Based on past system responses, I know the responses the system will likely give me."
    \item "The number of system responses I have seen is large enough to predict the system's behavior." ($>0.85$ inter-item correlation to previous item)
\end{itemize}

\subsection{Worked Scoring Example}\label{sec:appendix_psp_scoring_example}
To illustrate how the PSP scale can be scored, \Cref{tab:psps_scoring_example} shows one example response pattern. The primary score is the overall PSP score, computed as the mean across all six items. When a more granular analysis is warranted, the effective, epistemic, and aleatory subscale scores can additionally be computed as the means of the respective item pairs.

\ifarxivmode\begin{table}[ht]\else\begin{table}[h]\fi
    \centering
    \begin{tabular}{lc}
        \toprule
        \textbf{Item} & \textbf{Example rating} \\
        \midrule
        EF-1 & 4 \\
        EF-2 & 5 \\
        \addlinespace
        EP-1 & 3 \\
        EP-2 & 3 \\
        \addlinespace
        AL-1 & 6 \\
        AL-2 & 5 \\
        \midrule
        \textbf{Overall PSP} & \textbf{4.3} \\
        \textbf{Effective PSP (EF)} & \textbf{4.5} \\
        \textbf{Epistemic PSP (EP)} & \textbf{3.0} \\
        \textbf{Aleatory PSP (AL)} & \textbf{5.5} \\
        \bottomrule
    \end{tabular}
    \caption{Illustrative scoring example for the PSP scale. The overall PSP score is the mean across all six items, and the EF, EP, and AL subscale scores are the means of the corresponding item pairs.}
    \label{tab:psps_scoring_example}
\end{table}

\section{Scale Evaluation}
\subsubsection{Colored Shapes Experiment Interface.}\label{sec:appendix_psp_colored_shapes_scenarios}
In addition to the scenario depicted in \Cref{fig:scenario_mixed}, we included the scenarios depicted in \Cref{fig:scenario_mixed_less_certainty,fig:scenario_mixed_longer,fig:scenario_aleatory,fig:scenario_full} as described in \Cref{sec:humans_scale_development}.

\begin{figure}
    \centering
    \includegraphics[width=0.45\textwidth]{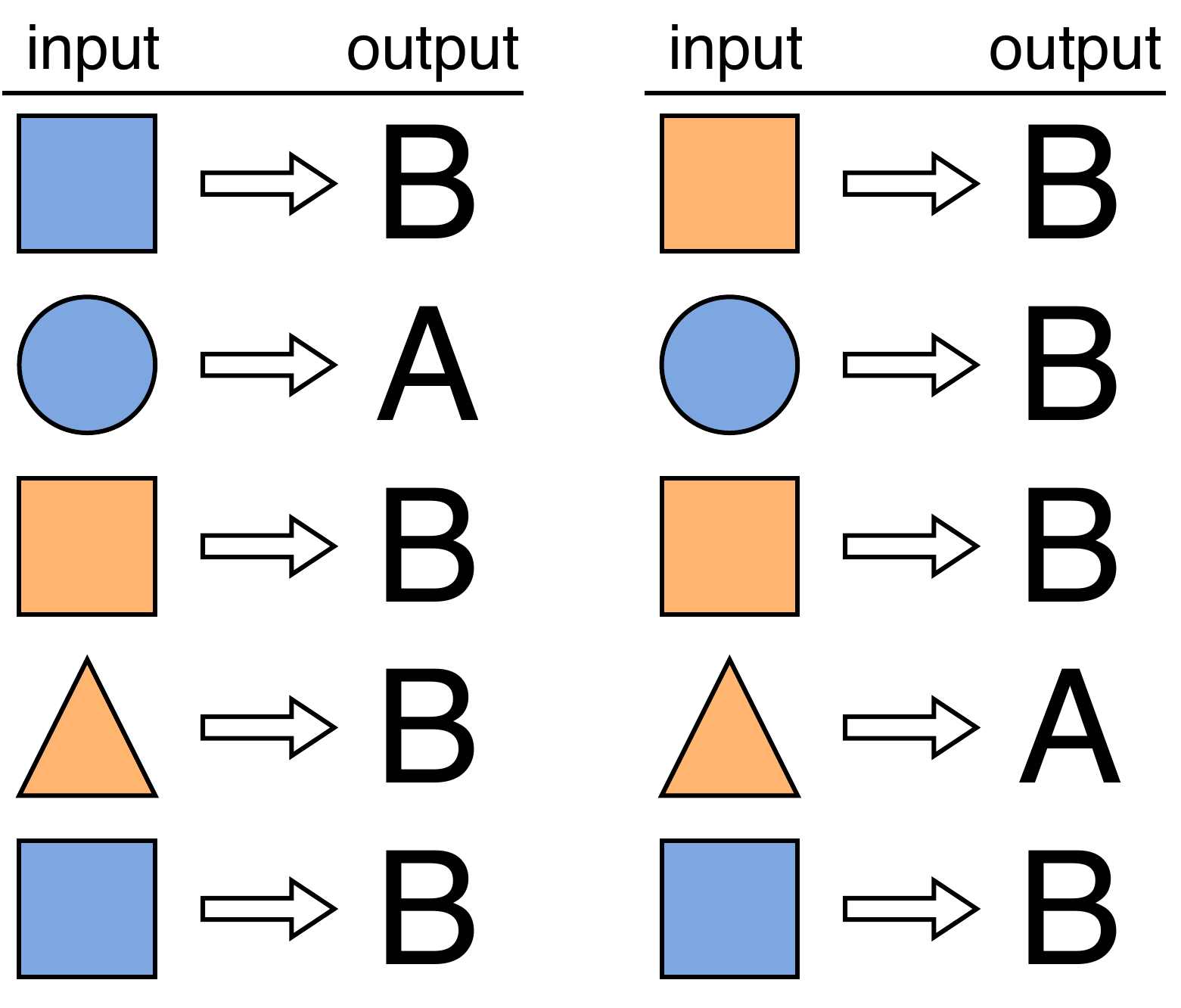}
    \caption{A scenario with mixed uncertainty, but slightly more aleatory uncertainty than the scenario shown in \Cref{fig:scenario_mixed} (i.e., less predictability). We refer to this scenario as \textsc{mixed-less-aleatory}.}
    \label{fig:scenario_mixed_less_certainty}
\end{figure}

\begin{figure}
    \centering
    \includegraphics[width=0.9\textwidth]{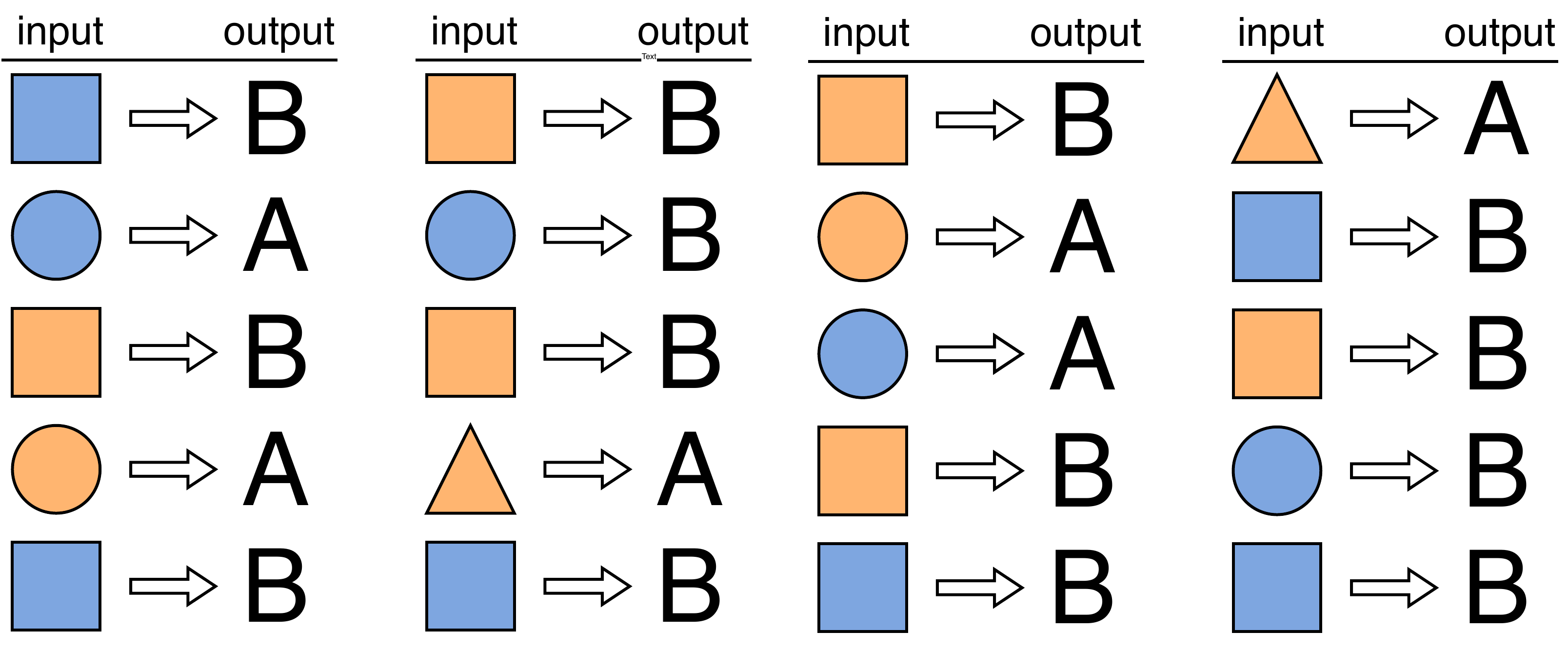}
    \caption{A scenario with mixed uncertainty, but twice the number of examples of the scenario shown in \Cref{fig:scenario_mixed} (i.e., more epistemic predictability). We refer to this scenario as \textsc{mixed-more-epistemic}.}
    \label{fig:scenario_mixed_longer}
\end{figure}

\begin{figure}
    \centering
    \includegraphics[width=0.9\textwidth]{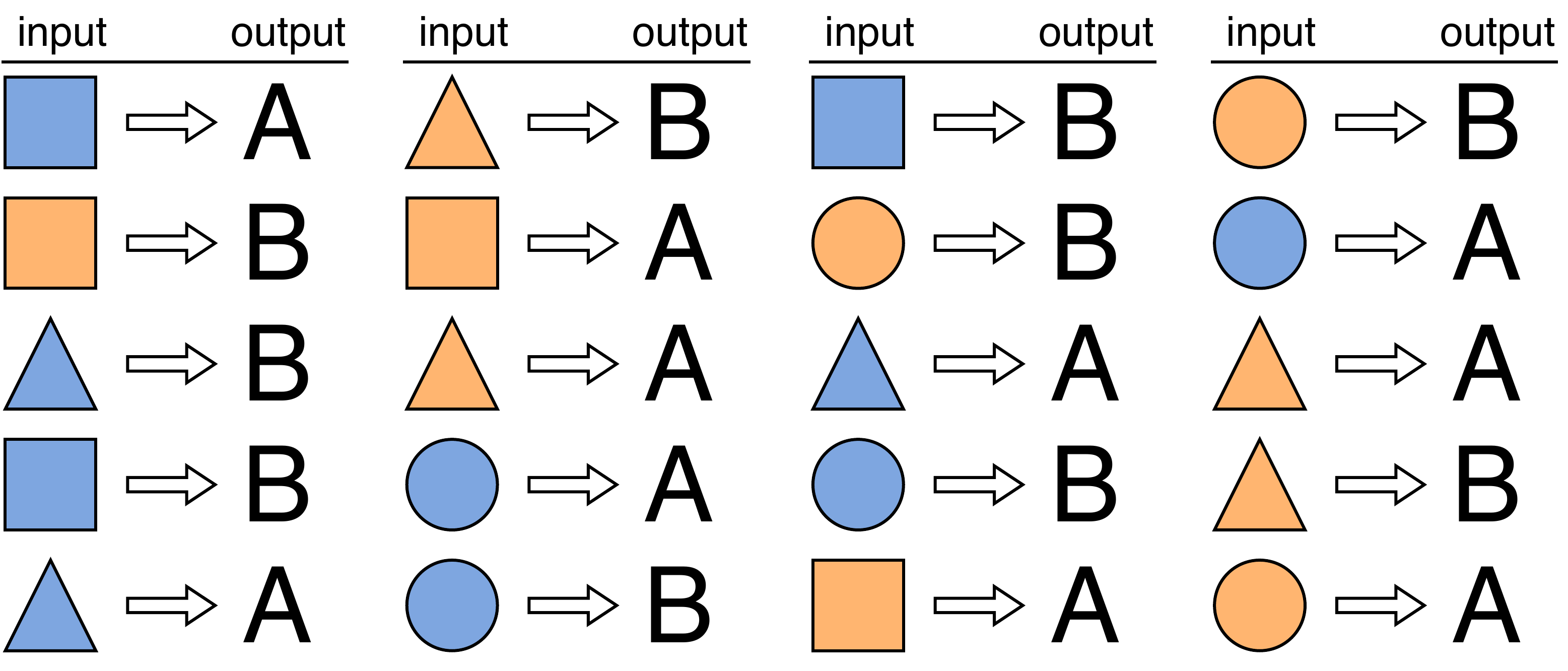}
    \caption{A scenario with strong aleatory uncertainty (i.e., low predictability). We refer to this scenario as \textsc{low-aleatory}.}
    \label{fig:scenario_aleatory}
\end{figure}

\begin{figure}
    \centering
    \includegraphics[width=0.9\textwidth]{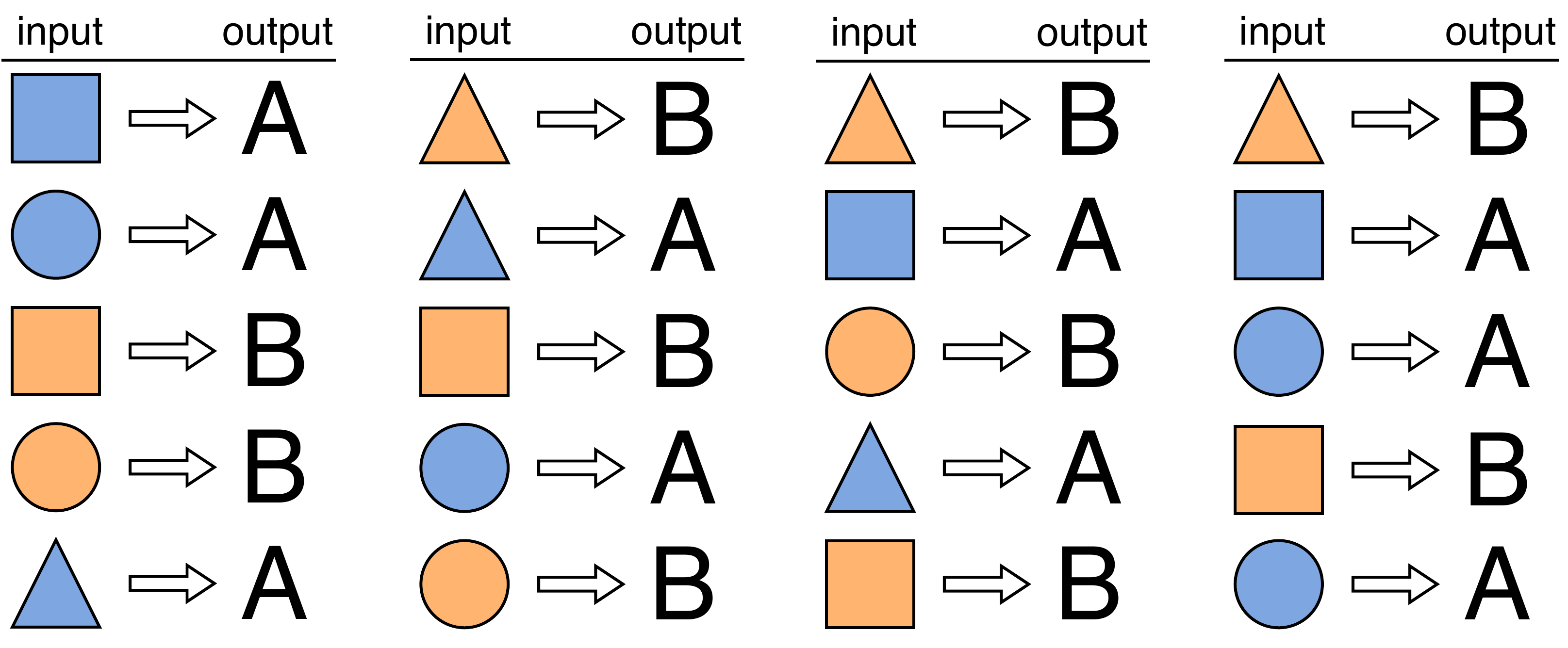}
    \caption{A scenario with a high degree of epistemic and aleatory certainty. We refer to this scenario as \textsc{high-both}.}
    \label{fig:scenario_full}
\end{figure}

\FloatBarrier

\subsection{Differentiation by Known Groups}\label{sec:appendix_psp_known_groups}
\Cref{tab:known_groups_hsd} shows details of the Tukey HSD post hoc test to determine significant differences between our scenarios.
\ifarxivmode\begin{table}[ht]\else\begin{table}[h]\fi
    \centering
    \resizebox{\linewidth}{!}{
    \begin{tabular}{ccccc}
        \toprule
        \textbf{Scenario pair} & \textbf{difference} & \textbf{CI-upper} & \textbf{CI-lower} & \textbf{p (adjusted)} \\
        \midrule
        \textbf{\textsc{high-both} vs. \textsc{low-aleatory}}  &                    2.025 & 1.177 & 2.873 & \textbf{$<$0.0001}\\
        \textbf{\textsc{mixed} vs. \textsc{low-aleatory} } &                   1.088  &0.240& 1.935 & \textbf{0.005}\\
        \textsc{mixed-less-aleatory} vs. \textsc{low-aleatory}&     0.250& -0.598 & 1.098 &0.927\\
        \textbf{\textsc{mixed-more-epistemic} vs. \textsc{low-aleatory} }             &  1.588  &0.740  &2.435 &\textbf{$<$0.0001}\\
        \textbf{\textsc{mixed} vs. \textsc{high-both} }      &                -0.938 &-1.785& -0.090 &\textbf{0.022}\\
        \textbf{\textsc{mixed-less-aleatory} vs. \textsc{high-both}}   &   -1.775 & -2.623 & -0.927 &\textbf{$<$0.0001} \\
        \textsc{mixed-more-epistemic} vs. \textsc{high-both}            &    -0.438 & -1.285 & 0.410& 0.615\\
        \textsc{mixed-less-aleatory} vs. \textsc{mixed}  &  -0.8375 & -1.685&  0.010 &0.0546\\
        \textsc{mixed-more-epistemic} vs. \textsc{mixed}          &      0.500 &-0.348&  1.348 &0.484\\
        \textbf{\textsc{mixed-more-epistemic} vs. \textsc{mixed-less-aleatory}}  &1.338 & 0.490 & 2.185 & \textbf{$<$0.001}\\
        \bottomrule
    \end{tabular}
    }
    \caption{Tukey HSD test result details for our perceived system predictability (PSP) scores between known groups, i.e. scenarios. Pairs with significant differences are highlighted in \textbf{bold} font.}
    \label{tab:known_groups_hsd}
\end{table}

\subsection{Confirmatory Factor Analysis}\label{sec:appendix_psp_cfa}
\Cref{tab:estimates_one_factor} and \Cref{tab:estimates_three_factor} show detailed parameter estimates of the one-factor and three-factor models.
\ifarxivmode\begin{table}[ht]\else\begin{table}[h]\fi
    \centering
    \resizebox{0.9\linewidth}{!}{
    \begin{tabular}{lllrrrrrr}
      \toprule
     \textbf{LHS} & \textbf{op} & \textbf{RHS} & \textbf{estimate} & \textbf{SE} & \textbf{z} & \textbf{p} & \textbf{CI-lower} & \textbf{CI-upper} \\ 
      \midrule
     predictability & =$\sim$ & EF1 & 1.00 & 0.00 &  &  & 1.00 & 1.00 \\ 
     predictability & =$\sim$ & EF2 & 0.96 & 0.05 & 19.62 & 0.00 & 0.86 & 1.06 \\ 
     predictability & =$\sim$ & EP1 & 1.04 & 0.05 & 19.95 & 0.00 & 0.94 & 1.14 \\ 
     predictability & =$\sim$ & EP2 & 1.01 & 0.05 & 22.28 & 0.00 & 0.92 & 1.10 \\ 
     predictability & =$\sim$ & AL1 & 0.98 & 0.05 & 18.41 & 0.00 & 0.88 & 1.09 \\ 
     predictability & =$\sim$ & AL2 & 1.00 & 0.05 & 21.58 & 0.00 & 0.91 & 1.09 \\ 
     EF1 & $\sim$$\sim$ & EF1 & 0.48 & 0.06 & 8.11 & 0.00 & 0.36 & 0.59 \\ 
     EF2 & $\sim$$\sim$ & EF2 & 0.60 & 0.07 & 8.62 & 0.00 & 0.46 & 0.73 \\ 
     EP1 & $\sim$$\sim$ & EP1 & 0.66 & 0.08 & 8.53 & 0.00 & 0.51 & 0.81 \\ 
     EP2 & $\sim$$\sim$ & EP2 & 0.40 & 0.05 & 7.67 & 0.00 & 0.29 & 0.50 \\ 
     AL1 & $\sim$$\sim$ & AL1 & 0.77 & 0.09 & 8.89 & 0.00 & 0.60 & 0.95 \\ 
     AL2 & $\sim$$\sim$ & AL2 & 0.44 & 0.06 & 7.99 & 0.00 & 0.33 & 0.55 \\ 
     predictability & $\sim$$\sim$ & predictability & 2.28 & 0.27 & 8.34 & 0.00 & 1.75 & 2.82 \\ 
       \bottomrule
    \end{tabular}
    }
    \caption{Detailed parameter estimates of the one-factor model.}
    \label{tab:estimates_one_factor}
\end{table}

\ifarxivmode\begin{table}[ht]\else\begin{table}[h]\fi
    \centering
    \resizebox{0.9\linewidth}{!}{
    \begin{tabular}{lllrrrrrr}
      \toprule
     \textbf{LHS} & \textbf{op} & \textbf{RHS} & \textbf{estimate} & \textbf{SE} & \textbf{z} & \textbf{p} & \textbf{CI-lower} & \textbf{CI-upper} \\ 
      \midrule
      effective & =$\sim$ & EF1 & 1.00 & 0.00 &  &  & 1.00 & 1.00 \\ 
      effective & =$\sim$ & EF2 & 0.96 & 0.05 & 18.99 & 0.00 & 0.86 & 1.06 \\ 
      epistemic & =$\sim$ & EP1 & 1.00 & 0.00 &  &  & 1.00 & 1.00 \\ 
      epistemic & =$\sim$ & EP2 & 0.97 & 0.05 & 21.34 & 0.00 & 0.88 & 1.06 \\ 
      aleatory & =$\sim$ & AL1 & 1.00 & 0.00 &  &  & 1.00 & 1.00 \\ 
      aleatory & =$\sim$ & AL2 & 1.02 & 0.05 & 19.11 & 0.00 & 0.92 & 1.13 \\ 
      predictability & =$\sim$ & epistemic & 1.00 & 0.00 &  &  & 1.00 & 1.00 \\ 
      predictability & =$\sim$ & aleatory & 0.94 & 0.06 & 16.83 & 0.00 & 0.83 & 1.05 \\ 
      predictability & =$\sim$ & effective & 0.98 & 0.05 & 19.70 & 0.00 & 0.88 & 1.08 \\ 
      EF1 & $\sim$$\sim$ & EF1 & 0.52 & 0.07 & 7.49 & 0.00 & 0.39 & 0.66 \\ 
      EF2 & $\sim$$\sim$ & EF2 & 0.64 & 0.08 & 8.31 & 0.00 & 0.49 & 0.79 \\ 
      EP1 & $\sim$$\sim$ & EP1 & 0.63 & 0.08 & 7.96 & 0.00 & 0.47 & 0.78 \\ 
      EP2 & $\sim$$\sim$ & EP2 & 0.36 & 0.06 & 6.28 & 0.00 & 0.25 & 0.48 \\ 
      AL1 & $\sim$$\sim$ & AL1 & 0.75 & 0.09 & 8.37 & 0.00 & 0.58 & 0.93 \\ 
      AL2 & $\sim$$\sim$ & AL2 & 0.39 & 0.06 & 6.05 & 0.00 & 0.26 & 0.52 \\ 
      effective & $\sim$$\sim$ & effective & -0.09 & 0.05 & -1.89 & 0.06 & -0.19 & 0.00 \\ 
      epistemic & $\sim$$\sim$ & epistemic & 0.07 & 0.05 & 1.28 & 0.20 & -0.04 & 0.17 \\ 
      aleatory & $\sim$$\sim$ & aleatory & 0.06 & 0.05 & 1.14 & 0.26 & -0.04 & 0.16 \\ 
      predictability & $\sim$$\sim$ & predictability & 2.43 & 0.31 & 7.88 & 0.00 & 1.83 & 3.03 \\ 
       \bottomrule
    \end{tabular}
    }
    \caption{Detailed parameter estimates of the three-factor model.}
    \label{tab:estimates_three_factor}
\end{table}

\section{Sentiment Classifier Experiments}\label{sec:appendix_psp_explanation_experiments}

\Cref{fig:screenshot_study_interface_understanding_saliencies} shows the full list of system prediction examples we showed to users.
While \Cref{fig:screenshot_study_interface_understanding_saliencies} shows examples of the heatmap conditions, we respectively use bar charts or no explanations in the other conditions.

 \begin{figure}
    \centering
    \arxivboxedincludegraphics{\textwidth}{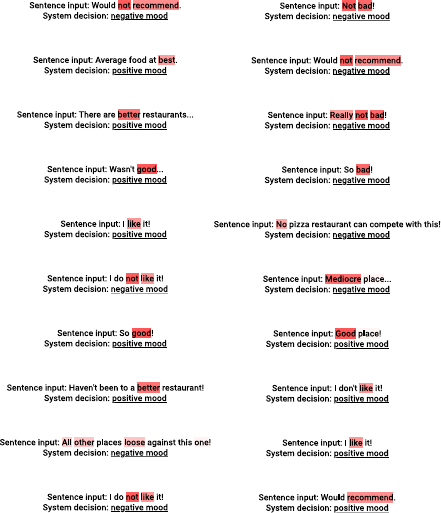}
    \caption{System predictions shown to users. Figure showing examples in the heatmap conditions, sentences were equal across conditions. Sentence order is randomized across participants.}
    \label{fig:screenshot_study_interface_understanding_saliencies}
\end{figure}

\FloatBarrier
The sentences we asked users to predict the system's decision for are (we provide the system predictions in parentheses and use \textit{italics} to highlight wrong model decisions):
\begin{itemize}
    \item I love the food at this place! (pos)
    \item Absolutely sensational! (pos)
    \item Quite nice! (pos)
    \item Tasty food! (pos)
    \item Super good place! (pos)
    \item Would not go there again. (neg)
    \item Pretty bad place. (neg)
    \item The food made me sick... (neg)
    \item Quite bad. (neg)
    \item Do not eat there! (neg)
    \item I don't like this restaurant very much! (\textit{pos})
    \item Wouldn't recommend. (\textit{pos})
    \item The water was the best part of the meal... (\textit{pos})
    \item Nice ads but didn't hold up the high expectations. (\textit{pos})
    \item I expected it to be better. (\textit{pos})
    \item Have not expected such a good place! (\textit{neg})
    \item I was so sad when I heard that they will close! (\textit{neg})
    \item I have not eaten at a better restaurant! (\textit{neg})
    \item Not too bad at all! (\textit{neg})
    \item Have not expected such a good place! (\textit{neg})
\end{itemize}
\FloatBarrier

\begin{table}
    \centering
    \resizebox{0.8\linewidth}{!}{
    \begin{tabular}{rrrrr}
      \toprule
     & \textbf{Estimate} & \textbf{Std. Error} & \textbf{t value} & \textbf{Pr($>$$|$t$|$)} \\ 
      \midrule
    (Intercept) & 4.78 & 0.42 & 11.41 & 0.00 \\ 
      bar charts & -0.08 & 0.10 & -0.78 & 0.44 \\ 
      saliency & 0.22 & 0.11 & 1.90 & 0.06 \\ 
      interactivity & 0.16 & 0.12 & 1.34 & 0.18 \\ 
      female & -0.18 & 0.42 & -0.43 & 0.67 \\ 
      male & -0.13 & 0.42 & -0.32 & 0.75 \\ 
      noise level L & 0.06 & 0.08 & 0.77 & 0.44 \\ 
      noise level Q & -0.04 & 0.08 & -0.44 & 0.66 \\ 
      bar charts : interactivity & 0.09 & 0.17 & 0.55 & 0.58 \\ 
      saliency : interactivity & -0.32 & 0.18 & -1.84 & 0.07 \\ 
       \bottomrule
    \end{tabular}
    }
    \caption{Parametric terms details for our model of PSP scores.}
        \label{tab:gam_results_psp_smooth_parametric_details}
\end{table}

\begin{table}
\centering
    \resizebox{0.8\linewidth}{!}{
    \begin{tabular}{rrrrr}
      \toprule
     & \textbf{Estimate} & \textbf{Std. Error} & \textbf{t value} & \textbf{Pr($>$$|$t$|$)} \\ 
      \midrule
    (Intercept) & 0.56 & 0.11 & 5.12 & 0.00 \\ 
      bar charts & 0.04 & 0.03 & 1.32 & 0.19 \\ 
      saliency & 0.05 & 0.03 & 1.53 & 0.13 \\ 
      interactivity & 0.02 & 0.03 & 0.74 & 0.46 \\ 
      female & 0.10 & 0.11 & 0.88 & 0.38 \\ 
      male & 0.11 & 0.11 & 0.97 & 0.33 \\ 
      noise level L & -0.12 & 0.02 & -5.88 & 0.00 \\ 
      noise level Q & 0.05 & 0.02 & 2.57 & 0.01 \\ 
      bar charts : interactivity & -0.04 & 0.04 & -0.96 & 0.34 \\ 
      saliency : interactivity & 0.01 & 0.05 & 0.29 & 0.77 \\ 
       \bottomrule
    \end{tabular}
    }
    \caption{Parametric terms details for our model of prediction correctness scores.}
        \label{tab:gam_results_correctness_smooth_details}
\end{table}

\newpage

\begin{table}
    \centering
    \resizebox{0.625\linewidth}{!}{
    \begin{tabular}{lrrr}
      \toprule
       & \textbf{df} & \textbf{F} & \textbf{p} \\ 
      \midrule
      \textbf{explanation format} & 2.00 & 5.15 & \textbf{0.01} \\
      interactivity & 1.00 & 0.12 & 0.73 \\ 
      \textbf{explanation format:interactivity} & 2.00 & 7.44 & \textbf{0.00} \\
      noise level & 2.00 & 2.66 & 0.07 \\ 
      identification & 2.00 & 0.11 & 0.90 \\ 
      \bottomrule
    \end{tabular}
    }
    \caption{Wald tests for the parametric terms in our model of FOST trust scores.}
    \label{tab:gam_results_trust_paramtric}
\end{table}

\begin{table}
    \centering
    \resizebox{0.65\linewidth}{!}{
    \begin{tabular}{rrrrr}
      \toprule
         & \textbf{edf} & \textbf{Ref.df} & \textbf{F} & \textbf{p} \\ 
          \midrule
          \textbf{s(prediction correctness)}  & 2.58 & 9.00 & 3.38 & \textbf{0.00} \\ 
          \textbf{s(PSP)} & 0.98 & 9.00 & 5.41 & \textbf{0.00} \\ 
          \textbf{s(completion time)} & 0.84 & 9.00 & 0.57 & \textbf{0.01} \\  
          s(SIPA) & 1.01 & 9.00 & 0.18 & 0.18 \\ 
          \textbf{s(NFC)} & 2.32 & 9.00 & 1.38 & \textbf{0.00} \\ 
          s(age) & 0.50 & 9.00 & 0.11 & 0.16 \\
       \bottomrule
    \end{tabular}
    }
    \caption{Wald tests for the smooth terms in our model of FOST trust scores.}
    \label{tab:gam_results_trust_smooth}
\end{table}

\end{document}